\documentclass[prb,superscriptaddress,amsfonts,amsmath,amssymb,showpacs,floatfix,reprint]{revtex4-1}

\usepackage[english]{babel}
\usepackage{graphicx}					
\usepackage[pdftex]{epsfig}
\usepackage{ragged2e}
\usepackage[caption=false]{subfig}
\usepackage{amsmath,amssymb}
\usepackage{times}
\usepackage{color}
\usepackage[table,xcdraw]{xcolor}
\usepackage{epstopdf}
\usepackage{physics}
\usepackage{soul}
\usepackage[export]{adjustbox}
\usepackage{dsfont}
\usepackage{float}
\usepackage[version=4]{mhchem}
\usepackage{siunitx}
\usepackage{mathtools}

\usepackage{todonotes}
\newcommand{\be}{\begin{equation}}
\newcommand{\ee}{\end{equation}}
\newcommand{\bee}{\begin{equation*}}
\newcommand{\eee}{\end{equation*}}

\newcommand{\id}{\mathds{1}}
\newcommand{\rr}{\mathbb{R}}
\newcommand{\bqp}{{\textsf{BQP}}}

\usepackage{amsthm}
  \newtheorem{theorem}{Theorem}
\definecolor{cadmiumgreen}{rgb}{0.0, 0.42, 0.24}

\usepackage{dsfont}

\renewcommand{\i}{\ensuremath\mathrm{i}}

\DeclareMathAlphabet{\mathpzcc}{OT1}{pzc}{m}{it}
\DeclareMathAlphabet{\mathpzc}{T1}{pzc}{m}{it}{\huge}
\DeclareMathAlphabet{\mathpzg}{T1}{pzc}{m}{n}{\huge}

\begin{document}

\title{Dynamical structure factors of dynamical quantum simulators}
\author{M.~L.\ Baez}
\email{Corresponding author: baez@pks.mpg.de}
\affiliation{Max Planck Institute for the Physics of Complex Systems, Dresden, Germany}
\affiliation{Dahlem Center for Complex Quantum Systems, Freie Universit\"{a}t Berlin, Germany}
\author{M.\ Goihl}
\affiliation{Dahlem Center for Complex Quantum Systems, Freie Universit\"{a}t Berlin, Germany}
\author{J.\ Haferkamp}
\affiliation{Dahlem Center for Complex Quantum Systems, Freie Universit\"{a}t Berlin, Germany}
\author{J.\ Bermejo-Vega}
\affiliation{University of Granada, Av. Fuentenueva s/n. 18071, Granada, Spain}
\author{M.\ Gluza}
\affiliation{Dahlem Center for Complex Quantum Systems, Freie Universit\"{a}t Berlin, Germany}
\author{J.\ Eisert}
\affiliation{Dahlem Center for Complex Quantum Systems, Freie Universit\"{a}t Berlin, Germany}
\affiliation{Helmholtz-Zentrum Berlin f{\"u}r Materialien und Energie,  Germany}

\date{\today}

\begin{abstract}
The dynamical structure factor is one of the experimental quantities crucial in scrutinizing the validity of the microscopic description of strongly correlated systems. 
However, despite its long-standing importance, it is exceedingly difficult in generic cases to numerically calculate it, ensuring that the necessary approximations involved yield a correct result. 
Acknowledging this practical difficulty, we discuss in what way results on the hardness of classically 
tracking time evolution under local Hamiltonians are precisely inherited by dynamical structure factors;
and hence offer in the same way the potential computational capabilities that dynamical quantum simulators do:
We argue that practically accessible variants of the dynamical structure factors are $\mathsf{BQP}$-hard for general 
local Hamiltonians.
Complementing these conceptual insights, we improve upon a novel, readily available, measurement setup allowing for the determination of the dynamical structure factor in different architectures, including arrays of ultra-cold atoms, trapped ions, Rydberg atoms, and superconducting qubits. 
Our results suggest that quantum simulations employing near-term noisy intermediate scale quantum devices should allow for the observation of  features of dynamical structure factors of correlated quantum matter in the presence of 
experimental imperfections,
for larger system sizes than what is achievable by classical simulation.
  \end{abstract}

\maketitle
\section{Introduction}

The field of condensed matter physics has seen a lot of successes aided by powerful computational tools.
Classical algorithms such as Monte Carlo techniques \cite{foulkes01}, exact diagonalization \cite{noack05}, tensor networks \cite{orus18} and more, have offered some of the greatest insights into the most surprising behaviour of many different systems.
While current numerical techniques are still extremely useful, in many cases the system sizes need to be constrained to a couple dozen atomic sites to obtain an efficient simulation, or the algorithms are just efficient for a narrow class of models. 
This arises from the fact that each one of these physical problems can be connected to a computational problem which belong to a (in many cases) well determined complexity class \cite{aaronson05}. 

Despite the field slowly pushing 
the boundaries of what is possible, the complexity boundary cannot be surpassed with classical algorithms. 
As long as the resource is a classical simulation, and considering certain assumptions believed to be true in the field of complexity theory~\cite{aaronson_bqp_2009,separation}, we know how far we can go.
For example, in higher dimensional frustrated quantum magnets or high-$T_c$ superconductors, we have no generic efficient way of calculating some of the most important quantum expectation values needed to understand the properties of a particular phase of interest.
For example, quantum Monte Carlo is a powerful method, but is affected by strong sign problems for frustrated and fermionic systems \cite{Curing1,Curing2, Hangleiter19}. 
Exact diagonalization can yield a plethora of useful results for many different physical systems, 
but the computational resources required scale exponentially in the system size. Other more sophisticated methods such as MPS, PEPS, MERA, etc.~are 
efficient 
for one dimensional short-range systems, but these methods are constrained by the amount of entanglement present in the system.

In this work, we propose dynamical analogue quantum simulators \cite{CiracZollerSimulation,jens15} as an
alternative method to simulate low energy excitations of strongly correlated
matter. In particular we suggest that \emph{dynamical structure factors}, which provide
key physical insights into quantum matter, can be accessed with
quantum simulators, while at the same time is a quantity which is significantly less accessible
with classical computers.

Large scale analogue quantum simulation platforms are unique systems in that they show exceptionally strong quantum effects and allow for measuring expectation values of microscopic observables \cite{Fukuhara2013,fukuhara13-1,cheneau12,simon11,Bernien17, Zhang17,Bohnet16,islam11}.
Among other platforms the propagation of excitations in XXZ models \cite{Fukuhara2013,fukuhara13-1}, Lieb-Robinson bounds \cite{cheneau12}, relaxation dynamics \cite{Trotzky12}, and phase diagrams of Fermi-Hubbard models \cite{simon11} have been  probed with ultra-cold atoms beyond capabilities of current classical algorithms.
At the same time, 
quantum simulations with trapped ions and Rydberg arrays have also seen several breakthroughs. As for example, 
the quantum dynamics of the long range transverse field Ising model, which have
recently been studied in systems of over fifty atoms via time dependent expectation values of single spin observables \cite{Bernien17, Zhang17,Bohnet16,islam11}.
Though a great body of observations has been assembled, a particular question arises: 
{\it Can quantum simulators provide qualitative dynamical quantities of systems
relevant in the condensed matter context,  for which there is evidence that in the regime discussed they are
inaccessible to classical algorithms?}

\begin{figure}[h!]
\centering
\includegraphics[scale=0.4]{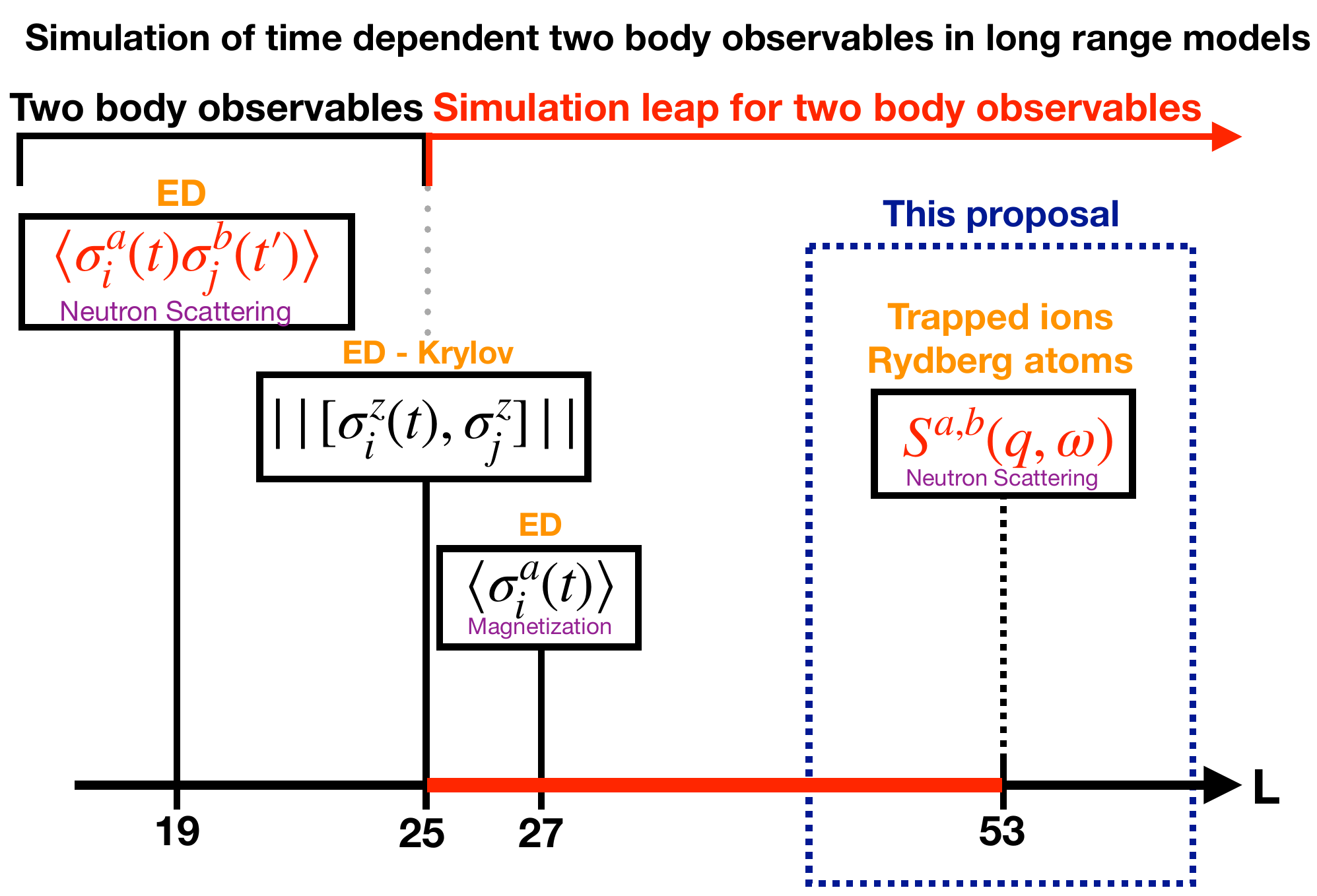}
\caption{
{\it  State of the art exact numerical algorithms to time evolve two body observables for the long range TFIM}. Larger system sizes which can access unequal time correlators (and by extension DSFs). In orange we show the methods that have been employed to calculate the different quantities. In purple, the experimental techniques that are used to measure those same quantities in material experiments. At 19 sites, the Pauli operators at different sites have been time evolved via exact diagonalization (ED) \cite{liu19}. Using Krylov space methods, the system size has been extended to 25 \cite{luitz19}. System sizes up to 27 (ED) and 128 (Lanczos and t-DRMG) sites have been obtained, but only single site observables have been accessed \cite{fratus16}. 
Here we show that our proposal offers a leap forward in terms of the system sizes that can be employed to study DSFs of long range models via quantum simulation. 
Please note that using variational methods, entanglement entropies up to 125 sites can be obtained \cite{hauke13}. 
}
\label{summary}
\end{figure}

We propose an answer to this question in form of the \emph{dynamical structure factor (DSF)}, 
a widely attainable experimental observable which gives information regarding  dynamical properties of a given system. 
In materials it is experimentally measured by inelastic neutron scattering \cite{Coldea10} and resonant inelastic X-ray scattering \cite{jia19}. 
Given the relative ease of measuring the DSF experimentally, 
an efficient way to simulate this quantity becomes imperative. 
We argue that the DSF can be accurately accessed with quantum simulators within the experimental level of accuracy currently available in the different architectures, and for system sizes beyond what current classical algorithms can achieve, as we show in Fig.~\ref{summary}.

The DSF is a quantity which can be considered stable to small perturbations of the microscopic model whose excitations it probes, given that the qualitative features of the DSF already provide a lot of information regarding those excitations.
 In this sense, we expect to see an inherent robustness in the DSF, finding that observing the signatures of low energy excitations is possible with state of the art setups in the presence of moderate experimental imperfections.
As a proof of principle we investigate the short and long range transverse field Ising model (TFIM).
The short range model is integrable \cite{sachdev11}, and allows us to study relatively big system sizes comparable to those achievable in trapped ions and Rydberg atoms simulators. 
We first study in detail the effects of experimental imperfections in the short range model, and the associated Fourier transform involved in the calculation of the DSF to give us an intuition of those effects. 
Once the short range model is well understood, we move to our application proposal.
The classical numerical calculation of unequal time correlation functions in long range
systems is constrained to system sizes much smaller than what current quantum
simulators can achieve\cite{Bernien17,Zhang17}. Thus, we propose {\it the measurement of the DSF for the long range transverse field Ising model as a practical application of quantum simulators in a quantity relevant for both condensed matter and material science}.

We study the long range transverse field Ising model under the same imperfections as for the short range model.
We show that the experimental imperfections currently present in quantum simulators, do not affect the DSF in a significant way, and that the scaling of these errors in the DSF is well controlled in the full range of system sizes studied here.

We also study the computational hardness of evaluating the DSF for general systems.
We find that the DSF can be likened to a \bqp-hard problem, 
meaning that any classical algorithm calculating it for general Hamiltonians efficiently would also efficiently solve all the tasks that a quantum computer can tackle efficiently.
The latter is regarded in the quantum computing community as a highly unlikely scenario. 
As such, realizing our proposal in practice would tackle a task hard for classical computers in a field of practical importance in condensed matter physics. 
While the specific proposal of this work is centred on a specific model, it is worth pointing out that the proof of hardness is valid for a wide range of Hamiltonians.
 It is our aim in this work to highlight a specific case in which the DSF can be experimentally achieved in the near term, but the protocol employed here, together with the error analysis and the study of the different architectures can be easily applied to other models, as for example the XY model in superconducting chips \cite{Roushan1175} or Rydberg atoms \cite{barredo15}.
As such, future advances in the field, where analogue quantum simulators implement further models in higher dimensions can make use of the study performed in this work to show a practical application of quantum simulators through the DSF in those models.    

\section{DSF in quantum simulators} \label{implementation}
In order to employ quantum simulators to study the DSF of solid state systems, we want to probe the fluctuations of their ground states or thermal states via \emph{unequal time correlation functions}.
For a spin system with lattice sites $i,j\in \Lambda$ (where $\Lambda$ is the collection of lattice sites), these are defined by
\be
C^{a,b}_{i,j}(t) =  \langle \sigma^a_i(0)\sigma^b_j(t)\rangle\,,
\label{out_of_time_corr}
\ee 
 we denote Pauli matrices by $\sigma^{a}$ with $a=x,y,z$. 
The Fourier transform of these quantities from real-space sites $\mathbf{x}_i$ to momentum 
$\mathbf q\in \rr^3$ and time- to frequency-domain $\omega\in \rr$  yields the DSF
\def\i{\mathrm{i}}
\be
S^{a,b}(\mathbf{q}, \omega) = \frac{1}{N}\sum_{i,j \in \Lambda}\int_{-\infty}^\infty \text{d}t\ e^{-\i\mathbf{q}\cdot(\mathbf{x}_i-\mathbf{x}_j)}e^{\i\omega t} C^{a,b}_{i,j}(t)\,,
\label{DSF}
\ee
where $N$ is the number of lattice sites.
There has been a recent proposal \cite{knap13} on how to measure retarded Green's functions
(which are related to the DSF in equilibrium via the fluctuation-dissipation
theorem) in cold atoms and trapped ion devices using \emph{Ramsey spectroscopy}, 
however a clear understanding of the feasibility of observing important physical effects 
and the DSF itself, when the proposal of Ref.~\cite{knap13} is applied to a quantum many-body system is still lacking. 

\begin{figure}[t]
\centering
\includegraphics[scale=0.5]{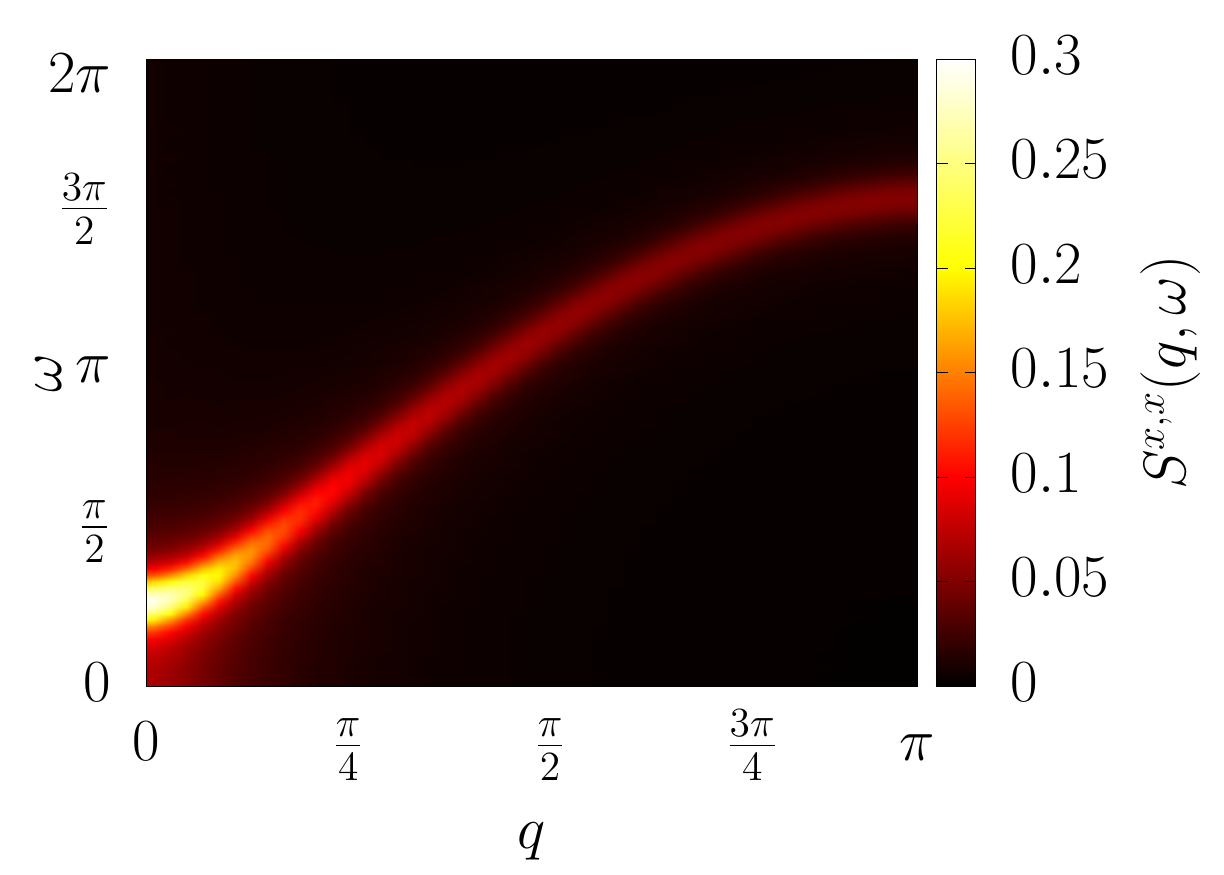}
\caption{{\it Dynamical structure factor for the transverse field Ising model}. We show the DSF away from the criticality, $J=1$ and $B=1.4$. We observe the gap around the $q=0$, $\omega = \pi/4$ point, and the $\omega$-dependent two particle continuum extending over the entire reciprocal space, in accordance with the exact solution of the transverse field Ising model \cite{sachdev11}.}
\label{PhT_dyn}
\end{figure}

In Fig.~\ref{PhT_dyn} we show a typical DSF for the transverse field Ising
model, one of the models we will study in detail in this work, away from
criticality. In the figure we observe a cosine shaped continuum, with a gap at
$q=0$. The goal of this work is to show that a DSF like the one in
Fig.~\ref{PhT_dyn} can be obtained from state of the art quantum simulations.
To obtain such a DSFs in quantum simulators, the crucial ingredient that needs to be supplemented beyond the existing techniques is a measurement protocol which gives access to unequal time correlation functions
as in
Eq.~(\ref{out_of_time_corr}). In the following, we propose a generalization of the  protocol proposed by \cite{knap13}, which can be employed in any setup where a single site spin rotation can be implemented. We extend this spectroscopy protocol via tomographic methods to systems which do not exhibit as many symmetries as Ref.~\cite{knap13} assumes. In this context, we offer a measurement protocol which can be implemented in many different architectures, as trapped ions, Rydberg atoms, and super-conducting qubit chips, and for a wide class of systems beyond Ising and XXZ as has previously been proposed \cite{knap13}. 
 
\subsection{DSF measurement protocol}

The DSF effectively probes low energy excitations of a given system, described by a particular Hamiltonian $H$. 
Given the definition of the DSF in \eqref{DSF}, the excitations to which it is sensitive are those related to observables of the form given in \eqref{out_of_time_corr}. The first step to obtain such a quantity then resides in the \emph{initialization} of the quantum simulator in a low energy state, ideally the ground state of $ H$. 
In this section we will assume, for the sake of simplicity, that the quantum simulator will be initialized in the unique 
ground state vector of $H$, which we refer to as  $\ket {\psi_0} $, though in principle, the protocol we employ can be used with any initial state be it in equilibrium or not, as exemplified in Ref.~\cite{yoshimura16} with the Ramsey spectroscopy technique.

Preparing such state can be achieved by adiabatic evolution.
 At the same time, the recently proposed \emph{quantum approximate optimization algorithms (QAOAs)} 
 can also been employed. These algorithms have recently been reported in trapped ions experiments \cite{pagano19}, achieving a very good approximation of the ground state of non-trivial Hamiltonians.
  It is worth pointing out that QAOAs have been shown to considerably reduce the experimental time required for ground state preparation in comparison to adiabatic evolutions in trapped ions, effectively extending the evolution times which can be achieved with this particular architecture.

\begin{figure}[t]
\centering
  \includegraphics[scale = 0.4]{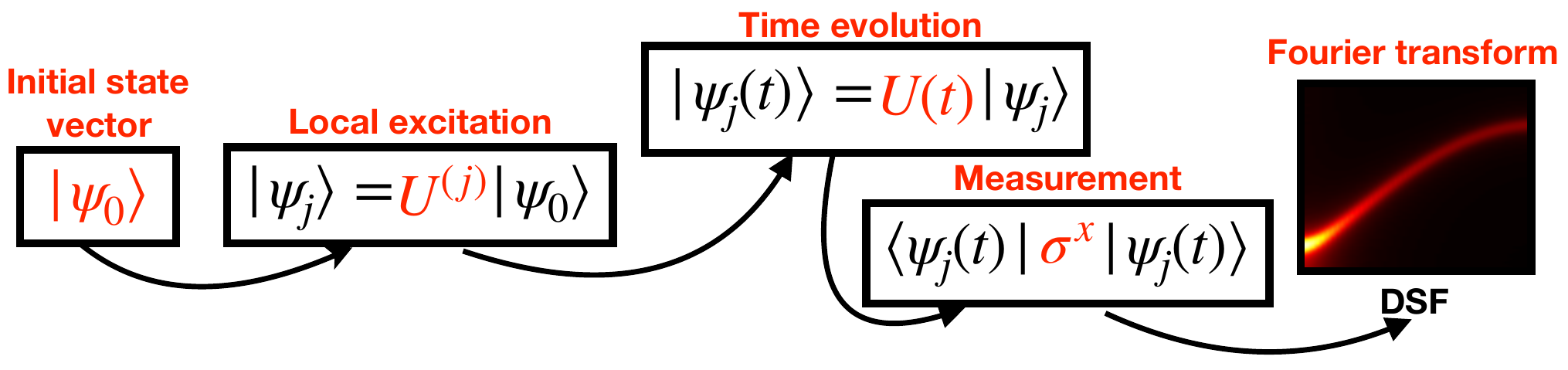}
\caption{{\it General scheme for measuring the DSF in quantum simulators:} {\it 1)} Prepare the initial state vector, which in this work is the ground state. {\it 2)} Excite it locally via a unitary single spin rotation at site $j$, $U^{(j)}$. {\it 3)} Evolve this state with the Hamiltonian H, $U(t)$. {\it 4)} Measure a local spin observable at a site different than $j$, which in our case is the local magnetization $\langle \sigma^x \rangle $. 
Processing such data via Fourier transforms allows us to obtain the DSF.
}
\label{protocol}
\end{figure}

Once the ground state is obtained, we then induce low energy excitations by applying a single \emph{spin rotation}. 
After exciting the system locally, the state is {\it evolved} with the Hamiltonian $H$. 
Finally, after the evolution, we \emph{measure} local spin operators with single-site resolution.
Once the unequal time correlators are measured, the DSF can be obtained via a spatial and temporal Fourier transform. We show this protocol in Fig.~\ref{protocol}.
\subsection{Measuring unequal time correlations}
Let us now discuss the crucial question at hand: How can we measure two point  \emph{unequal time} correlation functions if we can only perform unitary transformations and measure local spin operators?
The main insight of Ref.~\cite{knap13} (see also Ref.~\cite{yoshimura16} for a detailed study of the idea) has been
that the operator at initial time $\sigma^a_i(0)$ can be obtained as part of a unitary operation, the pulse of \emph{Ramsey interferometry}.

We begin by the basic, and at the same time most important, example of this idea:
Consider the unitary representing a $\frac \pi 4$-rotation of a spin at site $j\in \Lambda$ along of the $x$-axis
\be
U^{(j)}= \frac 1{\sqrt 2} ( \id -  \i\sigma_j^x)\ .
\label{rotation}
\ee
We would like to use it as an \emph{excitation} of a low-energy state vector $\ket {\psi_0}$ which then is probed by subsequent evolution $U(t)$ to time $t$ governed by the many-body Hamiltonian of the interacting system being investigated.
To keep the discussion simple let us assume that the expectation value of an odd number of spin operators vanishes for $\ket {\psi_0}$ (as is the case for the TFIM and arises from the symmetries of the Hamiltonian and initial state \cite{knap13}). 

Having these two ingredients at hand, we can consider the state vector
\begin{align}
  \ket{\psi} = U(t) U^{(j)} \ket{\psi_0}
\end{align}
which can be obtained by an appropriate \emph{unitary} single-qubit rotation $U^{(j)}$ that locally excites the system (as the one in \eqref{rotation}) and a subsequent time evolution of the system $U(t)$.
Observe that both operations are unitary and {thus} $ \ket {\psi} $ is a state vector.
If we measure the expectation value of $\sigma_i^x$ on this state we obtain
\begin{align}
 \bra\psi \sigma_i^x \ket \psi =&  
 \frac12 \bra{\psi_0}  \sigma_i^x(t)  \ket {\psi_0}
+  G_{x,x}^\text{ret}(i,j,t) + R(i,j,t)\ .
\label{full_exp}
\end{align}
with $R(i,j,t) = \frac12 \bra{\psi_0}\sigma_j^x   \sigma_i^x(t) \sigma_j^x \ket {\psi_0}$, and  $G_{x,x}^\text{ret}(i,j,t)$ the \emph{retarded Green function}
\begin{align}
  G_{x,x}^\text{ret}(i,j,t) = -\frac \i2 \langle\sigma_i^x(t) \sigma_j^x(0) - \sigma_j^x(0)\sigma_i^x(t) \rangle_0\ .
\end{align}
The first term in the last line of Eq.~(\ref{full_exp}) can be measured directly by simply omitting the excitation step and hence can be subtracted from the data if it is non-zero.
The last term, on the other hand, has a non-trivial unequal time dependence and hence must either vanish due to, 
e.g., symmetry arguments or has to be reconstructed.

The case considered in Ref.~\cite{knap13} is the one in which the Hamiltonian $\hat H$ has a unitary symmetry $\mathcal P$, such that the product of an odd number of Pauli operators vanish. From this it follows that $R(i,j,t)$ must vanish. As such, whenever a symmetry of this kind is present (as in the TFIM) we obtain the identity $\bra\psi \sigma_i^x \ket \psi  = G_{x,x}^{\text{ret}}(i,j,t)$.
Calculating this for all spin pairs $(i,j)$ we obtain the retarded Green function $G^{\mathrm{ret}}_{x,x}(i,j,t)$ and we can perform a Fourier transform in real space and time to obtain $G^{\mathrm{ret}}_{x,x}(\mathbf{q}, \omega)$. 
Finally, we can relate the retarded Green function, when linear response theory holds, to the dynamical structure factor via the fluctuation-dissipation theorem
\be 
S^{xx}(\mathbf{q},\omega) = - \frac{1}{\pi}[1 + n_B(\omega)]\mathrm{Im}[G^{\mathrm{ret}}_{x,x}(\mathbf{q}, \omega)]\,,
\label{eq:FDthm}
\ee
where $n_B(\omega) = 1/(e^{\omega/T} + 1)$.
This way, we get direct access to the dynamical structure factor by measuring the retarded Green's function via the above measurement protocol. 

There are two points which need to be made before we move on: First, while we study the zero temperature DSF, finite but small temperatures will broaden the features of the DSF but not change the overall behaviour, provided that $T$ is smaller than the smallest coupling of the model.
Second, note that the fluctuation-dissipation theorem holds when linear response theory is a good approximation, and its validity or lack of thereof away from equilibrium is a highly researched topic to the date \cite{jens15, baiesi09, non-equ-field-theory}.
As such, this measurement protocol for the DSF will be accurate when the system is close to thermal equilibrium in a practical sense.
\subsection{Tomographic recovery methods for unequal time correlation functions}
\label{app:DSF_tomography} 
If the symmetry argument can be relaxed, we can show how the term $R(i,j,t)$ can be extracted.
Let us define a modified Ramsey state vector which reads
\begin{align}
  \ket{\psi_\phi} = U(t) U^{(j)}(\phi) \ket{\psi_0}
\end{align}
where now we excite the ground state $\ket{\psi_0}$ with a $\phi$-rotation around the $x$-axis
\begin{align}
\begin{split}
   U^{(j)}(\phi)=e^{-\i \phi \sigma^x_j} &= \cos(\phi)\id -\i \sin(\phi) \sigma^x_j\\
   &=: c_\phi\id - \i s_\phi \sigma^x_j .
   \end{split}
\end{align}
For an analogous measurement to the case in the previous section we obtain
\begin{align}
\begin{split}
 \bra{\psi_\phi} \sigma_i^x \ket {\psi_\phi}=&
\, c^2_\phi \bra{\psi_0}  \sigma_i^x(t)  \ket {\psi_0}
+ 2 c_\phi s_\phi  G^\text{ret}_{x,x}(t)\\&
+  s_\phi^2 R(i,j,t)\ .
\end{split}
\label{tomo}
\end{align}
We now notice that we can directly measure the left hand side and the first term on last line of the expression above.
For a fixed angle $\phi$ we can write 
\begin{align}
 b_\phi  = \bra{\psi_\phi} \sigma_i^x \ket {\psi_\phi}-  c^2_\phi \bra{\psi_0}  \sigma_i^x(t)  \ket {\psi_0} \ .
\end{align}
Now, we can rewrite Eq.~(\ref{tomo}) as 
\begin{align}
  a_\phi^T v = b_\phi
\end{align}
where $v$ is the vector we want to reconstruct, given by 
\begin{align}
  v = [ G^\text{ret}_{x,x}(t),  R(i,j,t) ]^T
\end{align}
and $a_\phi = [ 2 s_\phi c_\phi, s_\phi^2 ]$.
If an experiment  measures $b_\phi$ using various angles $\phi$ then we can build a matrix $A$ using the different $a_\phi$'s as rows and in a corresponding fashion we can collect the measured $b_\phi$'s into a vector $b$.

The retarded Green's function can be reconstructed by noticing that 
\begin{align}
  v^\star = (A^TA)^{-1} A^T b
\end{align}
gives the value of $v$ that minimizes the least-square residue
\begin{align}
  \min_v \| Av - b\|_2\ .
\end{align}
Here we assume that one can choose the excitation angles $\phi$ in such a way that the matrix $A^TA$ is well conditioned as is done in typical tomographic schemes.
In order to measure the DSF this procedure must be performed  for all pairs of excitation and measurement positions $i,j\in \Lambda$ and the Fourier transform of the collection of reconstructed values $v_1^\star =  G^\text{ret}_{x,x}(i,j,t)$ will yield the DSF.
\section{On the computational complexity of the DSF}
\label{complexity}
Once we have formalized how dynamical quantum simulators can access the DSF, we will concentrate on answering the question {\it in what specific way is the calculation of the DSF a computationally hard problem?}
In the following, we formalize the statements about classical hardness and show that a
practically accessible variant of the dynamical structure factor is hard for the complexity class \bqp. 
To this end, we show that the building blocks of the DSF, the unequal time correlators $C^{a,b}_{i,j}(t)$ are \bqp-hard to compute. 
 
To start with, and without loss of generality, we show that $\left\langle \sigma_i^z(t)\sigma_j^z\right\rangle_{\psi}:=\langle \psi|\sigma_i^z(t)\sigma_j^z|\psi\rangle$ is \bqp-hard to compute for product state  vectors 
$|\psi\rangle$ and for ground states.
Then we use these observations to consider the DSF over a finite (but arbitrarily large) interval of time
\begin{equation}
S_{t_0,t_1}^{z,z}(q,\omega)=\frac{1}{N}\sum_{i,j}\int_{t_0}^{t_1}e^{-\mathrm{i}q(x_i-x_j)}e^{\mathrm{i}\omega t}\left\langle \sigma_i^z(t)\sigma_j^z\right\rangle_{\psi}\mathrm{d}t\ .
\label{disc_DSF}
\end{equation}
where $N$ is the system size. In particular, we prove the following.
\begin{theorem}[Hardness of computing the approximate dynamical structure factor]
	For $t_1-t_0=\mathrm{poly}(N)$, product states $|\psi\rangle$, and $2$-local Hamiltonians it is \bqp-hard to approximate $S^{z,z}_{t_0,t_1}(q,\omega)$ within an error $\varepsilon=\mathrm{poly'}^{-1}(N)$.
\end{theorem}
We consider the quantity $S^{z,z}_{t_0,t_1}$ instead of the full Fourier transform as it is the practically accessible one: Any time observation will necessarily be finite in practice. What is more, from a conceptual perspective, the latter is not even computable on a Turing machine due to arbitrarily large errors that are introduced by the Fourier transform: The continuous
Fourier transform is not Turing computable \cite{Turing}.
 
{\it Hardness for estimating correlators on ground states.}
For hardness of ground states, we observe that computing $C^{z,z}_{i,j}(t)=\langle \sigma_i^z(t)\sigma_j^z
\rangle_{\psi}$ for any $t$ is at least as hard as computing $C^{z,z}_{i,j}(0)=\langle \sigma_i^z\sigma_j^z
\rangle_{\psi}$.
 First, computing correlators up to constant additive errors on ground states of quasi-local Hamiltonians is \bqp-hard by the Feynman-Kitaev construction~\cite{kempe_complexity_2004}. Furthermore this remains true for several classes of local observables and local Hamiltonians, including one-local observables measured on ground states of nearest-neighbour two-local Hamiltonians on qubits~\cite{gosset_adiabatic_2015,lloyd_adiabatic_2016}, and two-local observables measured on ground states of translation invariant nearest-neighbour two-local Hamiltonians with local dimension three~\cite{ciani_hamiltonian_2018}.

{\it Hardness for out-of-time correlators.}
For the product states, we start with a general observation: 
Consider an arbitrary circuit $C_n=U_{n}\dots U_{1}$ consisting of $k$-local gates $U_i$. 
Evaluating the quantity $\left\langle \sigma_i^z(t)\sigma_j^z\right\rangle_{\psi}$ for product state vectors 
$|\psi\rangle$ within constant error is \bqp-hard.
Here, $0\leq t\leq n$ is an integer.
For $\Pr(1)$, the probability of measuring $1$, we obtain
\be
\Pr(1)=\langle \psi|C_t^{\dagger}\left(\frac{1+\sigma_i^z}{2}\right)C_t|\psi\rangle
= \frac12 \pm \frac12\left\langle \sigma_i^z(t)\sigma_j^z\right\rangle_{\psi}\ .\notag
\ee
Here, $|\psi\rangle$ is assumed to be in the $\sigma^z$-eigenbasis.
The sign in the above calculation can be immediately obtained from $|\psi\rangle$. 
Computing the above probability within a constant additive error 
suffices to yield a valid reduction to the output probabilities of quantum circuits.
We are interested in the case where the circuit $C_t$ is given by the time evolution $e^{\mathrm{i} t H}$ for some Hamiltonian $H$.

The definition of the DSF is given for continuous time (Eq.~(\ref{DSF})), but
quantum simulators (and also classical simulations) need to discretize time,
as the measurement protocols proposed cannot continuously measure
$C^{z,z}_{i,j}(t)$, but require a fresh preparation for each point in time.
In the following, we show that while this discretization leads to errors, they are bounded.

{\it The discretization error.}
Notice that there will always be an error from the discretization of time.
However, this can be bounded: For any differentiable function $f$ we can use the mean-value theorem to obtain
\begin{equation}
|f(t+\delta t)-f(t)|\leq \left|\max_{t'\in[t,t+\delta t]}\partial f(t')\right|\delta t.
\end{equation}
For $C^{z,z}_{i,j}(t)=\left\langle \sigma_i^z(t)\sigma_j^z\right\rangle_{\psi}$, we have
\begin{align}
\begin{split}
\left|\partial_t C_{i,j}(t)\right|=&\left|\left\langle\psi\left|\partial_t \left(\sigma_i^z(t)\sigma_j^z\right)\right|\psi\right\rangle\right|\\
=&\left|\left\langle\psi\left|\partial_t \left(e^{\mathrm{i} t H}\sigma_i^ze^{-\mathrm{i} t H}\sigma_j^z\right)\right|\psi\right\rangle\right|\\
=&|\mathrm{i}\left\langle\psi\left|e^{\mathrm{i}tH}[H,\sigma_i^z]e^{-\mathrm{i}tH}\sigma_j^z\right|\psi\right\rangle|
\leq  L'=\mathrm{const}~,
\end{split}
\end{align}
where we use the fact that we assume $H$ to be a (geometrically) local Hamiltonian and $L'$ is the Lipshitz constant.
Thus, $H=\sum_{i=1}^r h_i$ with $r=\mathrm{poly}(N)$ and $||h_i||_{\infty}\in \mathcal{O}(1)$ and furthermore, $\sigma_i^{z}$ commutes with all but constantly many summands $h_j$.
The inequality thus follows from the triangle inequality and the submultiplicativity of the operator norm.
It hence suffices to choose a constantly small discretization step to bound this error. 
In particular, this proves that $C^{z,z}_{i,j}(t)$ is Lipshitz continuous with size-independent Lipshitz constant.

{\it Hardness for a variant of the dynamical structure factor.}
The \textit{discrete dynamical structure factor} is defined as
\be
\tilde{S}^{z,z}(q,\omega)=\frac{1}{N}\sum_{i,j}\sum_{k=1}^M e^{-\mathrm{i} q(x_i-x_j)}e^{\mathrm{i}\omega (t_0+k\Delta  t)}
\left\langle \sigma_i^z(t_0+k\Delta t)\sigma_j^z\right\rangle
\ee
with $\Delta t=({t_1-t_0})/{M}$. 
Notice that this is the quantity that is usually approximated in numerical simulations.
Computing the discrete Fourier-transform can be done via the \textit{fast Fourier transform}, which runs in time $\mathcal{O}(\ln(M)M)$ for $M=\mathrm{poly}(N)$.
Hence if the correlators are \bqp-hard, the discrete dynamical structure factor is as well.

We can bound the error on the continuous dynamical structure factor as well if only a finite interval of time is involved.
We know that $C^{z,z}_{i,j}(t)=\left\langle \sigma_i^z(t)\sigma_j^z\right\rangle_{\psi}$ is a function with polynomially bounded Lipshitz constant.
For a bounded interval of time $[t_0,t_1]$, we consider the error that occurs by approximating the integral in Eq.\ref{disc_DSF}
with step functions
\be
S_{t_0,t_1}^{z,z}(q,\omega)\approx\frac{1}{N}\sum_{i,j}\sum_{k=1}^Me^{-\mathrm{i}q(x_i-x_j)}e^{\mathrm{i}\omega t_0+k\Delta t}
C^{z,z}_{i,j}(t_0+k\Delta t)\Delta t,
\ee
where $	\Delta t=t_1-t_0/M$. Integrating over the error made by the step function approximation gives us the cumulated error$
(t_1-t_0) L' \Delta t=(t_1-t_0)^2 L'/M,$
where $L'$ is the Lipshitz constant of the function $e^{-\mathrm{i}q(x_i-x_j)}e^{\mathrm{i}\omega t}C^{z,z}_{i,j}(t)$.
Hence, choosing $M$ to be constant and small suffices for an approximation within arbitrarily small constant error.
In essence, we have proven that $C^{a,b}_{i,j}(t)$ and $S^{a,b}(q,\omega)$ are \bqp-hard in a specific sense.
 Furthermore, since simulations both classical and quantum require a discretization of the time axis, we have shown that the possible errors from this are well behaved and controlled.

\bqp-hardness provides evidence against the existence of classical algorithms that compute dynamical structure factors in polynomial time.
However, it is important to point out that this is a 
so-called worst-case result, i.\,e.\,it only rules out an algorithm that solves \emph{all} cases in polynomial time. 
In general, subclasses of this problem are not necessarily hard in the complexity theoretic sense. 
For example, the time evolution of the nearest neighbour, short range, transverse Ising model is not expected to be universal for time evolution. 
\begin{figure*}
\centering
\includegraphics[scale=0.4]{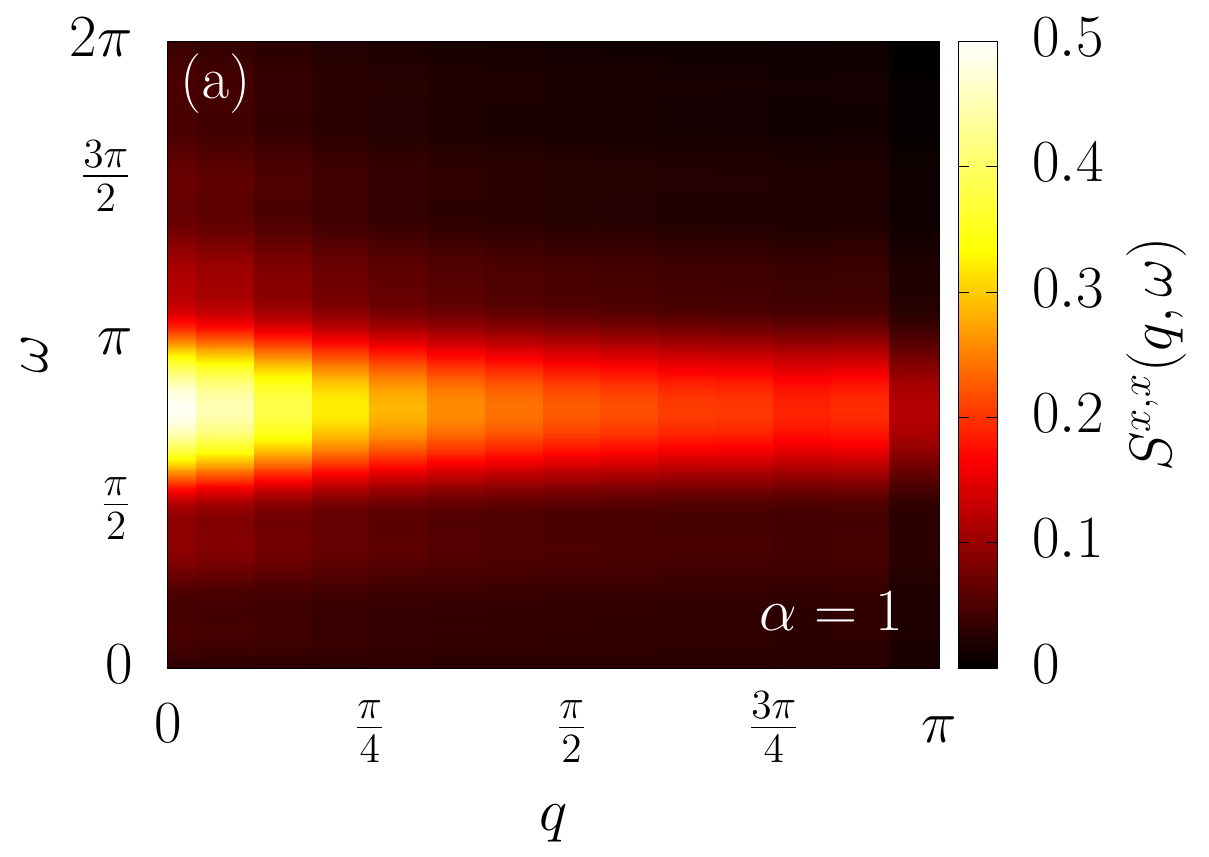}
\includegraphics[scale=0.4]{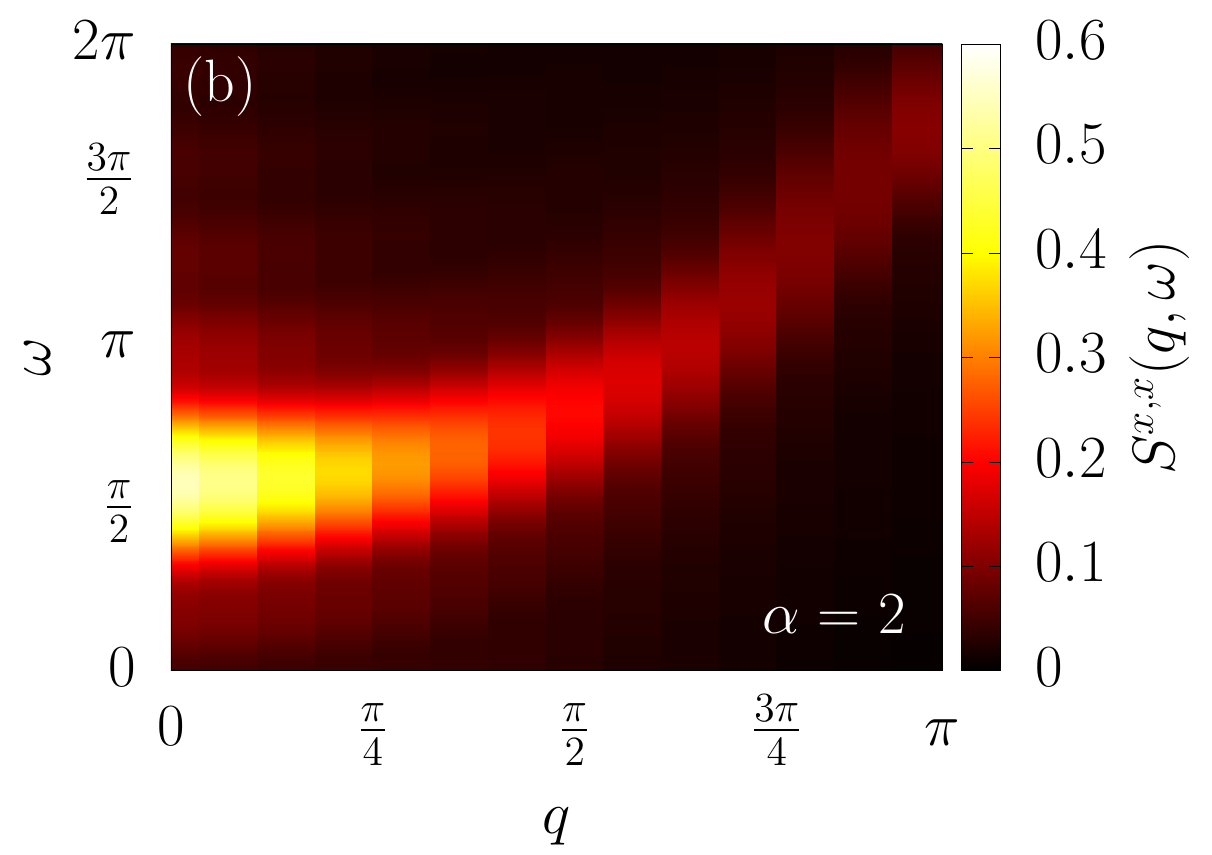}
\includegraphics[scale=0.4]{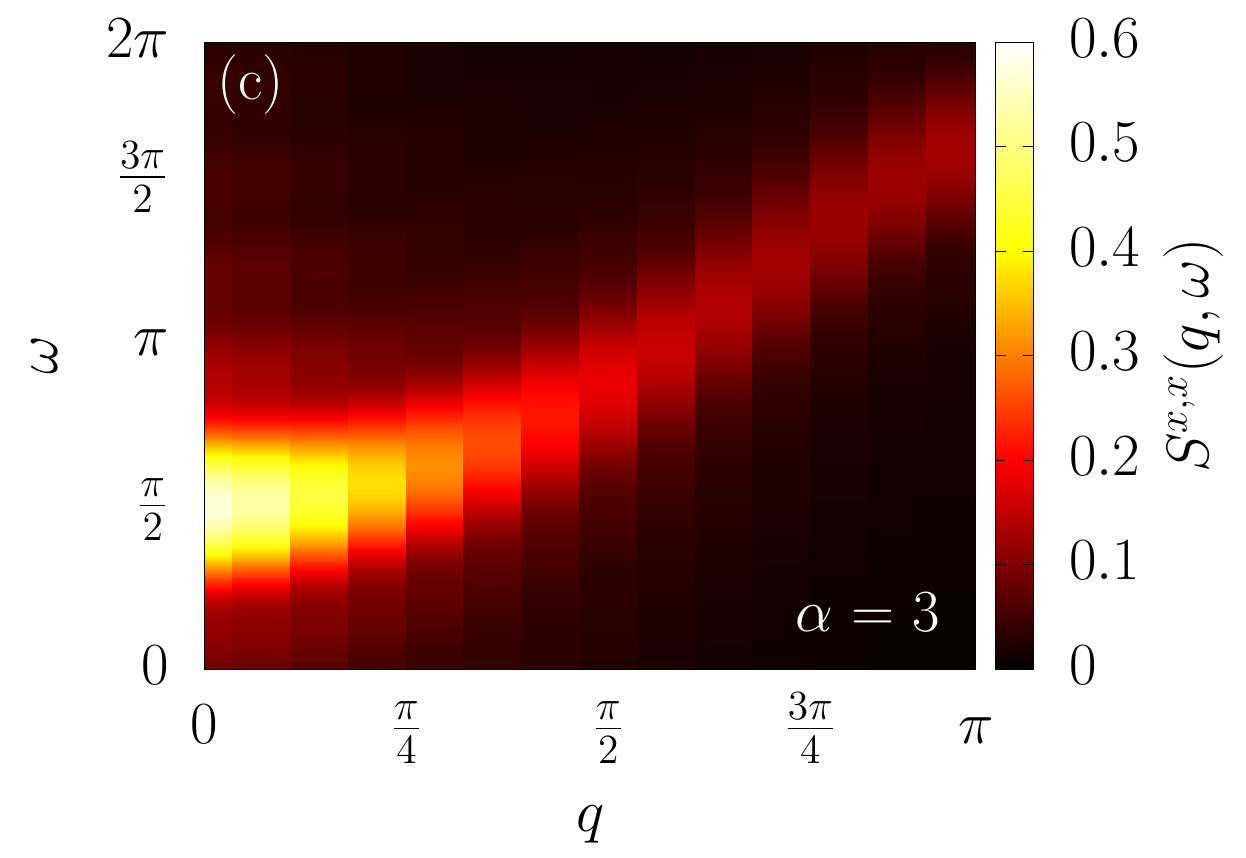}
\caption{{\it Dynamical structure factor for the long range transverse field Ising model, for the cases $\alpha = 1$ (a), $2$ (b), and $3$(c)}. For the case $\alpha = 1$ we see that the DSF does not vary in frequency, which is a possible signature of excitation confinement, in accordance with Refs.~\cite{liu19, lerose19}. For $\alpha > 2$ the two particle continuum is noticeable, and the gap is lowered. As the value of $\alpha$ is increased, the interactions become shorter range, and the gap approaches the value for the short range model. Accordingly, the continuum changes shape, from the absence of $\omega$ dependence for $\alpha = 1$ towards the cosine form at $\alpha = 3$. This cosine shape corresponds to the short range TFIM, obtained in the limit $\alpha \rightarrow \infty$. For comparison with the short range TFIM please refer to Fig.~\ref{PhT_dyn}.  }
\label{DSF_long_range_main}
\end{figure*}
 
 \section{Practical realization of DSFs in quantum simulators}
\label{realization}
As mentioned in the previous section, several near term quantum architectures can be employed to simulate DSFs.
So far we concentrated on how the previously mentioned measurement protocol can be employed to obtain DSFs, and on the complexity of this task.
To assess the degree of robustness of DSFs against experimental imperfections,  we will study the short and long range transverse fields Ising models (TFIM), in the presence of those imperfections.

The translational invariant 1D-TFIM is defined as
\be
H(J,B) = \sum_{i\in\Lambda} B_i \sigma^{z}_i-\sum_{i,j\in\Lambda}J_{i,j}\sigma^x_i\sigma^x_j\,.
\label{Ry_Ising}
\ee
%
The coupling parameters of the Ising term are $J_{i,j}$, and in principle can be site dependent. The strength of the magnetic field is given by $B_i$ and in this work we will consider it uniform throughout the chain, $B_i = B$. 
The spin-spin interaction can take the long range form $J_{i,j} = J/|i-j|^\alpha$ 
for analog quantum simulations in Rydberg arrays or trapped ions, where typically $\alpha\in[1,6]$ (see section I in the supplementary information).
In the case of digital simulation and optical lattices, one can study the short-range model \cite{simon11} with $J_{i,j} = J \delta_{i,j\pm1},$ which is exactly solvable by a mapping to non-interacting fermions \cite{sachdev11}.

While our proposal is focused on the long range model, the access to the DSF via quantum simulation for the short range case is of great importance for two main reasons. 
First, the short range model is much better understood than the long range counterpart, and as such, a study of its DSF can provide helpful insights on the effects of the different imperfection models, as well as on the accuracy of the measurement protocol which can be expected. 
Since the short range model is an easy instance of the time evolution problem, we perform a detailed study of the effects of the evolution imperfections in this case. 
This way we can provide sufficient understanding of the expected effects of these imperfections on the quantum simulation of the DSF. 
After this task is completed, we can move on to study the long range model and evaluate our practical proposal. 
Second, several architectures as optical lattices or Rydberg arrays can access the short range model, or the long range model at high values of $\alpha$, where the system effectively behaves short range. Our study of the short range model thus provides data which can be directly used to compare with experiments on those platforms.

\subsection{Universal properties of the short range TFIM}
The physics of the short range, nearest neighbour, TFIM has been studied in detail previously \cite{sachdev11}. Here we will briefly describe the low energy excitations of the TFIM and their signature in the DSF in terms of a two kink model. 

For the short range TFIM in the ferromagnetic phase, the ground state is given by a product state of spins fully polarized.
When the magnetic field and Ising coupling are at a finite value, fluctuations are induced in the system, in the form of fermionic pseudo-particles $\gamma$. 
These excitations, can be seen in the spin picture as spin flips, or kinks over the fully polarized state. Once a spin is flipped, it is free to move along the chain and create a domain. 
The walls of this domain can be regarded as the kinks (or equivalently, the $\gamma$-fermions) that interpolate between the two possible ground states connected by the $\mathbb{Z}_2$ symmetry of the model. 
When a domain is formed, the domain walls or kinks behave as free fermionic particles, that propagate through the chain. Since to create a domain we need at least two kinks (particles), the first contribution to the excitation spectrum will come from the two particle states, which will be described by their energy and momenta, $E = \epsilon_{q_1} + \epsilon_{q_2}$ and $ q = q_1 + q_2$. 
For a fixed $q$, the values of $q_1$ and $q_2$ can be chosen arbitrarily, which generates a continuum of excitations.

The spectrum of excitations will manifest in the dynamical structure factor: studying the longitudinal $xx$-structure factor, $S^{xx}(q, \omega)$, we observe the gap, and the continuum of excitations (the so called two particle continuum) that corresponds to the two particle states we mentioned previously. This observations have been previously shown, both numerically \cite{deryhko97} and experimentally via neutron scattering  \cite{Coldea10}.
In Fig.~\ref{PhT_dyn} we show the $xx$-DSF for the short range TFIM, 
as obtained from our free fermionic calculation for $J=1$ and $B = 1.4$, for $50$ sites. We clearly observe the two particle continuum which characterizes the low energy fluctuations, as well as the excitation gap at the point $q = 0$, $\omega \sim \pi/4$ (in units of $J$).

\subsection{Long range TFIM}

We can now 
concentrate on the case which is our test of a practical application:, the long range transverse field Ising model.
Models with these kind of long range interactions present considerable challenges to numerical studies.
The long range interactions severely constrain the system sizes which can be studied with exact diagonalization techniques based on sparse matrices.
Furthermore, studies of these systems employing finite size MPS based techniques are affected by severe finite size effects arising from the entanglement cutoffs required by these approaches. Recently, however, there has been success in studying the statics of long range models employing MPS algorithms, which directly act in the thermodynamic limit, such as iDMRG \cite{Saadatmand18, Crosswhite08}.

At the same time, several algorithms which can time evolve an MPS with long range interactions have been proposed \cite{Haegeman11,Zaletel15} to study the long range TFIM \cite{Hashizume18}. With the advent of these new techniques, and the state of the art of quantum simulators capable of implementing long range TFIM, the question of whether the DSF of this class of models can be accessed with these experimental architectures naturally arises. 

Unlike the short range model which has been thoroughly studied in the past, much less is known about the long range TFIM. Recent studies \cite{Buyskikh16, hauke13,luitz19} concentrate on the entanglement growth and the spread of correlations in this model as a function of the interaction length, $\alpha$, or in the thermalization of different initial states under this Hamiltonian \cite{fratus16}. 
Analyzing the light cones and possible Lieb-Robinson like bounds in the long range TFIM at zero temperature, these studies separate the dynamical behaviour of this model in three regions. 
For $\alpha > 3 $the system obeys the generalized Lieb-Robinson bound \cite{Hastings2006}, and the behaviour of the system mimics that of a short range model.
Via semi-classical arguments, the dispersion relation of excitations in the ground state (what we study here via the DSF) is found to approximately be a cosine, which coincides with the short range behaviour.
From this we can say that for quantum simulators, the behaviour of the DSF in the regime $\alpha > 3$ is expected to be very close that of the short range model.
On the other hand in the range $1<\alpha<3$ a broad light cone is observed and an excitation dispersion which is bounded.
This case is of special interest in this work, since trapped ion experiments can implement long range TFIMs in this range, but also given that it has recently been shown \cite{liu19, lerose19, verdel19} that in this regime the long-range interactions introduce an effective attractive force between a pair of domain walls.
This attractive force confines the excitations in bound states analogous to the confinement of mesons in high energy physics \cite{liu19, lerose19, verdel19}.
Since this exotic physics can be probed studying the confinement signatures in both the unequal time correlators and DSFs, our work opens the door to the study of these effects in quantum simulators. Finally we mention that for $\alpha < 1$ the light cone completely disappears and a virtually instantaneous spread of correlations is observed. 
In Fig.~\ref{DSF_long_range_main} we show the DSF of the long range TFIM, as obtained numerically from a full exact diagonalization of a system of 14 spins at zero temperature, for the interaction lengths $\alpha = 1$(a), $2$(b), and $3$(c).
In these figures we see that for $\alpha=1$ the DSF shows no $\omega$ dependence, which hints at the possibility of excitation confinement \cite{liu19, lerose19, verdel19} being evidenced through the DSF. For $\alpha > 2$ the $\omega$ dependence is recovered, slowly approaching the short range behaviour as $\alpha$ is increased.

\subsection{Imperfection models}
\label{error_models}
Three basic ingredients are needed to simulate DSFs on near term devices: First, we need to be able to {\it prepare} the ground state of the target Hamiltonian in a controlled way, and ideally with as high state fidelity as possible. 
Second, we need to be able to control the {\it time evolution} of the system, in such a way that the physics we desire to investigate is not severely mitigated by experimental imperfections.
And finally, we  want to employ the proposed {\it measurement} protocol to determine the unequal time Green functions.  Every one of these steps carries their own imperfections which we will consider separately.

For the {\it preparation} imperfections we will study the effect of measuring DSFs when the prepared state has a fidelity with respect to the ground state smaller than one, $F = \bra{\psi_\sigma}\ket{\psi_0} < 1$. The measurement protocol is not modified by this imperfection model, such that even if the prepared state is not the ground state, we can still recover the retarded Green's function via Eq.~(\ref{eq:FDthm}).     

In the case of {\it evolution imperfections} we will study three fundamental effects over the TFIM Hamiltonian. In the first case, we will study how a time dependent modulation of the Ising couplings affects the DSF.
 In this case the Hamiltonian couplings are modified to be time dependent and of the form
 \be
 J_{i,j}=\frac{J(0)}{|i-j|^{\alpha}}(1 + A \sin(w t) )\,, \quad  J_{i,i\pm 1} = J(0)(1 + A \sin(w t) ) .
 \label{eq:amp_drive1}
 \ee  
With $J(0) = J$, for the long and short range models respectively. We will study several modulation amplitudes, $A = 0.01$, $0.05$, $0.1$, and $0.5$, and different frequencies $\omega$ between $0.05$ and $25$. 

We will also study the case of random interactions and magnetic fields, related to lattice imperfections.
 In these cases the Hamiltonian takes the form in Eq.~(\ref{Ry_Ising}), but for the case of random interactions the Ising couplings take the form 
 \be\label{eq:shortrangerandom}
 J_{i,i\pm 1} = (J + A\xi_{i,j})\,, \quad   J_{i,j} = \frac{J + A\xi_{i,j}}{(i-j)^\alpha}
\ee
for the short and long range models respectively. While in the case of random transverse fields 
\be\label{eq:fieldrandom}
B_{i} = B + A\xi_i.
\ee
In all cases $\xi$ is drawn independently at random at each site, from a uniform distribution on the interval $[0.0,1.0)$ with $A = 0.01$, $0.04$, $0.1$, and $0.4$. We employ between 50 and 100 disorder realizations per data point.
    
\section{Effect of experimental imperfections on the DSF of the short range TFIM}
\label{short_range_noise}

In the following, we will demonstrate that the qualitative and quantitative features of the DSF for both the short (
and long 
range TFIM can be recovered from a quantum simulation even in the presence of experimental imperfections.
Here, we discuss the role of imperfections and their impact as such; the certification of the actual correctness of the
quantum simulation \cite{Certification} is a separate task. 

To phenomenologically and briefly summarize the results: We show that, at the experimental levels of control present in state of the art architectures, the errors which would be produced in a measurement of the DSF for both short and long range TFIM are small, and one can trust both the qualitative and quantitative results of such experiment. 
Since the overall behaviour of the DSF is what gives one information about low energy excitations of a given system, and how they behave, the errors studied show that, at the current level of experimental control  (when our imperfection parameter is set below $5\%$), the DSF is well behaved.
Thus, the overall form of the DSF does not change and one can safely extrapolate, from a quantum simulation via the DSF, what some of the low energy excitations of a given model are, and what their behaviour is.

\subsection{Quantifying imperfections}
\label{quantification}

To assess what the effect of experimental imperfections is on the DSF we will analyze two particular quantities based on the absolute error of the DSF. 
We define the absolute error as 
\be
\Delta S(q, \omega) = |S^{x,x}(q, \omega) - \tilde S^{x,x}(q, \omega)|\,,
\label{err_SQW}
\ee 
where $S^{x,x}(q, \omega)$ is the DSF obtained from the exact solution of the TFIM in the absence of imperfections.
 $\tilde S^{x,x}(q, \omega)$ is the DSF obtained from the exact solution with various perturbations in the Hamiltonian, arising from the different imperfection models.
If we integrate over frequency  (reciprocal space) we obtain the average error in reciprocal space (frequency),
\be
\Delta S(q) =\frac{1}{N_\omega} \sum_\omega  \Delta S(q, \omega),\quad
\Delta S(\omega) = \frac{1}{L^2} \sum_q  \Delta S(q, \omega) \,,
\label{error_DS1}
\ee
where $N_\omega$ is the number of frequencies, which depends on the discretization of the time evolution, and $L$ is the system size. We show the average error for different imperfection models in the supplementary information.  
The maximum of the absolute error, for fixed $\omega$ or $q$, will be denoted by
\be
\max_q [\Delta S(q, \omega)] ,\quad
\max_\omega [\Delta S(q, \omega)] \,. 
\label{error_DS2}
\ee

These errors can be understood in the following way: \eqref{err_SQW} is the absolute error of the DSF when imperfections are considered. 
If one makes a cut on the absolute error at a given value of reciprocal space, $q$, and integrates it over frequency, one obtains the frequency integrated error, $\Delta S(q)$.  This is equivalent for cuts at a given frequency $\omega$, to obtain  $\Delta S(\omega)$. 
If, on the other hand, one selects the maximal error at that value, one obtains \eqref{error_DS2}.  
The study of the imperfections in this way allows us to account for the effects in frequency and reciprocal space separately. 
If the imperfection models do not change the DSF, then these errors should be small and flat over the entire $q$ and $\omega$ range.
On the other hand, if these errors are not small, we can assess what their
effect is on the DSF by studying the shape of the quantities given in \eqref{error_DS1}
and \eqref{error_DS2}. 
For example, if one of these imperfection models were to close the gap, we would see errors towards small frequencies, but not on $q$-space. 

Since the Fourier transform is performed as data-processing over the
correlators, we will compare the error of the DSF to the error in the
correlators, as to assess the robustness of the Fourier transform. The error in the correlators, and the average and maximum over space (where space is indicated as $r = i-j$) are defined as 
\begin{align}
\Delta C_r(t) =  |C^{x,x}_r(t)& - \tilde C^{x,x}_r(t)|,\\
\Delta C(t) =  \frac{1}{L^2}\sum_r \Delta C_r(t),& \quad \max_r [\Delta C_r(t)]\,.
\end{align}
Finally, to determine the scaling properties of the long range model, we will study the integrated DSF error $\Delta S$ as a function of size and of the range of the interactions $\alpha$ where the integrated error is given by
\be
\Delta S  =\frac{1}{N_\omega} \frac{1}{L^2} \sum_\omega \sum_q  \Delta S(q, \omega)\,.
\label{integrated_error}
\ee

\subsection{Influence of state preparation imperfections on the DSF}
\begin{figure*}
\centering
\includegraphics[scale=0.4]{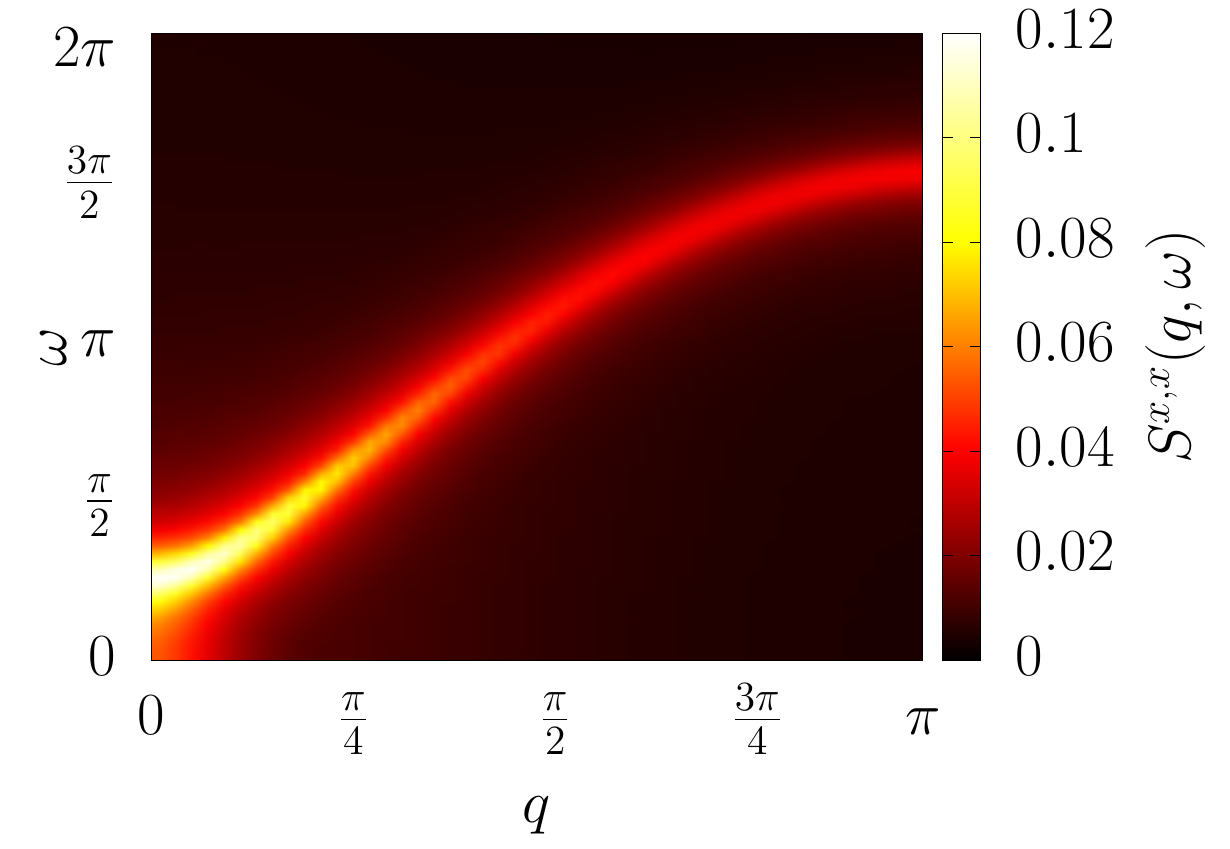}
\includegraphics[scale=0.4]{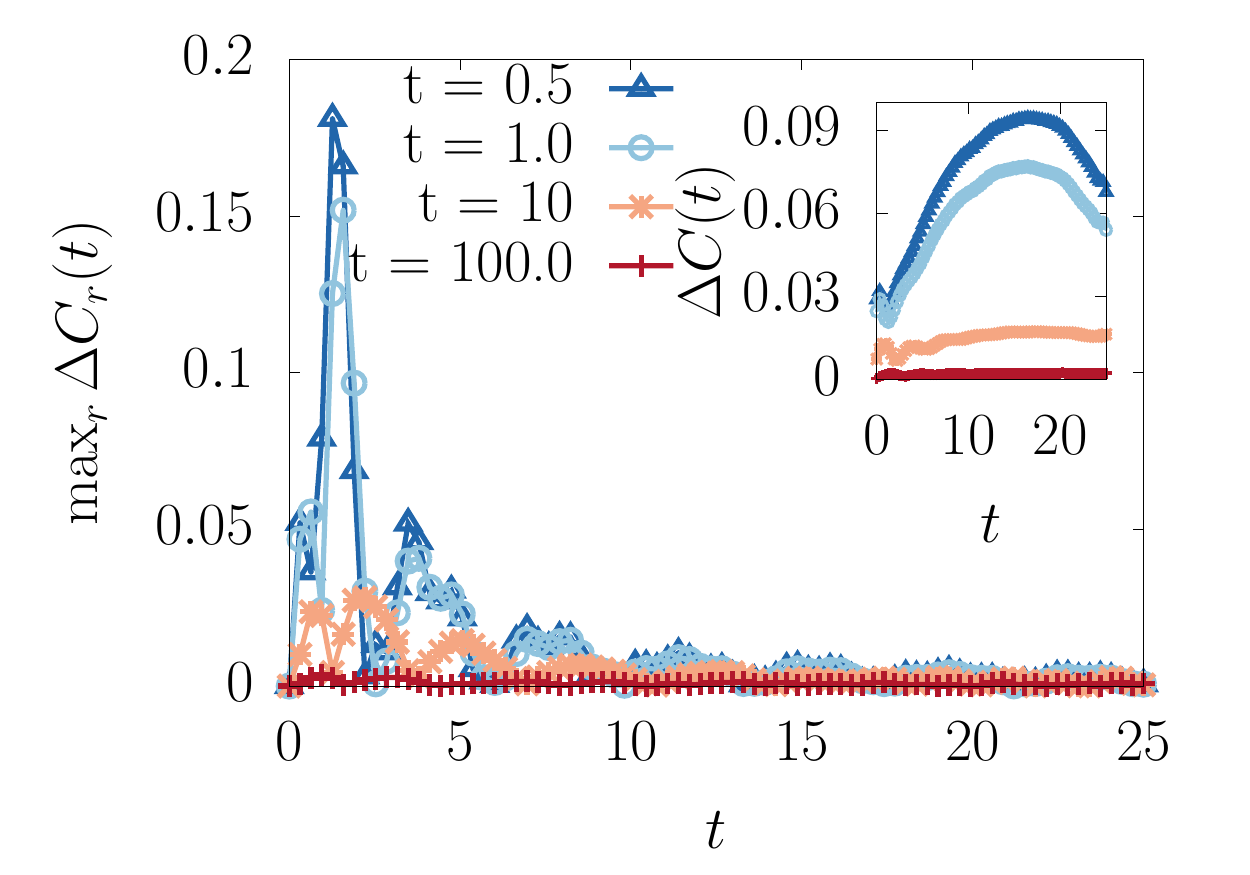}
\includegraphics[scale=0.4]{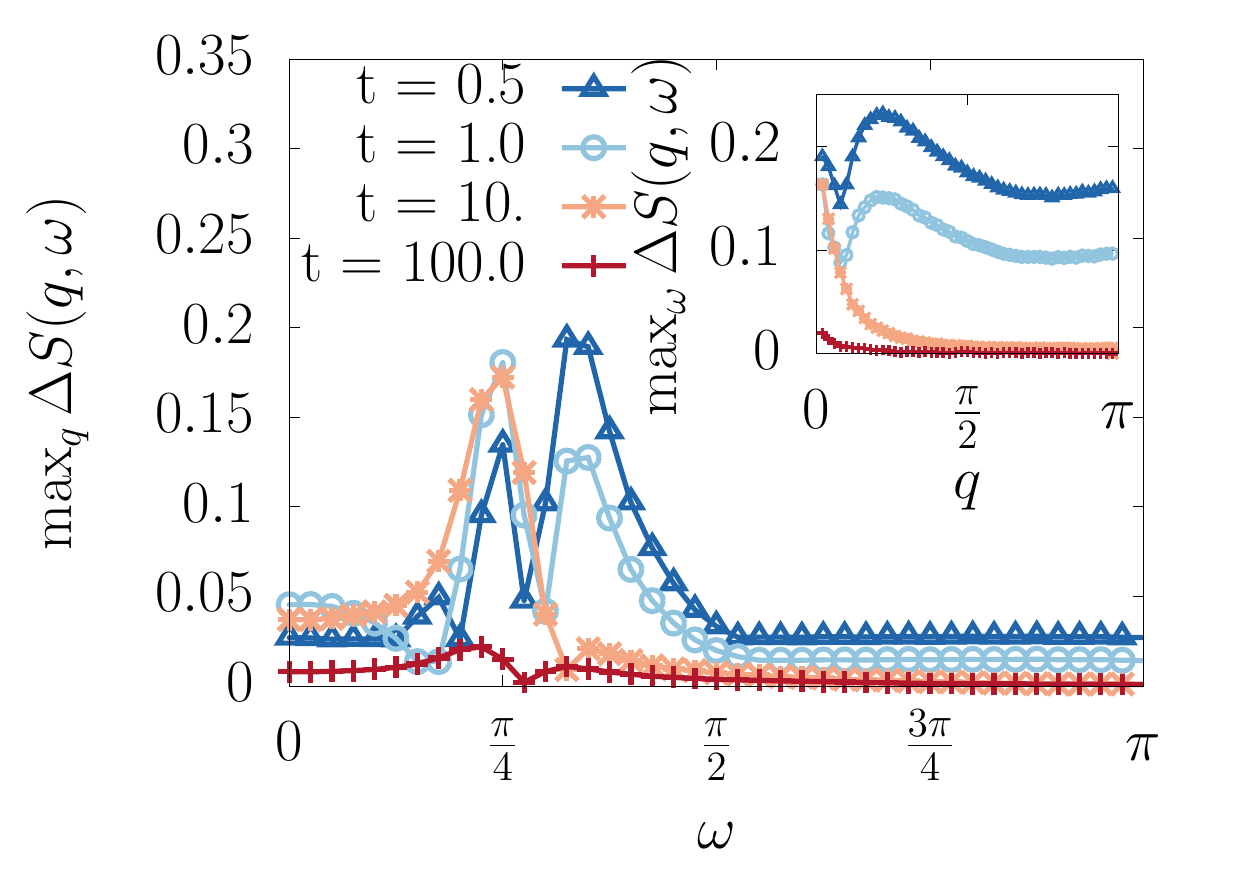}
\caption{{\it Effect of finite preparation times $\tau_Q$ on the DSF of the short range TFIM}. 
We show numerical results for the DSF and unequal time correlation functions of the short range TFIM subject to different preparation time.
{\it (Left panel)} DSF of the TFIM for a preparation time $\tau_Q = 0.5$. This can be directly compared to the imperfection free case shown in Fig.~\ref{PhT_dyn}. 
{\it (Middle panel)} Cuts of the absolute error for the unequal time correlation function, depicted 
for a quantitative comparison. At short preparation times $\tau_Q=0.5$ and $\tau_Q=1$ we find significant deviations of local correlation functions.  The inset shows the averaged error in the correlators, which indicates that the deviation seen in the maximal error persists, and in fact increases, at all times on the level of uniform real-space average.
{\it (Right panel)} Cuts in frequency and reciprocal space (inset) of the DSF absolute error. We quantitatively verify the intuition given by the left most panel.
 The DSF indeed encodes the correct physical information despite the deviations in real-space and the deformation of the low $\omega$ sector. 
The low error intensity away from $\omega = \pi/4$ and $q=0$ indicate that the gap remains open for $\tau_Q = 100$.
Absence of errors for long preparation times at $ q > 0 $ and $\omega > \pi/4$ indicate that the two particle continuum is not affected by these preparation times.}
\label{error_ad_evo}
\end{figure*} 
We will concentrate on two architectures, trapped ions and Rydberg atom arrays. 
Both of them can prepare the initial state via an adiabatic evolution. Furthermore, trapped ions can prepare it through quantum approximate optimization techniques \cite{pagano19}.
We will study here how the DSF is affected by different evolution times, when the final field value is far away from the quantum critical point, $J=1$ and $B = 1.4$.

\textit{Adiabatic time evolution.}
The question that motivates
us is how the features of the DSF change when the system is prepared for a time $\tau_Q$ (the preparation time) from an initial polarized state, (which corresponds to the $B \rightarrow \infty$ limit) to the final state $B = 1.4$. 
In the thermodynamic limit this preparation time diverges when one approaches the quantum critical point. 
For a finite system, it can be shown that the finite size gap destroys the divergence, and a finite bound on the preparation time can be obtained \cite{dziarmaga05,zurek05} within which the evolution remains adiabatic. 

Considering our previous discussion, it is imperative that we study how the properties of the DSF change when the preparation is not adiabatic.
 In this case, if the preparation time is not large enough to be in the adiabatic regime, the quantum simulation can still be able to obtain results which are  close to the physics that one desires to study. 
 This can be understood, and generalized to preparation protocols beyond adiabatic evolutions, in the context of prepared state fidelity. 
 If the fidelity is high, then the properties of the DSF evaluated over the prepared state will remain close to the properties of the DSF evaluated over the exact ground state. 
 Furthermore, if the preparation evolution is not adiabatic, but does not surpasses the quantum critical point, then there exists a time interval in which the transition 
 probabilities towards excited states is sufficiently small, such that the main contribution to the state 
 of the system is the ground state  \cite{dziarmaga05,zurek05,cincio07,cincio09,kennes18}. 
 At the parameter values studied here, the fidelity of the prepared state $F$ can be thought of as the probability that no extra domain walls (or equivalently, kinks) have been created during the preparation.
   Ref.~\cite{zurek05} calculates the fidelity of the final state with respect to the vacuum of excitations for a linear ramp, effectively probing the probability that no excitations have been created during the preparation.
  The authors find that the fidelity takes the form 
  \be
  \ln (1 - F) = -\pi \Delta^2/(4B/\tau_Q)\,. 
\ee
  With this in mind, we can simply ask the question of {\it how large does $\tau_Q$ need to be such that $F \sim 1$, and what are the effects on the DSF when $F < 1$?}.  

We  quantify the robustness of the DSF to preparation imperfections employing the error measures shown in before.
For this we  numerically calculate the DSF of the short range TFIM,when we prepare the state by a total time $\tau_Q$, starting with the field at $B_{\mathrm{ini}}\rightarrow\infty$ and finishing at $B_{\mathrm{final}}=1.4$. 
\begin{figure*}
\centering
\includegraphics[scale=0.4]{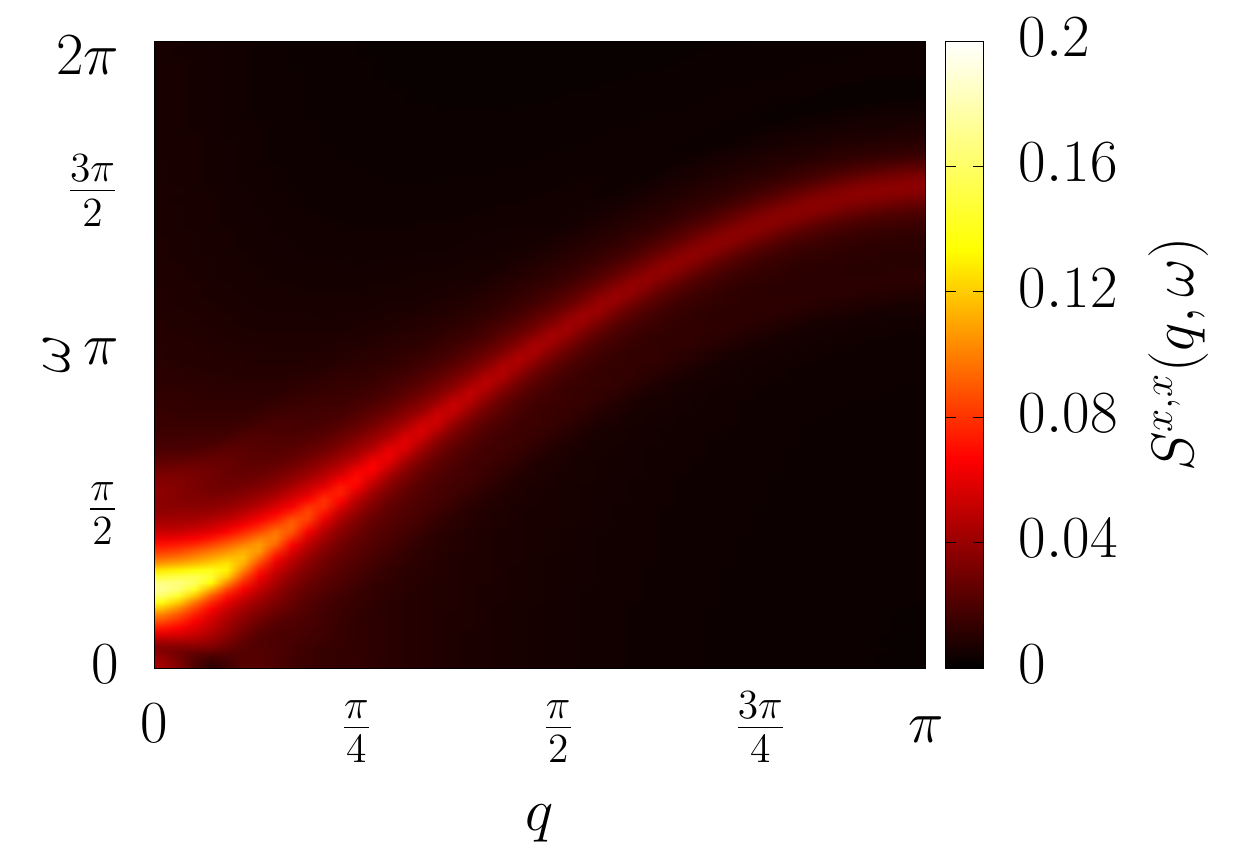}
\includegraphics[scale=0.4]{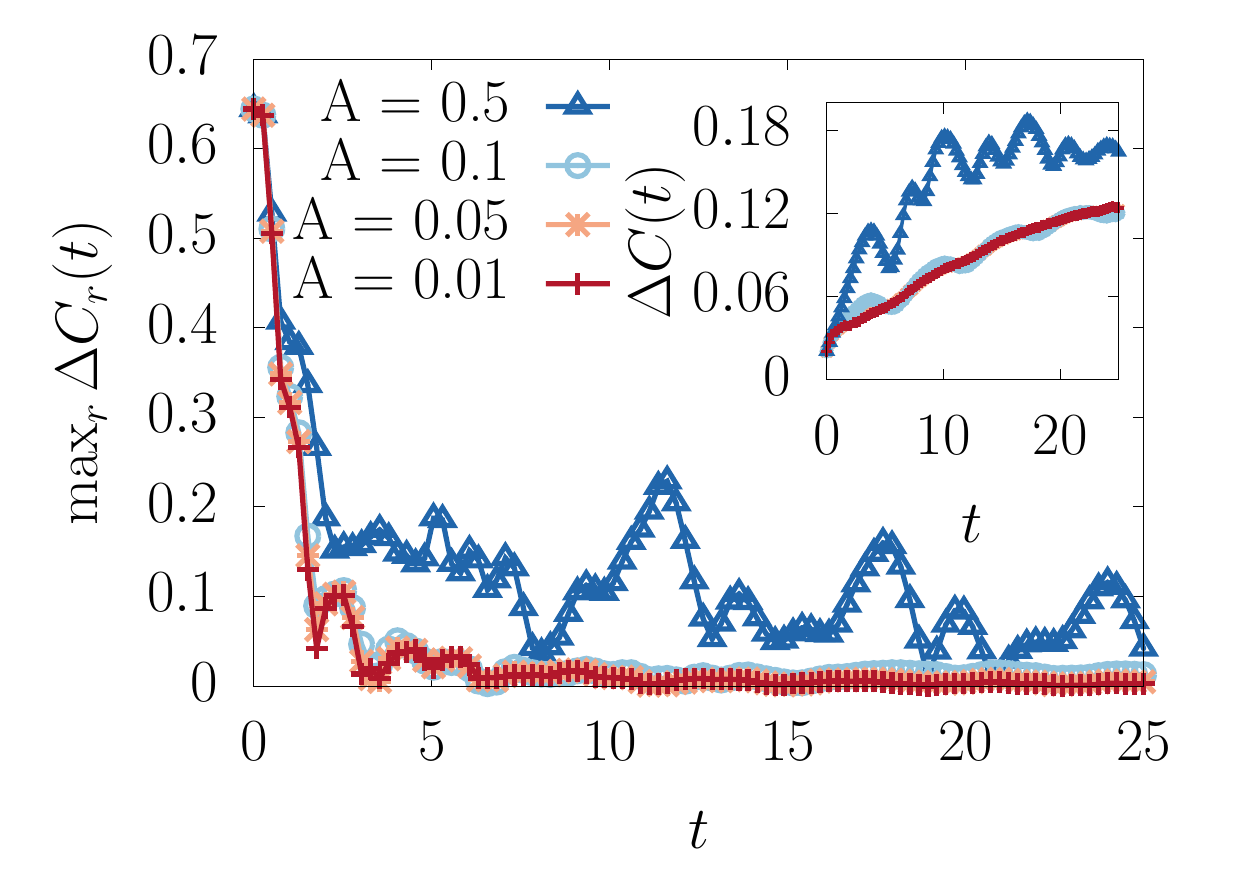}
\includegraphics[scale=0.4]{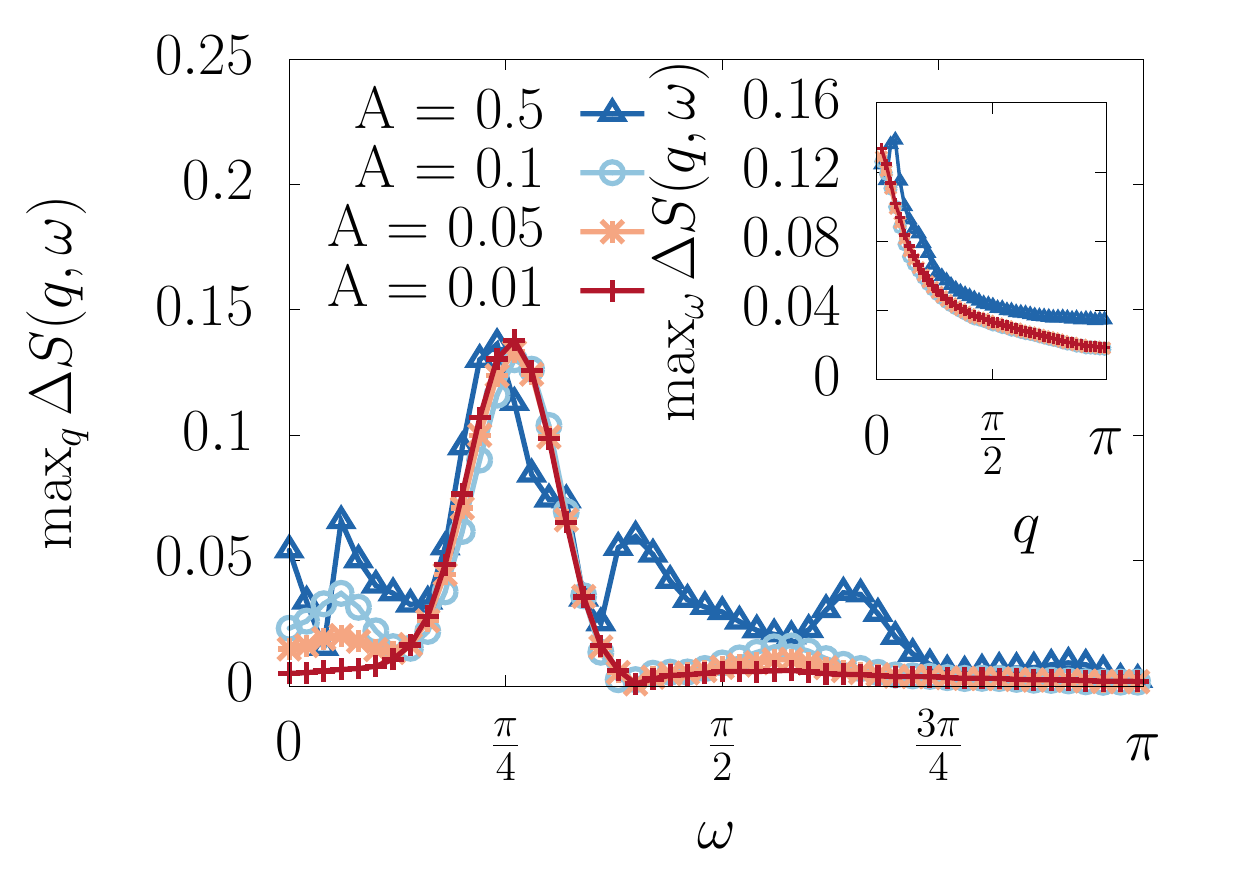}
\caption{{\it Effect of globally fluctuating
Ising couplings on the DSF of the short range TFIM.} 
We show numerical results for the DSF of the short range TFIM subject to the imperfection model
\eqref{eq:shortrangerandom} with Ising couplings harmonically modulated for amplitudes.
{\it (Left panel)} DSF of the TFIM for an Ising coupling modulated by $A=10\%$. This can be directly compared to the imperfection free case shown in Fig.~\ref{PhT_dyn}. 
{\it (Middle panel)} Cuts of the absolute error for the unequal time correlation function, shown for quantitative comparison. We see that at short times there are significant deviations of the local unequal time correlation functions. The inset shows that this deviation increases, at all times on the level of uniform real-space average. 
{\it (Right panel)} Cuts in frequency and reciprocal space (inset) of the DSF absolute error. We quantitatively verify the intuition given by the left most panel. The DSF indeed encodes the correct physical information despite the deviations in real-space and the deformation of the low $\omega$ sector. 
The gap remains open for the lowest imperfection level. The broadness of the maxima in the frequency cuts indicate that the gap is shifted with respect to the exact solution.
The inset shows that the effect over momentum space does not decay to zero, coming from the lower intensity in the DSF signal in comparison to the clean solution, and not from a deformation of the two particle continuum.}
\label{error_g_fluct}
\end{figure*}

The adiabatic evolution has been performed for different evolution times, ranging from $ \tau_Q = 0.005$ to $\tau_Q = 3000$. 
In Fig.~\ref{error_ad_evo} we show the error analysis for preparation times $\tau_Q = 0.5$, $1$, $10$, and $100$. 
Following the calculations of Ref.~\cite{zurek05} we estimate that the fidelity of the prepared state (assuming no other error sources) will correspond to $F \sim 0.043$ for a preparation time of $\tau_Q = 0.5$,  to $F \sim 0.59$ for $\tau_Q = 10$, and $F \sim 0.99$ for $\tau_Q = 100$. 
Our numerical error analysis of the DSF and correlators coincides with these fidelity estimates. 
We tackle the DSF first: in the left most panel of Fig.~\ref{error_ad_evo} we show a typical DSF for a preparation time $\tau_Q = 0.5$, and in the right most panel we show the maximum error of the DSF over frequency (main figure) and reciprocal space (inset).
For $\tau_Q=100$ the maximum error is below $5\%$ and mostly flat over the entire $(\omega, q)$ space, indicating that this $\tau_Q$ is enough to obtain an accurate DSF. This can be confirmed by comparing the left most panel of Fig.~\ref{error_ad_evo} and Fig.~\ref{PhT_dyn}.
From these figures we notice that the discrepancy in the DSF between the exact case and the one studied in this section appears around the point $(q = 0, \omega = \pi/4)$ which corresponds to the position of the gap in the clean case.
Even for $\tau_Q = 0.5$, most of the error is constrained around the gap, indicating that the main contribution to the error in the DSF is a qualitative change in the overall broadness of the low $q$ low $\omega$ sector, even though the overall shape of the DSF does not change ( as is seen in the left most panel).

In the case of unequal time correlators, shown in the middle panel of Fig.~\ref{error_ad_evo}, we see that the maximal (average in the inset) error in this case, for $\tau_Q=100$, is also below $5\%$ ($1\%$), but we can see how the average error increases over time. 
While looking at the correlators directly could also be a way to study the ground state fluctuation of the system (given that the effect of the imperfections is small), the interpretation of the data as a function of time can be much more challenging, especially for long times. This can be understood by considering the propagation of errors as a function of time, which takes place with a maximal velocity consistent	 with the Lieb-Robinson bounds. We show in Fig.5 in the supplementary information the propagation of errors in the correlators as a function of time.
With this in mind, we can note that the Fourier transform leading to the DSF allows one to account for all the spacial and temporal data of the correlators, as well as understand and deal with errors arising from this imperfection model in a much simpler way.

\subsection{Influence of  evolution imperfections on DSF}
In trapped ions architectures, the spin-spin interactions are created by coupling the spin states to the normal modes of motion of the ions by  laser beams \cite{kim10,edwards10} (see section I in the supplementary information), obtaining a coupling strength directly proportional to the Rabi frequency of the ions.
 The lasers employed present intensity and phase oscillations which can be currently controlled up to a certain threshold \cite{schneider98}. This induces a variation of the Rabi frequencies across the chain, resulting in interactions which are not uniform over time along the chain. 
\begin{figure*}
\centering
\includegraphics[scale=0.4]{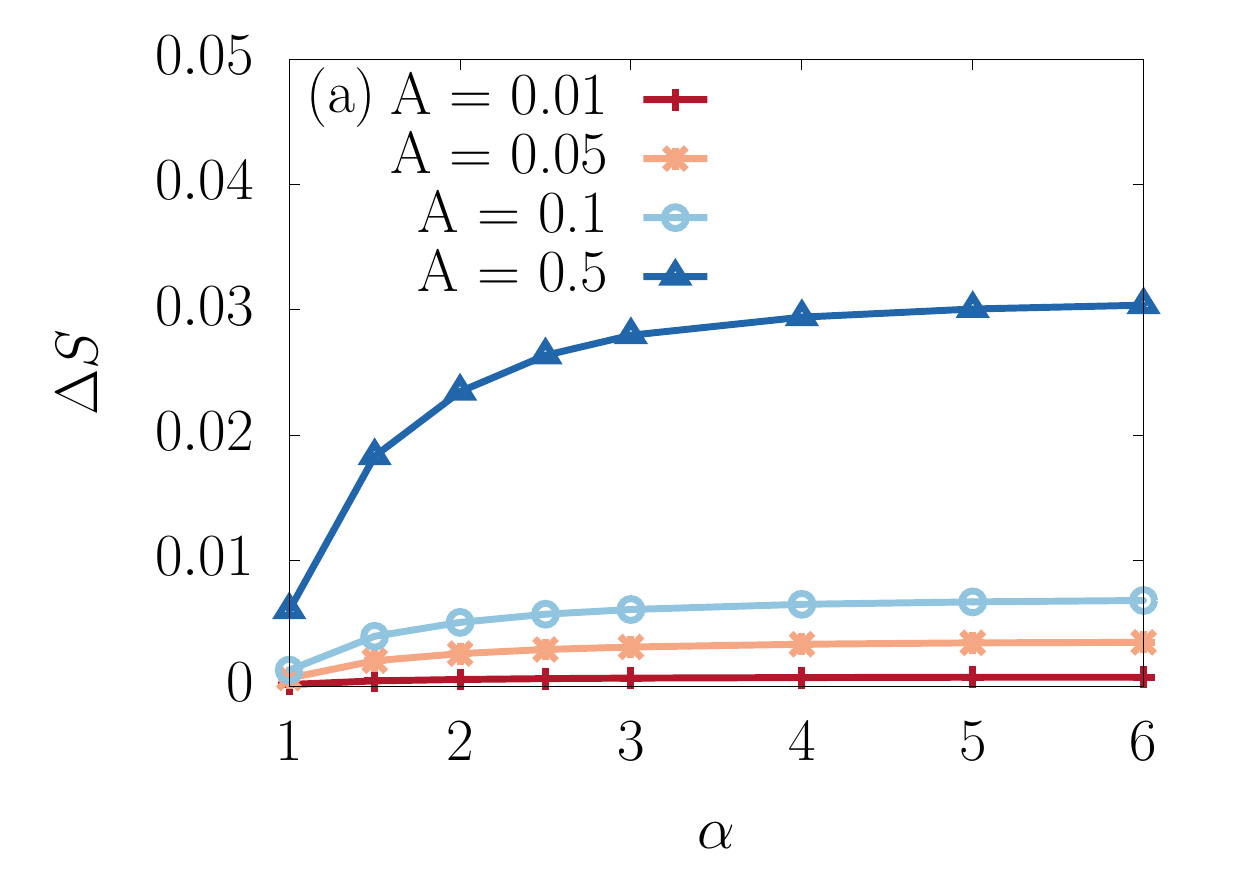}
\includegraphics[scale=0.4]{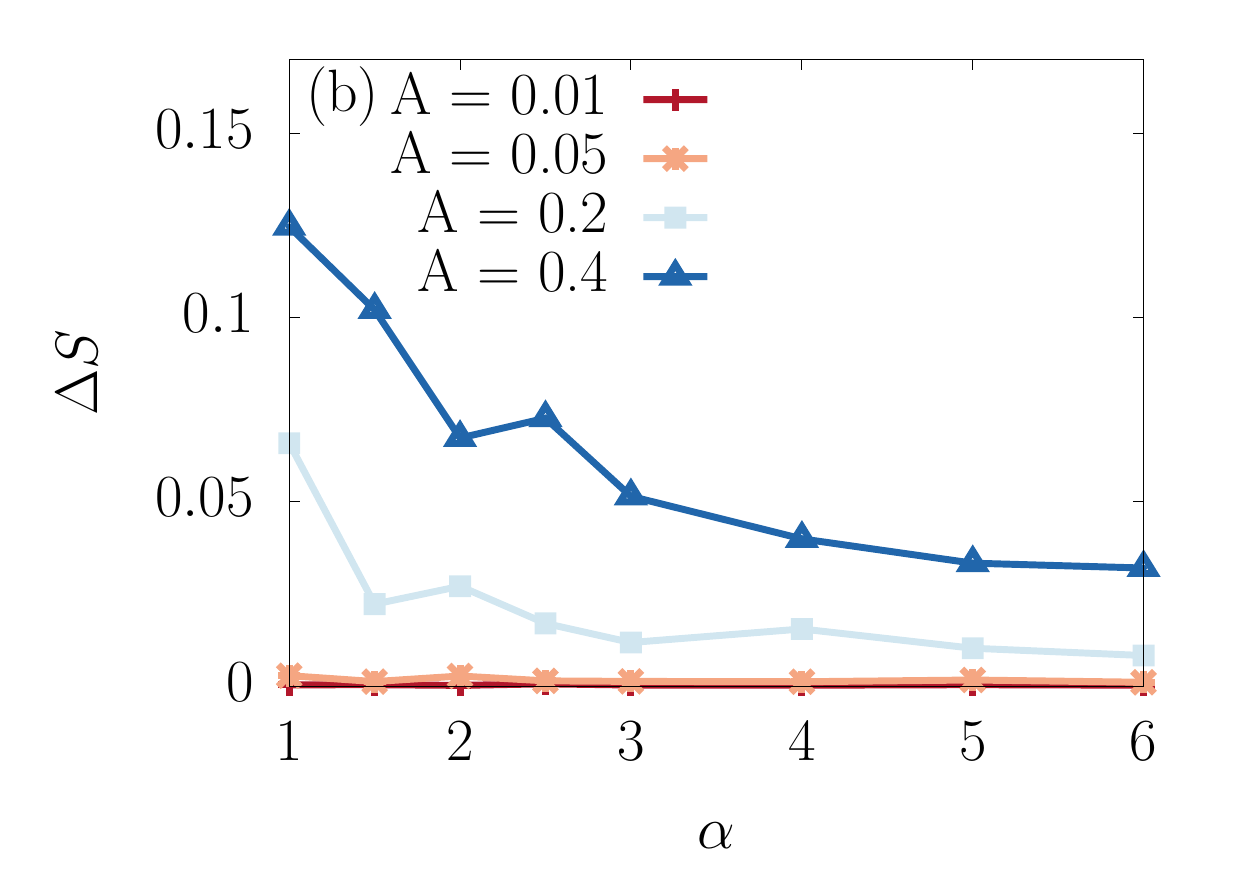}
\includegraphics[scale=0.4]{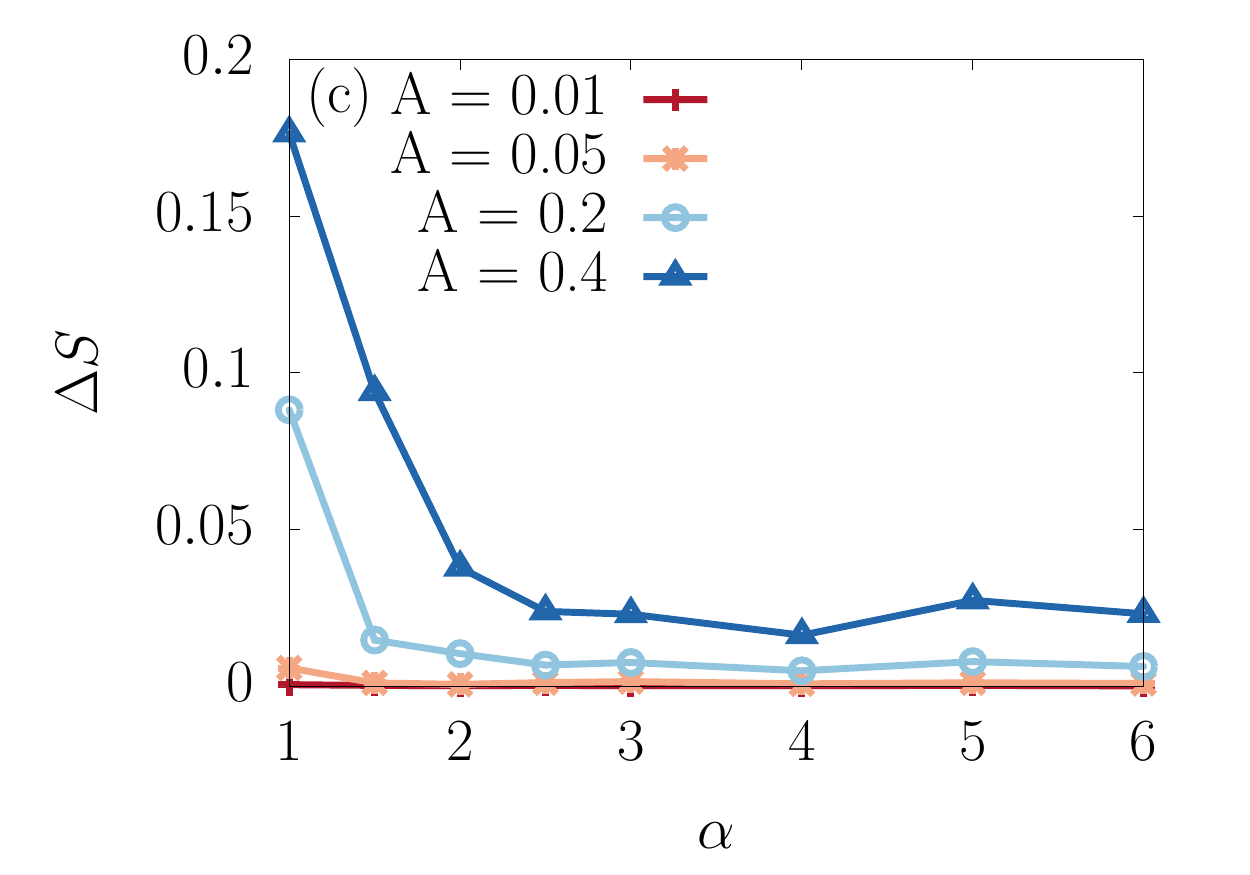}
\caption{ {\it Effect of experimental imperfections in the DSF for the long range TFIM}. Average error, $\Delta S$ as a function of the interaction range, $\alpha$, for $L=14$ sites. 
{\it (a)} Effects of Ising couplings harmonically modulated, Eq.~(\ref{eq:shortrangerandom}). The error is minimal at $\alpha = 1$, within the confined phase, and saturating at $\alpha > 3$, when the system approaches the short range TFIM. 
{\it (b), (c)} Effect of lattice imperfection for random fields (b), and random interactions (c). 
For both these cases the effect is the same, the DSF is highly susceptible to randomness at low values of $\alpha$, and it monotonically becomes more robust as $\alpha$ is increased, recovering the short range behaviour for $\alpha \rightarrow \infty$.
A physical interpretation of the effect on the excitations of these imperfection models on the DSF is beyond the scope of this work, as the presence of excitation confinement in the long range model has been proposed recently \cite{liu19, lerose19,verdel19}. At the experimental levels of control currently available, the integrated error is minimal, and the overall shape of the DSF is unchanged, indicating that quantum simulators can probe the regime of interactions studied in this work, $1<\alpha<6$, and obtain accurate DSFs for system sizes bigger than what state of the art classical algorithms can achieve.  
}
\label{alpha_long_range}
\end{figure*}
\textit{Globally fluctuating Ising coupling.}
We will study the particular case in which the intensity fluctuations of the lasers directly induce periodic fluctuations of the spin-spin interactions.
 We will model these evolution imperfections by modulating the Ising coupling as in ~\eqref{eq:shortrangerandom}, with different amplitudes $A$ and frequencies $w$. 

In Fig.~\ref{error_g_fluct} we show the error analysis of the DSF and unequal time correlation functions for the case of modulated Ising couplings.
We have studied a range of frequencies from $w = 0.05$ to $w =25$ and intensities in the range $A \in [0.01,1]$.
 Here, where $J = 1$, a coupling intensity $A = 0.01$ correspond to a $1\%$ fluctuation in the Ising coupling.
 Current experimental capabilities can constrain these parameters within the $1\%$ threshold \cite{friedenauer08}.  

The effects are mainly noticed as a function of the coupling $A$. 
Concentrating on the error of the DSF shown in the left and right most panels of Fig.~\ref{error_g_fluct}, we see that even the lowest coupling studied $A = 1\% = 0.01$ can induce a maximal error of $15\%$ in the DSF.
The error is mostly concentrated around gap. 
Already for $A = 0.05$ we see that small broad peaks appear for  $\omega > \pi/4$, while for $A = 0.5$ the shape of the DSF is changed, as indicated by the large errors all along the frequency axis in the left most panel of Fig.~\ref{error_g_fluct}.
This can be understood by looking at the left most panel of Fig.~\ref{error_g_fluct}, comparing it with Fig.~\ref{PhT_dyn}.
 Besides the overall decrease in intensity, the low frequency shape of the two particle continuum has changed, giving the maximum in the right most panel of Fig.~\ref{error_g_fluct} at $\omega = \pi/4$.
 There is also an increase of intensity at small $q$ for a range of frequencies up to $\omega = \pi/2$.

In the case of the unequal time correlators, the middle panel of Fig.~\ref{error_g_fluct}, the error intensity is much higher, with a maximum of $65\%$ at small times which decays to close to zero, except for $A = 0.5$. 
This is an artefact generated by the error propagation in a Lieb-Robinson cone. Even if the maximal error is small, there is an overall error which increases with time, which indicates that long measurement times lead to the propagation of errors and to an, in average, very large inaccuracy in the correlators. 
As for the case of preparation imperfections, the error in the correlators can make for a hard determination of the propagation of excitations through the system. 
The DSF allows us to study these effects even in the presence of imperfections, given that the errors in this quantity are localized close to the maxima, and the overall shape of the two particle continuum is minimally changed for small intensities of the fluctuating coupling.

It has to be pointed out, that for slightly higher intensities of the fluctuating coupling, even for $A = 5\%$, small changes in the DSF are seen both in the frequency and reciprocal space axis. 
This indicates that this imperfection model has to be dealt with carefully in an experimental setup, as small increases in $A$ can lead to appreciable effects in both the DSF and the unequal time correlators.

\textit{Lattice imperfections.}
In a Rydberg atom setup spin-spin interactions can be generated by applying a spin-dependent optical dipole-force ~\citep{Zhang17,Bernien17} (see section I in the supplementary material ). 
Since the Rydberg atoms are not in the ground state once in the local trap, and the experiment is carried at a finite temperature, fluctuations in the atomic positions for each atom in each cycle of the experiment are introduced  \cite{Zhang17,Bernien17}, which will affect the Ising interaction (see Eq. (2) in the supplementary information). 
In a typical experiment the fluctuation of the position lead to a change in the Ising coupling between $0.1\%$ and $0.2\%$ 
\cite{Zhang17,Bernien17} from shot to shot. This can be empirically modelled as a random Ising interactions as in \eqref{eq:shortrangerandom}.  
 On the other hand the Rabi frequency is also not uniform along the chain. Since this frequency gives rise to the transverse field, this type of evolution imperfection can be studied as a random transverse fields as in ~\eqref{eq:fieldrandom}.

In a setup where ions are trapped by a linear Paul trap, spin-spin interactions can be obtained applying off-resonant laser beams \cite{kim10,edwards10}. The Rabi frequency $\Omega_i$ can also vary across the chain from shot to shot (see Eq. (3) in the supplementary material), inducing random Ising interactions.

For these cases, random transverse fields and Ising interactions, we show the results in the section II of the supplementary material, since they are very similar as those found for the preparation imperfections.
In both cases we see that the majority of the imperfections are concentrated around the maximum of the DSF, where the gap is located. 
Strong random Ising interactions tend to close the gap, as can be seen in Figs.1 and 3(c) in the section II of the supplementary material. 
On the other hand, random transverse fields tend to open it 
(see Figs.3 and 4(d) in the supplementary material). 
In both these cases, for the experimentally tolerable imperfections of around  $1\%$, the errors in the DSF and correlators are both reduced, leading to no noticeable effects in the DSF. 
While the errors in the correlators (Figs.1 and 2 in the supplementary material) are also small, the average error increases with time. In this sense, long measurement times can lead to very large errors, rendering the results analyzed purely via correlation functions highly unreliable.  

We note that the regime in which randomness is large is interesting in itself, as it offers the chance to directly probe the effect of random disorder in spin chains via time dependent observables, and the DSF in particular, in near term quantum devices. 

\begin{figure*}
\centering
\includegraphics[scale=0.4]{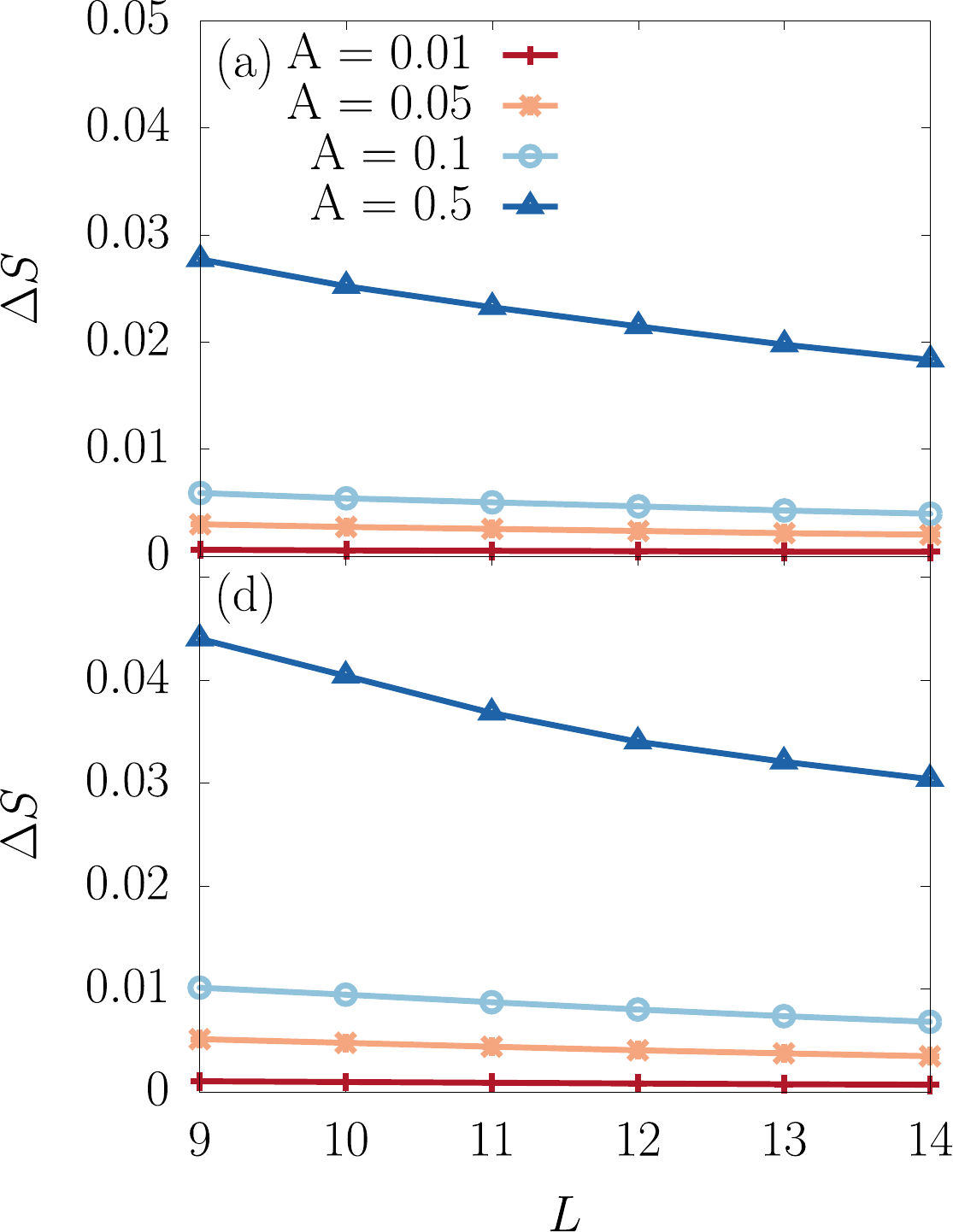}
\includegraphics[scale=0.4]{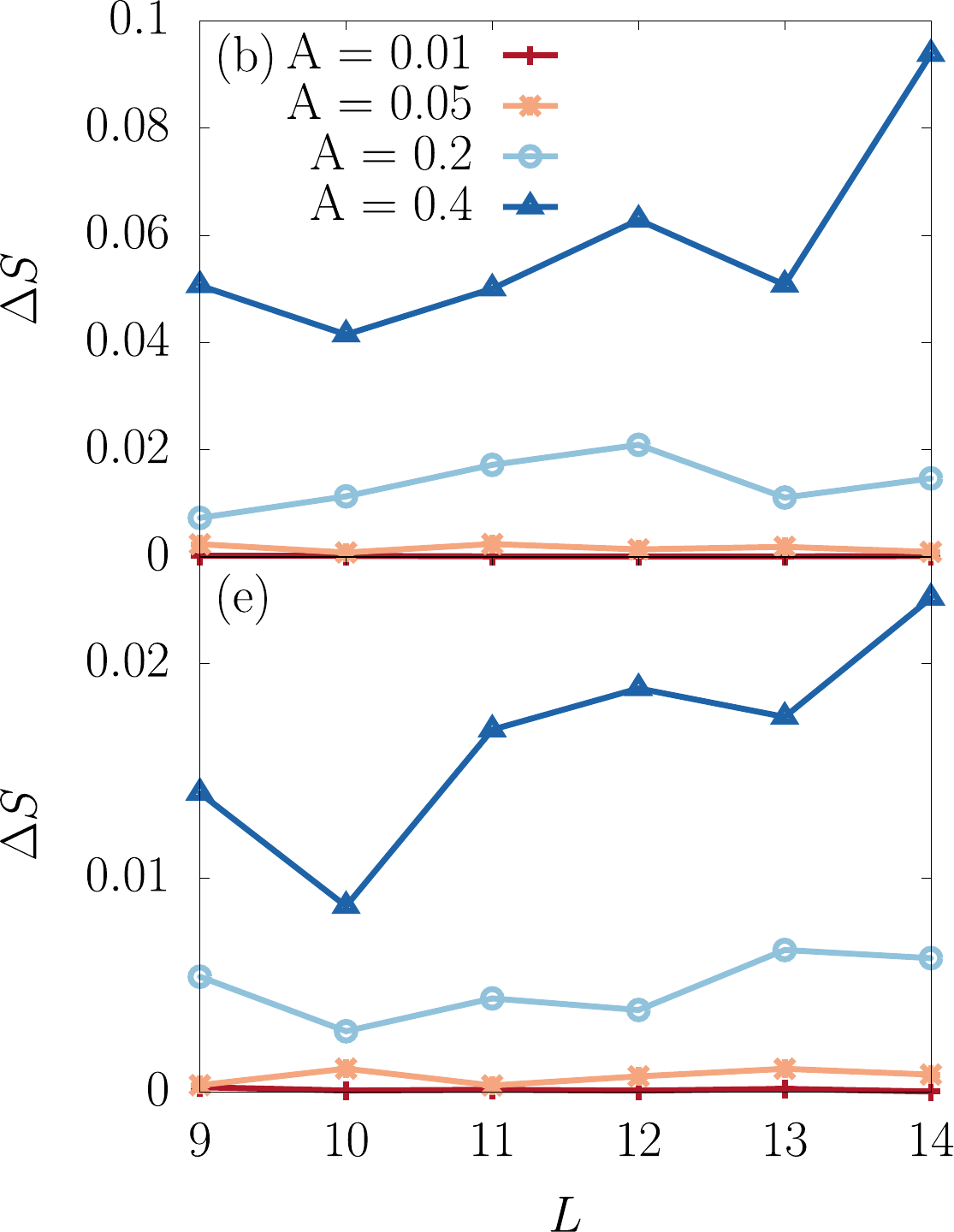}
\includegraphics[scale=0.4]{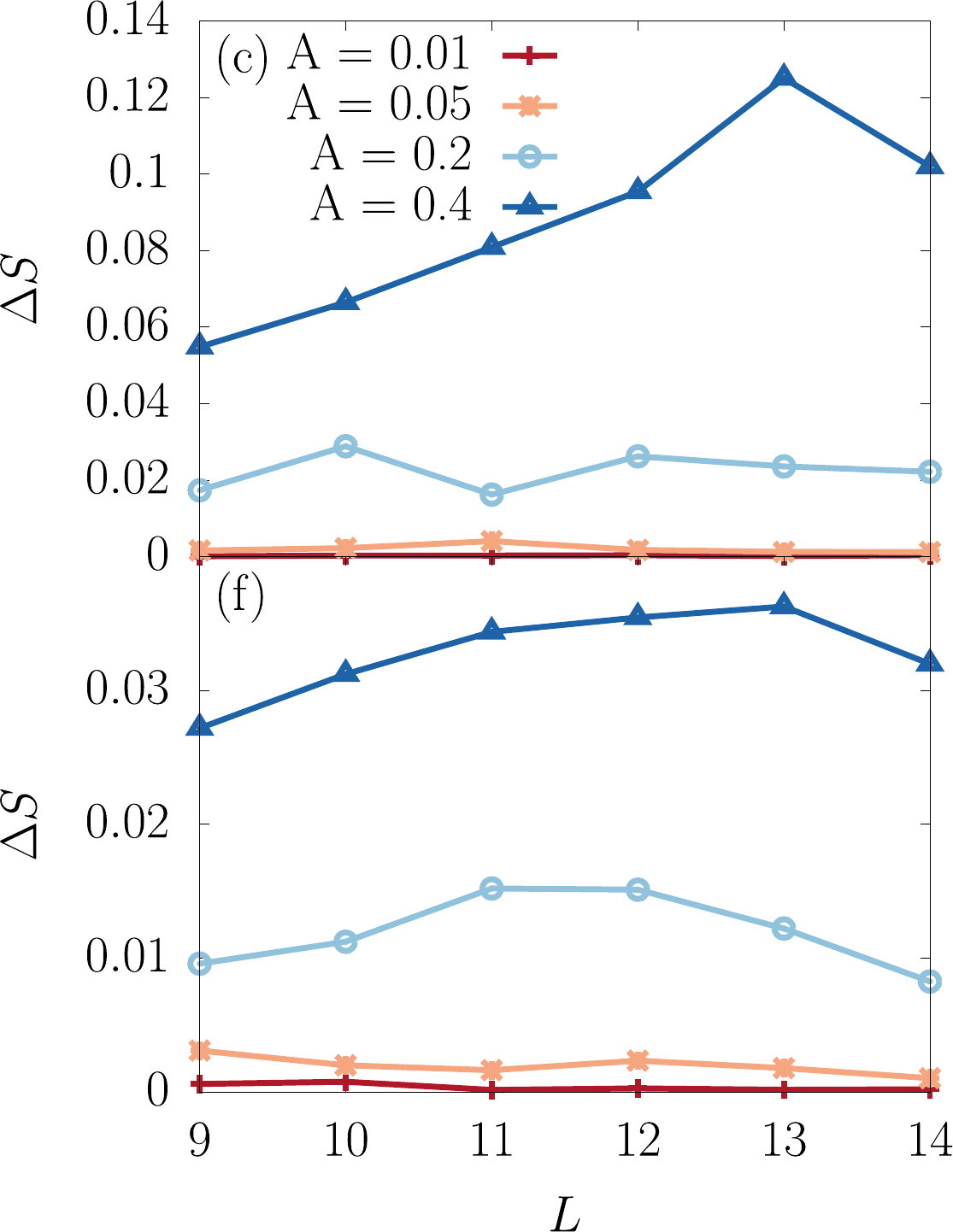}
\caption{{\it Average DSF error as a function of system size for the long range TFIM}.
Average error in the DSF, $\Delta S$, arising from the different imperfection models. We show the numerical results for two interaction ranges. In the top panels, {\it (a)}, {\it (b)}, {\it (c)} we show the results for the interaction exponent $\alpha = 1.5$, while in panels {\it (d)}, {\it (e)}, and {\it (f)} we show the results for  $\alpha = 6$.
{\it (a), (d)}, We show the effects of harmonically modulated Ising interactions. {\it (b), (e)} Average error in the DSF for the case of random interactions. {\it (c), (f)} Average error in the DSF arising from random transverse fields.
For all these cases, at the experimental level of control over the different imperfections, $A < 5\%$, the error is small and constant along the whole range of sizes. 
When the imperfection level is below $20\%$, it becomes negligible for all system sizes and interaction ranges. 
At the current levels of experimental control, the error remain constant and small even at the smallest size studied here, with the DSF remaining unchanged through the entire $\alpha$ range employed in this work. }
\label{scale_long_range}
\end{figure*}
\section{Influence of experimental imperfections on the DSF of the long range TFIM}
\label{long_range_noise}

In the previous section we have assessed the accuracy of a DSF measurement for the short range TFIM using quantum simulators,
and have shown that the DSF is well behaved in the presence of experimental imperfections even at large system sizes. 
Now, we will put forward the idea that {\it quantum simulators can recover the DSF of long range models accurately for system sizes 
larger than state of the art classical simulations can treat}. 
To show this, we will numerically study the long range TFIM in the presence of the same  experimental imperfections as for the short range TFIM.

In Fig.~\ref{DSF_long_range_main} we show the DSF in the absence of imperfections for three different values of $\alpha$. 
In section IV of the supplementary material we show the results for $\alpha = 2$, $3$, and $6$. 
Furthermore, in section V of the supplementary information we show the heat maps for the unequal time correlations with respect to the middle of the chain,  $C^{x,x}_{i,5}(t)$. 
In the regime $1 < \alpha < 2$ our results exhibit a particular signature in the DSF which has no $\omega-$dependence. For higher values of $\alpha$ remnants of this behaviour are noticed, but  an $\omega-$dependence is recovered. For  $\alpha > 3$ we recover the cosine shaped two particle continuum. 
These results, especially the absence of $\omega$-dependence for $\alpha < 2$, indicates that the signatures of excitation confinement, which have been recently proposed \cite{liu19, lerose19,verdel19} can be observed in dynamical quantum simulators via the DSF employing our proposed method. 

In section IV in the supplementary material  
we show a typical case for the the maximal error as a function of frequency (reciprocal space) Eq.~(\ref{error_DS1}) 
(Eq.~(\ref{error_DS2})) in the case of random transverse fields. 
There we can see that the overall behaviour of the error is very similar to that for the short range TFIM. There is a large error around the gap, with small fluctuations at other values of $\omega$ for strong imperfections. For small imperfection levels ($1\% - 5\%$) the error in the DSF is negligible, for all imperfections models, as it was found for the short range TFIM.

In Fig.~\ref{alpha_long_range} we show the integrated error (\ref{integrated_error}) as a function of the interaction range $\alpha$ for the models corresponding to evolution imperfections. 
Fig.~\ref{alpha_long_range}(a) show the error for the case of laser intensity fluctuations, while Fig.~\ref{alpha_long_range}(b) and (c) show the random fields and random interactions respectively.
 In both cases we see two regimes, where the error drastically changes for $1< \alpha < 3$, while it stabilizes for $\alpha > 3$. For the laser intensity fluctuations, the error monotonically increases in the first regime, and saturates in the second.
On the other hand, the opposite behaviour is observed for the lattice imperfections, where the error decreases as a function of $\alpha$. 

While the errors change as the value of $\alpha$ is modified, at the imperfection levels present in the current architectures the integrated error is negligible,
indicating that the DSF at all values of $\alpha$ can be probed using these setups and the measurement would yield accurate results. 

\subsection{System size scaling for long range models}

Now we concentrate on the scaling properties of the DSF of the long range TFIM. We study system sizes ranging from $L = 9$ to $L=14$ sites employing full exact diagonalization, and analyze how the integrated error changes with size. 
Since current architectures can simulate up to approximately 50 sites \cite{Zhang17, Bernien17}, this is a playground in which the DSF can be employed to explore the potential of dynamical analogue quantum simulators.
We show the scaling properties of the error for $\alpha = 1.5$ and $\alpha = 6$. 
In Fig.~\ref{scale_long_range} we show the integrated error originating from the evolution imperfections, as a function of system size, for the two aforementioned values of $\alpha$. 

For imperfection levels below $10\%$, the integrated error is relatively constant over the full range of system sizes studied here. 
When the imperfection level is reduced further, below $5\%$ ( within current experimental capabilities), it becomes negligible for all system sizes and interaction ranges. 
Our data suggests that the error remains constant and small at even the smallest sizes studied here, indicates that the integrated DSF error is intensive with respect to the system sizes, and as such we expect that the scaling properties will be maintained for larger chains. 
Furthermore, our data indicates that a dynamical quantum simulator can measure the DSF of the long range TFIM accurately, even in the presence of realistic experimental imperfections, for system sizes considerably bigger than what is currently achievable with numerical simulations, thus paving the way towards a practical quantum advantage. 

\section{Conclusions}

In this work, we propose the observation of dynamical structure factors as a
practical application of dynamical quantum simulators. We have shown that the \bqp-hardness of general
local Hamiltonian time evolution is inherited by the dynamical structure factor, suggesting that 
its efficient classical computation might be infeasible also for practically relevant instances. In this 
endeavour, we build on the measurement protocol of Ref.~\cite{knap13} and tomographic ideas, allowing 
to measure the DSF in several different quantum architectures. These architectures include those of 
trapped ion, Rydberg atoms, cold atoms in optical lattices, and superconducting qubits.
 
 To emphasize the feasibility of this approach, we study the robustness of the DSF against several meaningful
models of experimental imperfections for the short and long range transverse field Ising model (TFIM).
%
Our results for the short range TFIM indicate that the overall features of the DSF are preserved when one considers state of the art setups, their associated experimental imperfections, and the current level of control over them. 
For the long range model, 
we observe that the effects of imperfections at the current experimental levels and for the system sizes studied in this work do not change the DSF.
We have brought our findings into contact with signatures of the exotic physics in the long range TFIM,
%
in particular the confinement of excitations which has been recently reported. 
Unlike previous studies, we observe the signatures of confinement in both the DSF and correlators in equilibrium, i.e., without the need to quench the Hamiltonian, with equilibrium being a fundamental requirement of our proposed DSF measurement protocol.
Following the study of these imperfections we carry out a system size scaling, which indicates that the errors of the DSF induced by these imperfections are controlled over the whole range of sizes -- remaining small and constant. 
This indicates that for the imperfections considered in this work, a quantum simulation experiment 
with system sizes considerably bigger than what state of the art classical algorithms can achieve is expected to yield accurate results.
We therefore argue that the measurement of DSFs in quantum simulators provides a useful tool to assess
 time dependent quantities of key importance in condensed matter physics, and further place
 quantum simulators into the realm of quantum technological devices \cite{Roadmap}. We hope that the present work 
 stimulates further assessments of this quantity in other physical contexts.

\section{Acknowledgements}
M.~L.~B. wishes to thank Dante Kennes, Markus Heyl, David Luitz, Pedram Roushan, Pietro Silvi, Jan Carl Budich, Bela Bauer, and Thierry Giamarchi for fruitful discussions.  J.~E.~ acknowledges discussions with Holger Boche.
This work has been supported by the ERC  (TAQ),  the  Templeton  Foundation,  the FQXi, and  the  DFG  (EI 519/14-1,  EI 519/15-1,  CRC 183, FOR 2724) and MATH+.  This work has also received funding from the European Unions Horizon 2020 research and innovation programme under grant agreement No 817482 (PASQuanS). 
J.~B.~V.~acknowledges funding from the European Union’s Horizon 2020 research and innovation programme under the Marie Skłodowska-Curie grant agreement Nº 754446 and UGR Research and Knowledge Transfer Found – Athenea3i.

\bibliography{ref}

\begin{thebibliography}{62}%
\makeatletter
\providecommand \@ifxundefined [1]{%
 \@ifx{#1\undefined}
}%
\providecommand \@ifnum [1]{%
 \ifnum #1\expandafter \@firstoftwo
 \else \expandafter \@secondoftwo
 \fi
}%
\providecommand \@ifx [1]{%
 \ifx #1\expandafter \@firstoftwo
 \else \expandafter \@secondoftwo
 \fi
}%
\providecommand \natexlab [1]{#1}%
\providecommand \enquote  [1]{``#1''}%
\providecommand \bibnamefont  [1]{#1}%
\providecommand \bibfnamefont [1]{#1}%
\providecommand \citenamefont [1]{#1}%
\providecommand \href@noop [0]{\@secondoftwo}%
\providecommand \href [0]{\begingroup \@sanitize@url \@href}%
\providecommand \@href[1]{\@@startlink{#1}\@@href}%
\providecommand \@@href[1]{\endgroup#1\@@endlink}%
\providecommand \@sanitize@url [0]{\catcode `\\12\catcode `\$12\catcode
  `\&12\catcode `\#12\catcode `\^12\catcode `\_12\catcode `\%12\relax}%
\providecommand \@@startlink[1]{}%
\providecommand \@@endlink[0]{}%
\providecommand \url  [0]{\begingroup\@sanitize@url \@url }%
\providecommand \@url [1]{\endgroup\@href {#1}{\urlprefix }}%
\providecommand \urlprefix  [0]{URL }%
\providecommand \Eprint [0]{\href }%
\providecommand \doibase [0]{http://dx.doi.org/}%
\providecommand \selectlanguage [0]{\@gobble}%
\providecommand \bibinfo  [0]{\@secondoftwo}%
\providecommand \bibfield  [0]{\@secondoftwo}%
\providecommand \translation [1]{[#1]}%
\providecommand \BibitemOpen [0]{}%
\providecommand \bibitemStop [0]{}%
\providecommand \bibitemNoStop [0]{.\EOS\space}%
\providecommand \EOS [0]{\spacefactor3000\relax}%
\providecommand \BibitemShut  [1]{\csname bibitem#1\endcsname}%
\let\auto@bib@innerbib\@empty
\bibitem [{\citenamefont {Foulkes}\ \emph {et~al.}(2001)\citenamefont
  {Foulkes}, \citenamefont {Mitas}, \citenamefont {Needs},\ and\ \citenamefont
  {Rajagopal}}]{foulkes01}%
  \BibitemOpen
  \bibfield  {author} {\bibinfo {author} {\bibfnamefont {W.~M.~C.}\
  \bibnamefont {Foulkes}}, \bibinfo {author} {\bibfnamefont {L.}~\bibnamefont
  {Mitas}}, \bibinfo {author} {\bibfnamefont {R.~J.}\ \bibnamefont {Needs}}, \
  and\ \bibinfo {author} {\bibfnamefont {G.}~\bibnamefont {Rajagopal}},\ }\href
  {\doibase 10.1103/RevModPhys.73.33} {\bibfield  {journal} {\bibinfo
  {journal} {Rev. Mod. Phys.}\ }\textbf {\bibinfo {volume} {73}},\ \bibinfo
  {pages} {33} (\bibinfo {year} {2001})}\BibitemShut {NoStop}%
\bibitem [{\citenamefont {Noack}\ and\ \citenamefont
  {Manmana}(2005)}]{noack05}%
  \BibitemOpen
  \bibfield  {author} {\bibinfo {author} {\bibfnamefont {R.~M.}\ \bibnamefont
  {Noack}}\ and\ \bibinfo {author} {\bibfnamefont {S.~R.}\ \bibnamefont
  {Manmana}},\ }\href {\doibase 10.1063/1.2080349} {\bibfield  {journal}
  {\bibinfo  {journal} {AIP Conf. Proc.}\ }\textbf {\bibinfo {volume} {789}},\
  \bibinfo {pages} {93} (\bibinfo {year} {2005})}\BibitemShut {NoStop}%
\bibitem [{\citenamefont {{Orus}}(2018)}]{orus18}%
  \BibitemOpen
  \bibfield  {author} {\bibinfo {author} {\bibfnamefont {R.}~\bibnamefont
  {{Orus}}},\ }\href@noop {} {\bibfield  {journal} {\bibinfo  {journal} {arXiv
  e-prints}\ } (\bibinfo {year} {2018})},\ \Eprint
  {http://arxiv.org/abs/1812.04011} {arXiv:1812.04011} \BibitemShut {NoStop}%
\bibitem [{\citenamefont {{Aaronson}}(2005)}]{aaronson05}%
  \BibitemOpen
  \bibfield  {author} {\bibinfo {author} {\bibfnamefont {S.}~\bibnamefont
  {{Aaronson}}},\ }\href@noop {} {\bibfield  {journal} {\bibinfo  {journal}
  {arXiv e-prints}\ } (\bibinfo {year} {2005})},\ \Eprint
  {http://arxiv.org/abs/quant-ph/0502072} {arXiv:quant-ph/0502072 [quant-ph]}
  \BibitemShut {NoStop}%
\bibitem [{\citenamefont {Aaronson}(2009)}]{aaronson_bqp_2009}%
  \BibitemOpen
  \bibfield  {author} {\bibinfo {author} {\bibfnamefont {S.}~\bibnamefont
  {Aaronson}},\ }\href@noop {} {\  (\bibinfo {year} {2009})},\ \Eprint
  {http://arxiv.org/abs/arXiv:0910.4698} {arXiv:0910.4698} \BibitemShut
  {NoStop}%
\bibitem [{\citenamefont {Ran}\ and\ \citenamefont {Tal}(2019)}]{separation}%
  \BibitemOpen
  \bibfield  {author} {\bibinfo {author} {\bibfnamefont {R.}~\bibnamefont
  {Ran}}\ and\ \bibinfo {author} {\bibfnamefont {A.}~\bibnamefont {Tal}},\
  }\href {\doibase 10.1145/3313276.3316315} {\bibfield  {journal} {\bibinfo
  {journal} {Proc. STOC 51st Ann. ACM SIGACT Symp. Th. Comp.}\ ,\ \bibinfo
  {pages} {13}} (\bibinfo {year} {2019})}\BibitemShut {NoStop}%
\bibitem [{\citenamefont {Marvian}\ \emph {et~al.}(2019)\citenamefont
  {Marvian}, \citenamefont {Lidar},\ and\ \citenamefont {Hen}}]{Curing1}%
  \BibitemOpen
  \bibfield  {author} {\bibinfo {author} {\bibfnamefont {M.}~\bibnamefont
  {Marvian}}, \bibinfo {author} {\bibfnamefont {D.~A.}\ \bibnamefont {Lidar}},
  \ and\ \bibinfo {author} {\bibfnamefont {I.}~\bibnamefont {Hen}},\ }\href
  {\doibase 10.1038/s41467-019-09501-6} {\bibfield  {journal} {\bibinfo
  {journal} {Nature Comm.}\ }\textbf {\bibinfo {volume} {10}},\ \bibinfo
  {pages} {1571} (\bibinfo {year} {2019})}\BibitemShut {NoStop}%
\bibitem [{\citenamefont {Klassen}\ \emph {et~al.}()\citenamefont {Klassen},
  \citenamefont {Marvian}, \citenamefont {Piddock}, \citenamefont {Ioannou},
  \citenamefont {Hen},\ and\ \citenamefont {Terhal}}]{Curing2}%
  \BibitemOpen
  \bibfield  {author} {\bibinfo {author} {\bibfnamefont {J.}~\bibnamefont
  {Klassen}}, \bibinfo {author} {\bibfnamefont {M.}~\bibnamefont {Marvian}},
  \bibinfo {author} {\bibfnamefont {S.}~\bibnamefont {Piddock}}, \bibinfo
  {author} {\bibfnamefont {M.}~\bibnamefont {Ioannou}}, \bibinfo {author}
  {\bibfnamefont {I.}~\bibnamefont {Hen}}, \ and\ \bibinfo {author}
  {\bibfnamefont {B.}~\bibnamefont {Terhal}},\ }\href@noop {} {\ }\Eprint
  {http://arxiv.org/abs/1906.08800} {arXiv:1906.08800} \BibitemShut {NoStop}%
\bibitem [{\citenamefont {{Hangleiter}}\ \emph {et~al.}(2019)\citenamefont
  {{Hangleiter}}, \citenamefont {{Roth}}, \citenamefont {{Nagaj}},\ and\
  \citenamefont {{Eisert}}}]{Hangleiter19}%
  \BibitemOpen
  \bibfield  {author} {\bibinfo {author} {\bibfnamefont {D.}~\bibnamefont
  {{Hangleiter}}}, \bibinfo {author} {\bibfnamefont {I.}~\bibnamefont
  {{Roth}}}, \bibinfo {author} {\bibfnamefont {D.}~\bibnamefont {{Nagaj}}}, \
  and\ \bibinfo {author} {\bibfnamefont {J.}~\bibnamefont {{Eisert}}},\
  }\href@noop {} {\bibfield  {journal} {\bibinfo  {journal} {arXiv e-prints}\ }
  (\bibinfo {year} {2019})},\ \Eprint {http://arxiv.org/abs/1906.02309}
  {arXiv:1906.02309 [quant-ph]} \BibitemShut {NoStop}%
\bibitem [{\citenamefont {Cirac}\ and\ \citenamefont
  {Zoller}(2012)}]{CiracZollerSimulation}%
  \BibitemOpen
  \bibfield  {author} {\bibinfo {author} {\bibfnamefont {J.~I.}\ \bibnamefont
  {Cirac}}\ and\ \bibinfo {author} {\bibfnamefont {P.}~\bibnamefont {Zoller}},\
  }\href {\doibase doi.org/10.1038/nphys2275} {\bibfield  {journal} {\bibinfo
  {journal} {Nature Phys.}\ }\textbf {\bibinfo {volume} {8}},\ \bibinfo {pages}
  {264} (\bibinfo {year} {2012})}\BibitemShut {NoStop}%
\bibitem [{\citenamefont {Eisert}\ \emph {et~al.}(2015)\citenamefont {Eisert},
  \citenamefont {Friesdorf},\ and\ \citenamefont {Gogolin}}]{jens15}%
  \BibitemOpen
  \bibfield  {author} {\bibinfo {author} {\bibfnamefont {J.}~\bibnamefont
  {Eisert}}, \bibinfo {author} {\bibfnamefont {M.}~\bibnamefont {Friesdorf}}, \
  and\ \bibinfo {author} {\bibfnamefont {C.}~\bibnamefont {Gogolin}},\ }\href
  {\doibase 10.1038/nphys3215} {\bibfield  {journal} {\bibinfo  {journal}
  {Nature Phys.}\ }\textbf {\bibinfo {volume} {11}},\ \bibinfo {pages} {124}
  (\bibinfo {year} {2015})}\BibitemShut {NoStop}%
\bibitem [{\citenamefont {Fukuhara}\ \emph
  {et~al.}(2013{\natexlab{a}})\citenamefont {Fukuhara}, \citenamefont
  {Schau{\ss}}, \citenamefont {Endres}, \citenamefont {Hild}, \citenamefont
  {Cheneau}, \citenamefont {Bloch},\ and\ \citenamefont
  {Gross}}]{Fukuhara2013}%
  \BibitemOpen
  \bibfield  {author} {\bibinfo {author} {\bibfnamefont {T.}~\bibnamefont
  {Fukuhara}}, \bibinfo {author} {\bibfnamefont {P.}~\bibnamefont
  {Schau{\ss}}}, \bibinfo {author} {\bibfnamefont {M.}~\bibnamefont {Endres}},
  \bibinfo {author} {\bibfnamefont {S.}~\bibnamefont {Hild}}, \bibinfo {author}
  {\bibfnamefont {M.}~\bibnamefont {Cheneau}}, \bibinfo {author} {\bibfnamefont
  {I.}~\bibnamefont {Bloch}}, \ and\ \bibinfo {author} {\bibfnamefont
  {C.}~\bibnamefont {Gross}},\ }\href {\doibase 10.1038/nature12541} {\bibfield
   {journal} {\bibinfo  {journal} {Nature}\ }\textbf {\bibinfo {volume}
  {502}},\ \bibinfo {pages} {76} (\bibinfo {year}
  {2013}{\natexlab{a}})}\BibitemShut {NoStop}%
\bibitem [{\citenamefont {Fukuhara}\ \emph
  {et~al.}(2013{\natexlab{b}})\citenamefont {Fukuhara}, \citenamefont
  {Kantian}, \citenamefont {Endres}, \citenamefont {Cheneau}, \citenamefont
  {Schau{\ss}}, \citenamefont {Hild}, \citenamefont {Bellem}, \citenamefont
  {Schollw{\"o}ck}, \citenamefont {Giamarchi}, \citenamefont {Gross},
  \citenamefont {Bloch},\ and\ \citenamefont {Kuhr}}]{fukuhara13-1}%
  \BibitemOpen
  \bibfield  {author} {\bibinfo {author} {\bibfnamefont {T.}~\bibnamefont
  {Fukuhara}}, \bibinfo {author} {\bibfnamefont {A.}~\bibnamefont {Kantian}},
  \bibinfo {author} {\bibfnamefont {M.}~\bibnamefont {Endres}}, \bibinfo
  {author} {\bibfnamefont {M.}~\bibnamefont {Cheneau}}, \bibinfo {author}
  {\bibfnamefont {P.}~\bibnamefont {Schau{\ss}}}, \bibinfo {author}
  {\bibfnamefont {S.}~\bibnamefont {Hild}}, \bibinfo {author} {\bibfnamefont
  {D.}~\bibnamefont {Bellem}}, \bibinfo {author} {\bibfnamefont
  {U.}~\bibnamefont {Schollw{\"o}ck}}, \bibinfo {author} {\bibfnamefont
  {T.}~\bibnamefont {Giamarchi}}, \bibinfo {author} {\bibfnamefont
  {C.}~\bibnamefont {Gross}}, \bibinfo {author} {\bibfnamefont
  {I.}~\bibnamefont {Bloch}}, \ and\ \bibinfo {author} {\bibfnamefont
  {S.}~\bibnamefont {Kuhr}},\ }\href {\doibase 10.1038/nphys2561} {\bibfield
  {journal} {\bibinfo  {journal} {Nature Phys.}\ }\textbf {\bibinfo {volume}
  {9}},\ \bibinfo {pages} {235} (\bibinfo {year}
  {2013}{\natexlab{b}})}\BibitemShut {NoStop}%
\bibitem [{\citenamefont {Cheneau}\ \emph {et~al.}(2012)\citenamefont
  {Cheneau}, \citenamefont {Barmettler}, \citenamefont {Poletti}, \citenamefont
  {Endres}, \citenamefont {Schau{\ss}}, \citenamefont {Fukuhara}, \citenamefont
  {Gross}, \citenamefont {Bloch}, \citenamefont {Kollath},\ and\ \citenamefont
  {Kuhr}}]{cheneau12}%
  \BibitemOpen
  \bibfield  {author} {\bibinfo {author} {\bibfnamefont {M.}~\bibnamefont
  {Cheneau}}, \bibinfo {author} {\bibfnamefont {P.}~\bibnamefont {Barmettler}},
  \bibinfo {author} {\bibfnamefont {D.}~\bibnamefont {Poletti}}, \bibinfo
  {author} {\bibfnamefont {M.}~\bibnamefont {Endres}}, \bibinfo {author}
  {\bibfnamefont {P.}~\bibnamefont {Schau{\ss}}}, \bibinfo {author}
  {\bibfnamefont {T.}~\bibnamefont {Fukuhara}}, \bibinfo {author}
  {\bibfnamefont {C.}~\bibnamefont {Gross}}, \bibinfo {author} {\bibfnamefont
  {I.}~\bibnamefont {Bloch}}, \bibinfo {author} {\bibfnamefont
  {C.}~\bibnamefont {Kollath}}, \ and\ \bibinfo {author} {\bibfnamefont
  {S.}~\bibnamefont {Kuhr}},\ }\href {https://doi.org/10.1038/nature10748}
  {\bibfield  {journal} {\bibinfo  {journal} {Nature}\ }\textbf {\bibinfo
  {volume} {481}},\ \bibinfo {pages} {484} (\bibinfo {year}
  {2012})}\BibitemShut {NoStop}%
\bibitem [{\citenamefont {Simon}\ \emph {et~al.}(2011)\citenamefont {Simon},
  \citenamefont {Bakr}, \citenamefont {Ma}, \citenamefont {Tai}, \citenamefont
  {Preiss},\ and\ \citenamefont {Greiner}}]{simon11}%
  \BibitemOpen
  \bibfield  {author} {\bibinfo {author} {\bibfnamefont {J.}~\bibnamefont
  {Simon}}, \bibinfo {author} {\bibfnamefont {W.~S.}\ \bibnamefont {Bakr}},
  \bibinfo {author} {\bibfnamefont {R.}~\bibnamefont {Ma}}, \bibinfo {author}
  {\bibfnamefont {M.~E.}\ \bibnamefont {Tai}}, \bibinfo {author} {\bibfnamefont
  {P.~M.}\ \bibnamefont {Preiss}}, \ and\ \bibinfo {author} {\bibfnamefont
  {M.}~\bibnamefont {Greiner}},\ }\href {\doibase 10.1038/nature09994}
  {\bibfield  {journal} {\bibinfo  {journal} {Nature}\ }\textbf {\bibinfo
  {volume} {472}},\ \bibinfo {pages} {307} (\bibinfo {year}
  {2011})}\BibitemShut {NoStop}%
\bibitem [{\citenamefont {Bernien}\ \emph {et~al.}(2017)\citenamefont
  {Bernien}, \citenamefont {Schwartz}, \citenamefont {Keesling}, \citenamefont
  {Levine}, \citenamefont {Omran}, \citenamefont {Pichler}, \citenamefont
  {Choi}, \citenamefont {Zibrov}, \citenamefont {Endres}, \citenamefont
  {Greiner}, \citenamefont {Vuleti{\'c}},\ and\ \citenamefont
  {Lukin}}]{Bernien17}%
  \BibitemOpen
  \bibfield  {author} {\bibinfo {author} {\bibfnamefont {H.}~\bibnamefont
  {Bernien}}, \bibinfo {author} {\bibfnamefont {S.}~\bibnamefont {Schwartz}},
  \bibinfo {author} {\bibfnamefont {A.}~\bibnamefont {Keesling}}, \bibinfo
  {author} {\bibfnamefont {H.}~\bibnamefont {Levine}}, \bibinfo {author}
  {\bibfnamefont {A.}~\bibnamefont {Omran}}, \bibinfo {author} {\bibfnamefont
  {H.}~\bibnamefont {Pichler}}, \bibinfo {author} {\bibfnamefont
  {S.}~\bibnamefont {Choi}}, \bibinfo {author} {\bibfnamefont {A.~S.}\
  \bibnamefont {Zibrov}}, \bibinfo {author} {\bibfnamefont {M.}~\bibnamefont
  {Endres}}, \bibinfo {author} {\bibfnamefont {M.}~\bibnamefont {Greiner}},
  \bibinfo {author} {\bibfnamefont {V.}~\bibnamefont {Vuleti{\'c}}}, \ and\
  \bibinfo {author} {\bibfnamefont {M.~D.}\ \bibnamefont {Lukin}},\ }\href
  {\doibase /doi.org/10.1038/nature24622} {\bibfield  {journal} {\bibinfo
  {journal} {Nature}\ }\textbf {\bibinfo {volume} {551}},\ \bibinfo {pages}
  {579} (\bibinfo {year} {2017})}\BibitemShut {NoStop}%
\bibitem [{\citenamefont {Zhang}\ \emph {et~al.}(2017)\citenamefont {Zhang},
  \citenamefont {Pagano}, \citenamefont {Hess}, \citenamefont {Kyprianidis},
  \citenamefont {Becker}, \citenamefont {Kaplan}, \citenamefont {Gorshkov},
  \citenamefont {Gong},\ and\ \citenamefont {Monroe}}]{Zhang17}%
  \BibitemOpen
  \bibfield  {author} {\bibinfo {author} {\bibfnamefont {J.}~\bibnamefont
  {Zhang}}, \bibinfo {author} {\bibfnamefont {G.}~\bibnamefont {Pagano}},
  \bibinfo {author} {\bibfnamefont {P.~W.}\ \bibnamefont {Hess}}, \bibinfo
  {author} {\bibfnamefont {A.}~\bibnamefont {Kyprianidis}}, \bibinfo {author}
  {\bibfnamefont {P.}~\bibnamefont {Becker}}, \bibinfo {author} {\bibfnamefont
  {H.}~\bibnamefont {Kaplan}}, \bibinfo {author} {\bibfnamefont {A.~V.}\
  \bibnamefont {Gorshkov}}, \bibinfo {author} {\bibfnamefont {Z.~X.}\
  \bibnamefont {Gong}}, \ and\ \bibinfo {author} {\bibfnamefont
  {C.}~\bibnamefont {Monroe}},\ }\href {\doibase 10.1038/nature24654}
  {\bibfield  {journal} {\bibinfo  {journal} {Nature}\ }\textbf {\bibinfo
  {volume} {551}},\ \bibinfo {pages} {601} (\bibinfo {year}
  {2017})}\BibitemShut {NoStop}%
\bibitem [{\citenamefont {Bohnet}\ \emph {et~al.}(2016)\citenamefont {Bohnet},
  \citenamefont {Sawyer}, \citenamefont {Britton}, \citenamefont {Wall},
  \citenamefont {Rey}, \citenamefont {Foss-Feig},\ and\ \citenamefont
  {Bollinger}}]{Bohnet16}%
  \BibitemOpen
  \bibfield  {author} {\bibinfo {author} {\bibfnamefont {J.~G.}\ \bibnamefont
  {Bohnet}}, \bibinfo {author} {\bibfnamefont {B.~C.}\ \bibnamefont {Sawyer}},
  \bibinfo {author} {\bibfnamefont {J.~W.}\ \bibnamefont {Britton}}, \bibinfo
  {author} {\bibfnamefont {M.~L.}\ \bibnamefont {Wall}}, \bibinfo {author}
  {\bibfnamefont {A.~M.}\ \bibnamefont {Rey}}, \bibinfo {author} {\bibfnamefont
  {M.}~\bibnamefont {Foss-Feig}}, \ and\ \bibinfo {author} {\bibfnamefont
  {J.~J.}\ \bibnamefont {Bollinger}},\ }\href {\doibase
  10.1126/science.aad9958} {\bibfield  {journal} {\bibinfo  {journal}
  {Science}\ }\textbf {\bibinfo {volume} {352}},\ \bibinfo {pages} {1297}
  (\bibinfo {year} {2016})}\BibitemShut {NoStop}%
\bibitem [{\citenamefont {Islam}\ \emph {et~al.}(2011)\citenamefont {Islam},
  \citenamefont {Edwards}, \citenamefont {Kim}, \citenamefont {Korenblit},
  \citenamefont {Noh}, \citenamefont {Carmichael}, \citenamefont {Lin},
  \citenamefont {Duan}, \citenamefont {Joseph~Wang}, \citenamefont
  {Freericks},\ and\ \citenamefont {Monroe}}]{islam11}%
  \BibitemOpen
  \bibfield  {author} {\bibinfo {author} {\bibfnamefont {R.}~\bibnamefont
  {Islam}}, \bibinfo {author} {\bibfnamefont {E.~E.}\ \bibnamefont {Edwards}},
  \bibinfo {author} {\bibfnamefont {K.}~\bibnamefont {Kim}}, \bibinfo {author}
  {\bibfnamefont {S.}~\bibnamefont {Korenblit}}, \bibinfo {author}
  {\bibfnamefont {C.}~\bibnamefont {Noh}}, \bibinfo {author} {\bibfnamefont
  {H.}~\bibnamefont {Carmichael}}, \bibinfo {author} {\bibfnamefont {G.~D.}\
  \bibnamefont {Lin}}, \bibinfo {author} {\bibfnamefont {L.~M.}\ \bibnamefont
  {Duan}}, \bibinfo {author} {\bibfnamefont {C.~C.}\ \bibnamefont
  {Joseph~Wang}}, \bibinfo {author} {\bibfnamefont {J.~K.}\ \bibnamefont
  {Freericks}}, \ and\ \bibinfo {author} {\bibfnamefont {C.}~\bibnamefont
  {Monroe}},\ }\href {\doibase 10.1038/ncomms1374} {\bibfield  {journal}
  {\bibinfo  {journal} {Nature Comm.}\ }\textbf {\bibinfo {volume} {2}},\
  \bibinfo {pages} {377} (\bibinfo {year} {2011})}\BibitemShut {NoStop}%
\bibitem [{\citenamefont {Trotzky}\ \emph {et~al.}(2012)\citenamefont
  {Trotzky}, \citenamefont {Chen}, \citenamefont {Flesch}, \citenamefont
  {McCulloch}, \citenamefont {Schollw{\"o}ck}, \citenamefont {Eisert},\ and\
  \citenamefont {Bloch}}]{Trotzky12}%
  \BibitemOpen
  \bibfield  {author} {\bibinfo {author} {\bibfnamefont {S.}~\bibnamefont
  {Trotzky}}, \bibinfo {author} {\bibfnamefont {Y.-A.}\ \bibnamefont {Chen}},
  \bibinfo {author} {\bibfnamefont {A.}~\bibnamefont {Flesch}}, \bibinfo
  {author} {\bibfnamefont {I.~P.}\ \bibnamefont {McCulloch}}, \bibinfo {author}
  {\bibfnamefont {U.}~\bibnamefont {Schollw{\"o}ck}}, \bibinfo {author}
  {\bibfnamefont {J.}~\bibnamefont {Eisert}}, \ and\ \bibinfo {author}
  {\bibfnamefont {I.}~\bibnamefont {Bloch}},\ }\href {\doibase
  10.1038/nphys2232} {\bibfield  {journal} {\bibinfo  {journal} {Nature Phys.}\
  }\textbf {\bibinfo {volume} {8}},\ \bibinfo {pages} {325} (\bibinfo {year}
  {2012})}\BibitemShut {NoStop}%
\bibitem [{\citenamefont {Coldea}\ \emph {et~al.}(2010)\citenamefont {Coldea},
  \citenamefont {Tennant}, \citenamefont {Wheeler}, \citenamefont {Wawrzynska},
  \citenamefont {Prabhakaran}, \citenamefont {Telling}, \citenamefont
  {Habicht}, \citenamefont {Smeibidl},\ and\ \citenamefont
  {Kiefer}}]{Coldea10}%
  \BibitemOpen
  \bibfield  {author} {\bibinfo {author} {\bibfnamefont {R.}~\bibnamefont
  {Coldea}}, \bibinfo {author} {\bibfnamefont {D.~A.}\ \bibnamefont {Tennant}},
  \bibinfo {author} {\bibfnamefont {E.~M.}\ \bibnamefont {Wheeler}}, \bibinfo
  {author} {\bibfnamefont {E.}~\bibnamefont {Wawrzynska}}, \bibinfo {author}
  {\bibfnamefont {D.}~\bibnamefont {Prabhakaran}}, \bibinfo {author}
  {\bibfnamefont {M.}~\bibnamefont {Telling}}, \bibinfo {author} {\bibfnamefont
  {K.}~\bibnamefont {Habicht}}, \bibinfo {author} {\bibfnamefont
  {P.}~\bibnamefont {Smeibidl}}, \ and\ \bibinfo {author} {\bibfnamefont
  {K.}~\bibnamefont {Kiefer}},\ }\href {\doibase 10.1126/science.1180085}
  {\bibfield  {journal} {\bibinfo  {journal} {Science}\ }\textbf {\bibinfo
  {volume} {327}},\ \bibinfo {pages} {177} (\bibinfo {year}
  {2010})}\BibitemShut {NoStop}%
\bibitem [{\citenamefont {Jia}\ \emph {et~al.}(2014)\citenamefont {Jia},
  \citenamefont {Nowadnick}, \citenamefont {Wohlfeld}, \citenamefont {Kung},
  \citenamefont {Chen}, \citenamefont {Johnston}, \citenamefont {Tohyama},
  \citenamefont {Moritz},\ and\ \citenamefont {Devereaux}}]{jia19}%
  \BibitemOpen
  \bibfield  {author} {\bibinfo {author} {\bibfnamefont {C.~J.}\ \bibnamefont
  {Jia}}, \bibinfo {author} {\bibfnamefont {E.~A.}\ \bibnamefont {Nowadnick}},
  \bibinfo {author} {\bibfnamefont {K.}~\bibnamefont {Wohlfeld}}, \bibinfo
  {author} {\bibfnamefont {Y.~F.}\ \bibnamefont {Kung}}, \bibinfo {author}
  {\bibfnamefont {C.~C.}\ \bibnamefont {Chen}}, \bibinfo {author}
  {\bibfnamefont {S.}~\bibnamefont {Johnston}}, \bibinfo {author}
  {\bibfnamefont {T.}~\bibnamefont {Tohyama}}, \bibinfo {author} {\bibfnamefont
  {B.}~\bibnamefont {Moritz}}, \ and\ \bibinfo {author} {\bibfnamefont {T.~P.}\
  \bibnamefont {Devereaux}},\ }\href {https://doi.org/10.1038/ncomms4314}
  {\bibfield  {journal} {\bibinfo  {journal} {Nature Comm.}\ }\textbf {\bibinfo
  {volume} {5}},\ \bibinfo {pages} {3314} (\bibinfo {year} {2014})}\BibitemShut
  {NoStop}%
\bibitem [{\citenamefont {Sachdev}(2011)}]{sachdev11}%
  \BibitemOpen
  \bibfield  {author} {\bibinfo {author} {\bibfnamefont {S.}~\bibnamefont
  {Sachdev}},\ }\href {\doibase 10.1017/CBO9780511973765} {\emph {\bibinfo
  {title} {Quantum phase transitions}}},\ \bibinfo {edition} {2nd}\ ed.\
  (\bibinfo  {publisher} {Cambridge University Press},\ \bibinfo {year}
  {2011})\BibitemShut {NoStop}%
\bibitem [{\citenamefont {Roushan}\ \emph {et~al.}(2017)\citenamefont
  {Roushan}, \citenamefont {Neill}, \citenamefont {Tangpanitanon},
  \citenamefont {Bastidas}, \citenamefont {Megrant}, \citenamefont {Barends},
  \citenamefont {Chen}, \citenamefont {Chen}, \citenamefont {Chiaro},
  \citenamefont {Dunsworth}, \citenamefont {Fowler}, \citenamefont {Foxen},
  \citenamefont {Giustina}, \citenamefont {Jeffrey}, \citenamefont {Kelly},
  \citenamefont {Lucero}, \citenamefont {Mutus}, \citenamefont {Neeley},
  \citenamefont {Quintana}, \citenamefont {Sank}, \citenamefont {Vainsencher},
  \citenamefont {Wenner}, \citenamefont {White}, \citenamefont {Neven},
  \citenamefont {Angelakis},\ and\ \citenamefont {Martinis}}]{Roushan1175}%
  \BibitemOpen
  \bibfield  {author} {\bibinfo {author} {\bibfnamefont {P.}~\bibnamefont
  {Roushan}}, \bibinfo {author} {\bibfnamefont {C.}~\bibnamefont {Neill}},
  \bibinfo {author} {\bibfnamefont {J.}~\bibnamefont {Tangpanitanon}}, \bibinfo
  {author} {\bibfnamefont {V.~M.}\ \bibnamefont {Bastidas}}, \bibinfo {author}
  {\bibfnamefont {A.}~\bibnamefont {Megrant}}, \bibinfo {author} {\bibfnamefont
  {R.}~\bibnamefont {Barends}}, \bibinfo {author} {\bibfnamefont
  {Y.}~\bibnamefont {Chen}}, \bibinfo {author} {\bibfnamefont {Z.}~\bibnamefont
  {Chen}}, \bibinfo {author} {\bibfnamefont {B.}~\bibnamefont {Chiaro}},
  \bibinfo {author} {\bibfnamefont {A.}~\bibnamefont {Dunsworth}}, \bibinfo
  {author} {\bibfnamefont {A.}~\bibnamefont {Fowler}}, \bibinfo {author}
  {\bibfnamefont {B.}~\bibnamefont {Foxen}}, \bibinfo {author} {\bibfnamefont
  {M.}~\bibnamefont {Giustina}}, \bibinfo {author} {\bibfnamefont
  {E.}~\bibnamefont {Jeffrey}}, \bibinfo {author} {\bibfnamefont
  {J.}~\bibnamefont {Kelly}}, \bibinfo {author} {\bibfnamefont
  {E.}~\bibnamefont {Lucero}}, \bibinfo {author} {\bibfnamefont
  {J.}~\bibnamefont {Mutus}}, \bibinfo {author} {\bibfnamefont
  {M.}~\bibnamefont {Neeley}}, \bibinfo {author} {\bibfnamefont
  {C.}~\bibnamefont {Quintana}}, \bibinfo {author} {\bibfnamefont
  {D.}~\bibnamefont {Sank}}, \bibinfo {author} {\bibfnamefont {A.}~\bibnamefont
  {Vainsencher}}, \bibinfo {author} {\bibfnamefont {J.}~\bibnamefont {Wenner}},
  \bibinfo {author} {\bibfnamefont {T.}~\bibnamefont {White}}, \bibinfo
  {author} {\bibfnamefont {H.}~\bibnamefont {Neven}}, \bibinfo {author}
  {\bibfnamefont {D.~G.}\ \bibnamefont {Angelakis}}, \ and\ \bibinfo {author}
  {\bibfnamefont {J.}~\bibnamefont {Martinis}},\ }\href {\doibase
  10.1126/science.aao1401} {\bibfield  {journal} {\bibinfo  {journal}
  {Science}\ }\textbf {\bibinfo {volume} {358}},\ \bibinfo {pages} {1175}
  (\bibinfo {year} {2017})}\BibitemShut {NoStop}%
\bibitem [{\citenamefont {Barredo}\ \emph {et~al.}(2015)\citenamefont
  {Barredo}, \citenamefont {Labuhn}, \citenamefont {Ravets}, \citenamefont
  {Lahaye}, \citenamefont {Browaeys},\ and\ \citenamefont {Adams}}]{barredo15}%
  \BibitemOpen
  \bibfield  {author} {\bibinfo {author} {\bibfnamefont {D.}~\bibnamefont
  {Barredo}}, \bibinfo {author} {\bibfnamefont {H.}~\bibnamefont {Labuhn}},
  \bibinfo {author} {\bibfnamefont {S.}~\bibnamefont {Ravets}}, \bibinfo
  {author} {\bibfnamefont {T.}~\bibnamefont {Lahaye}}, \bibinfo {author}
  {\bibfnamefont {A.}~\bibnamefont {Browaeys}}, \ and\ \bibinfo {author}
  {\bibfnamefont {C.~S.}\ \bibnamefont {Adams}},\ }\href {\doibase
  10.1103/PhysRevLett.114.113002} {\bibfield  {journal} {\bibinfo  {journal}
  {Phys. Rev. Lett.}\ }\textbf {\bibinfo {volume} {114}},\ \bibinfo {pages}
  {113002} (\bibinfo {year} {2015})}\BibitemShut {NoStop}%
\bibitem [{\citenamefont {Liu}\ \emph {et~al.}(2019)\citenamefont {Liu},
  \citenamefont {Lundgren}, \citenamefont {Titum}, \citenamefont {Pagano},
  \citenamefont {Zhang}, \citenamefont {Monroe},\ and\ \citenamefont
  {Gorshkov}}]{liu19}%
  \BibitemOpen
  \bibfield  {author} {\bibinfo {author} {\bibfnamefont {F.}~\bibnamefont
  {Liu}}, \bibinfo {author} {\bibfnamefont {R.}~\bibnamefont {Lundgren}},
  \bibinfo {author} {\bibfnamefont {P.}~\bibnamefont {Titum}}, \bibinfo
  {author} {\bibfnamefont {G.}~\bibnamefont {Pagano}}, \bibinfo {author}
  {\bibfnamefont {J.}~\bibnamefont {Zhang}}, \bibinfo {author} {\bibfnamefont
  {C.}~\bibnamefont {Monroe}}, \ and\ \bibinfo {author} {\bibfnamefont {A.~V.}\
  \bibnamefont {Gorshkov}},\ }\href {\doibase 10.1103/PhysRevLett.122.150601}
  {\bibfield  {journal} {\bibinfo  {journal} {Phys. Rev. Lett.}\ }\textbf
  {\bibinfo {volume} {122}},\ \bibinfo {pages} {150601} (\bibinfo {year}
  {2019})}\BibitemShut {NoStop}%
\bibitem [{\citenamefont {Luitz}\ and\ \citenamefont
  {Bar~Lev}(2019)}]{luitz19}%
  \BibitemOpen
  \bibfield  {author} {\bibinfo {author} {\bibfnamefont {D.~J.}\ \bibnamefont
  {Luitz}}\ and\ \bibinfo {author} {\bibfnamefont {Y.}~\bibnamefont
  {Bar~Lev}},\ }\href {\doibase 10.1103/PhysRevA.99.010105} {\bibfield
  {journal} {\bibinfo  {journal} {Phys. Rev. A}\ }\textbf {\bibinfo {volume}
  {99}},\ \bibinfo {pages} {010105(R)} (\bibinfo {year} {2019})}\BibitemShut
  {NoStop}%
\bibitem [{\citenamefont {{Fratus}}\ and\ \citenamefont
  {{Srednicki}}(2016)}]{fratus16}%
  \BibitemOpen
  \bibfield  {author} {\bibinfo {author} {\bibfnamefont {K.~R.}\ \bibnamefont
  {{Fratus}}}\ and\ \bibinfo {author} {\bibfnamefont {M.}~\bibnamefont
  {{Srednicki}}},\ }\href@noop {} {\  (\bibinfo {year} {2016})},\ \Eprint
  {http://arxiv.org/abs/1611.03992} {arXiv:1611.03992} \BibitemShut {NoStop}%
\bibitem [{\citenamefont {Hauke}\ and\ \citenamefont
  {Tagliacozzo}(2013)}]{hauke13}%
  \BibitemOpen
  \bibfield  {author} {\bibinfo {author} {\bibfnamefont {P.}~\bibnamefont
  {Hauke}}\ and\ \bibinfo {author} {\bibfnamefont {L.}~\bibnamefont
  {Tagliacozzo}},\ }\href {\doibase 10.1103/PhysRevLett.111.207202} {\bibfield
  {journal} {\bibinfo  {journal} {Phys. Rev. Lett.}\ }\textbf {\bibinfo
  {volume} {111}},\ \bibinfo {pages} {207202} (\bibinfo {year}
  {2013})}\BibitemShut {NoStop}%
\bibitem [{\citenamefont {Knap}\ \emph {et~al.}(2013)\citenamefont {Knap},
  \citenamefont {Kantian}, \citenamefont {Giamarchi}, \citenamefont {Bloch},
  \citenamefont {Lukin},\ and\ \citenamefont {Demler}}]{knap13}%
  \BibitemOpen
  \bibfield  {author} {\bibinfo {author} {\bibfnamefont {M.}~\bibnamefont
  {Knap}}, \bibinfo {author} {\bibfnamefont {A.}~\bibnamefont {Kantian}},
  \bibinfo {author} {\bibfnamefont {T.}~\bibnamefont {Giamarchi}}, \bibinfo
  {author} {\bibfnamefont {I.}~\bibnamefont {Bloch}}, \bibinfo {author}
  {\bibfnamefont {M.~D.}\ \bibnamefont {Lukin}}, \ and\ \bibinfo {author}
  {\bibfnamefont {E.}~\bibnamefont {Demler}},\ }\href {\doibase
  10.1103/PhysRevLett.111.147205} {\bibfield  {journal} {\bibinfo  {journal}
  {Phys. Rev. Lett.}\ }\textbf {\bibinfo {volume} {111}},\ \bibinfo {pages}
  {147205} (\bibinfo {year} {2013})}\BibitemShut {NoStop}%
\bibitem [{\citenamefont {Yoshimura}\ and\ \citenamefont
  {Freericks}(2016)}]{yoshimura16}%
  \BibitemOpen
  \bibfield  {author} {\bibinfo {author} {\bibfnamefont {B.~T.}\ \bibnamefont
  {Yoshimura}}\ and\ \bibinfo {author} {\bibfnamefont {J.~K.}\ \bibnamefont
  {Freericks}},\ }\href {\doibase 10.1103/PhysRevA.93.052314} {\bibfield
  {journal} {\bibinfo  {journal} {Phys. Rev. A}\ }\textbf {\bibinfo {volume}
  {93}},\ \bibinfo {pages} {052314} (\bibinfo {year} {2016})}\BibitemShut
  {NoStop}%
\bibitem [{\citenamefont {{Pagano}}\ \emph {et~al.}(2019)\citenamefont
  {{Pagano}}, \citenamefont {{Bapat}}, \citenamefont {{Becker}}, \citenamefont
  {{Collins}}, \citenamefont {{De}}, \citenamefont {{Hess}}, \citenamefont
  {{Kaplan}}, \citenamefont {{Kyprianidis}}, \citenamefont {{Tan}},
  \citenamefont {{Baldwin}}, \citenamefont {{Brady}}, \citenamefont
  {{Deshpande}}, \citenamefont {{Liu}}, \citenamefont {{Jordan}}, \citenamefont
  {{Gorshkov}},\ and\ \citenamefont {{Monroe}}}]{pagano19}%
  \BibitemOpen
  \bibfield  {author} {\bibinfo {author} {\bibfnamefont {G.}~\bibnamefont
  {{Pagano}}}, \bibinfo {author} {\bibfnamefont {A.}~\bibnamefont {{Bapat}}},
  \bibinfo {author} {\bibfnamefont {P.}~\bibnamefont {{Becker}}}, \bibinfo
  {author} {\bibfnamefont {K.~S.}\ \bibnamefont {{Collins}}}, \bibinfo {author}
  {\bibfnamefont {A.}~\bibnamefont {{De}}}, \bibinfo {author} {\bibfnamefont
  {P.~W.}\ \bibnamefont {{Hess}}}, \bibinfo {author} {\bibfnamefont {H.~B.}\
  \bibnamefont {{Kaplan}}}, \bibinfo {author} {\bibfnamefont {A.}~\bibnamefont
  {{Kyprianidis}}}, \bibinfo {author} {\bibfnamefont {W.~L.}\ \bibnamefont
  {{Tan}}}, \bibinfo {author} {\bibfnamefont {C.}~\bibnamefont {{Baldwin}}},
  \bibinfo {author} {\bibfnamefont {L.~T.}\ \bibnamefont {{Brady}}}, \bibinfo
  {author} {\bibfnamefont {A.}~\bibnamefont {{Deshpande}}}, \bibinfo {author}
  {\bibfnamefont {F.}~\bibnamefont {{Liu}}}, \bibinfo {author} {\bibfnamefont
  {S.}~\bibnamefont {{Jordan}}}, \bibinfo {author} {\bibfnamefont {A.~V.}\
  \bibnamefont {{Gorshkov}}}, \ and\ \bibinfo {author} {\bibfnamefont
  {C.}~\bibnamefont {{Monroe}}},\ }\href@noop {} {\bibfield  {journal}
  {\bibinfo  {journal} {arXiv e-prints}\ ,\ \bibinfo {eid} {arXiv:1906.02700}}
  (\bibinfo {year} {2019})},\ \Eprint {http://arxiv.org/abs/1906.02700}
  {arXiv:1906.02700 [quant-ph]} \BibitemShut {NoStop}%
\bibitem [{\citenamefont {Baiesi}\ \emph {et~al.}(2009)\citenamefont {Baiesi},
  \citenamefont {Maes},\ and\ \citenamefont {Wynants}}]{baiesi09}%
  \BibitemOpen
  \bibfield  {author} {\bibinfo {author} {\bibfnamefont {M.}~\bibnamefont
  {Baiesi}}, \bibinfo {author} {\bibfnamefont {C.}~\bibnamefont {Maes}}, \ and\
  \bibinfo {author} {\bibfnamefont {B.}~\bibnamefont {Wynants}},\ }\href
  {\doibase 10.1103/PhysRevLett.103.010602} {\bibfield  {journal} {\bibinfo
  {journal} {Phys. Rev. Lett.}\ }\textbf {\bibinfo {volume} {103}},\ \bibinfo
  {pages} {010602} (\bibinfo {year} {2009})}\BibitemShut {NoStop}%
\bibitem [{\citenamefont {{Calzetta}}\ and\ \citenamefont
  {{Hu}}(2008)}]{non-equ-field-theory}%
  \BibitemOpen
  \bibfield  {author} {\bibinfo {author} {\bibfnamefont {E.~A.}\ \bibnamefont
  {{Calzetta}}}\ and\ \bibinfo {author} {\bibfnamefont {B.-L.~B.}\ \bibnamefont
  {{Hu}}},\ }\href {\doibase 10.1017/CBO9780511535123} {\emph {\bibinfo {title}
  {{Nonequilibrium quantum field theory}}}}\ (\bibinfo {year}
  {2008})\BibitemShut {NoStop}%
\bibitem [{\citenamefont {Boche}\ and\ \citenamefont
  {M{\"o}nich}(2019)}]{Turing}%
  \BibitemOpen
  \bibfield  {author} {\bibinfo {author} {\bibfnamefont {H.}~\bibnamefont
  {Boche}}\ and\ \bibinfo {author} {\bibfnamefont {U.~J.}\ \bibnamefont
  {M{\"o}nich}},\ }\href@noop {} {\  (\bibinfo {year} {2019})},\ \bibinfo
  {note} {in preparation}\BibitemShut {NoStop}%
\bibitem [{\citenamefont {Kempe}\ \emph {et~al.}(2004)\citenamefont {Kempe},
  \citenamefont {Kitaev},\ and\ \citenamefont {Regev}}]{kempe_complexity_2004}%
  \BibitemOpen
  \bibfield  {author} {\bibinfo {author} {\bibfnamefont {J.}~\bibnamefont
  {Kempe}}, \bibinfo {author} {\bibfnamefont {A.}~\bibnamefont {Kitaev}}, \
  and\ \bibinfo {author} {\bibfnamefont {O.}~\bibnamefont {Regev}},\ }\href
  {\doibase 10.1137/S0097539704445226} {\bibfield  {journal} {\bibinfo
  {journal} {SIAM J. Comp.}\ }\textbf {\bibinfo {volume} {35}},\ \bibinfo
  {pages} {1070} (\bibinfo {year} {2004})}\BibitemShut {NoStop}%
\bibitem [{\citenamefont {Gosset}\ \emph {et~al.}(2015)\citenamefont {Gosset},
  \citenamefont {Terhal},\ and\ \citenamefont
  {Vershynina}}]{gosset_adiabatic_2015}%
  \BibitemOpen
  \bibfield  {author} {\bibinfo {author} {\bibfnamefont {D.}~\bibnamefont
  {Gosset}}, \bibinfo {author} {\bibfnamefont {B.~M.}\ \bibnamefont {Terhal}},
  \ and\ \bibinfo {author} {\bibfnamefont {A.}~\bibnamefont {Vershynina}},\
  }\href {\doibase 10.1103/PhysRevLett.114.140501} {\bibfield  {journal}
  {\bibinfo  {journal} {Phys. Rev. Lett.}\ }\textbf {\bibinfo {volume} {114}},\
  \bibinfo {pages} {140501} (\bibinfo {year} {2015})}\BibitemShut {NoStop}%
\bibitem [{\citenamefont {Lloyd}\ and\ \citenamefont
  {Terhal}(2016)}]{lloyd_adiabatic_2016}%
  \BibitemOpen
  \bibfield  {author} {\bibinfo {author} {\bibfnamefont {S.}~\bibnamefont
  {Lloyd}}\ and\ \bibinfo {author} {\bibfnamefont {B.~M.}\ \bibnamefont
  {Terhal}},\ }\href {\doibase 10.1088/1367-2630/18/2/023042} {\bibfield
  {journal} {\bibinfo  {journal} {New J. Phys.}\ }\textbf {\bibinfo {volume}
  {18}},\ \bibinfo {pages} {023042} (\bibinfo {year} {2016})}\BibitemShut
  {NoStop}%
\bibitem [{\citenamefont {Ciani}\ \emph {et~al.}(2019)\citenamefont {Ciani},
  \citenamefont {Terhal},\ and\ \citenamefont
  {DiVinzenzo}}]{ciani_hamiltonian_2018}%
  \BibitemOpen
  \bibfield  {author} {\bibinfo {author} {\bibfnamefont {A.}~\bibnamefont
  {Ciani}}, \bibinfo {author} {\bibfnamefont {B.}~\bibnamefont {Terhal}}, \
  and\ \bibinfo {author} {\bibfnamefont {D.}~\bibnamefont {DiVinzenzo}},\
  }\href {\doibase 10.1088/2058-9565/ab18dd} {\bibfield  {journal} {\bibinfo
  {journal} {Quant. Sc. Tech.}\ }\textbf {\bibinfo {volume} {4}},\ \bibinfo
  {pages} {035002} (\bibinfo {year} {2019})}\BibitemShut {NoStop}%
\bibitem [{\citenamefont {Lerose}\ \emph {et~al.}(2019)\citenamefont {Lerose},
  \citenamefont {\ifmmode \check{Z}\else
  \v{Z}\fi{}unkovi\ifmmode~\check{c}\else \v{c}\fi{}}, \citenamefont {Silva},\
  and\ \citenamefont {Gambassi}}]{lerose19}%
  \BibitemOpen
  \bibfield  {author} {\bibinfo {author} {\bibfnamefont {A.}~\bibnamefont
  {Lerose}}, \bibinfo {author} {\bibfnamefont {B.}~\bibnamefont {\ifmmode
  \check{Z}\else \v{Z}\fi{}unkovi\ifmmode~\check{c}\else \v{c}\fi{}}}, \bibinfo
  {author} {\bibfnamefont {A.}~\bibnamefont {Silva}}, \ and\ \bibinfo {author}
  {\bibfnamefont {A.}~\bibnamefont {Gambassi}},\ }\href {\doibase
  10.1103/PhysRevB.99.121112} {\bibfield  {journal} {\bibinfo  {journal} {Phys.
  Rev. B}\ }\textbf {\bibinfo {volume} {99}},\ \bibinfo {pages} {121112(R)}
  (\bibinfo {year} {2019})}\BibitemShut {NoStop}%
\bibitem [{\citenamefont {Derzhko}\ and\ \citenamefont
  {Krokhmalskii}(1997)}]{deryhko97}%
  \BibitemOpen
  \bibfield  {author} {\bibinfo {author} {\bibfnamefont {O.}~\bibnamefont
  {Derzhko}}\ and\ \bibinfo {author} {\bibfnamefont {T.}~\bibnamefont
  {Krokhmalskii}},\ }\href {\doibase 10.1103/PhysRevB.56.11659} {\bibfield
  {journal} {\bibinfo  {journal} {Phys. Rev. B}\ }\textbf {\bibinfo {volume}
  {56}},\ \bibinfo {pages} {11659} (\bibinfo {year} {1997})}\BibitemShut
  {NoStop}%
\bibitem [{\citenamefont {Saadatmand}\ \emph {et~al.}(2018)\citenamefont
  {Saadatmand}, \citenamefont {Bartlett},\ and\ \citenamefont
  {McCulloch}}]{Saadatmand18}%
  \BibitemOpen
  \bibfield  {author} {\bibinfo {author} {\bibfnamefont {S.~N.}\ \bibnamefont
  {Saadatmand}}, \bibinfo {author} {\bibfnamefont {S.~D.}\ \bibnamefont
  {Bartlett}}, \ and\ \bibinfo {author} {\bibfnamefont {I.~P.}\ \bibnamefont
  {McCulloch}},\ }\href {\doibase 10.1103/PhysRevB.97.155116} {\bibfield
  {journal} {\bibinfo  {journal} {Phys. Rev. B}\ }\textbf {\bibinfo {volume}
  {97}},\ \bibinfo {pages} {155116} (\bibinfo {year} {2018})}\BibitemShut
  {NoStop}%
\bibitem [{\citenamefont {Crosswhite}\ \emph {et~al.}(2008)\citenamefont
  {Crosswhite}, \citenamefont {Doherty},\ and\ \citenamefont
  {Vidal}}]{Crosswhite08}%
  \BibitemOpen
  \bibfield  {author} {\bibinfo {author} {\bibfnamefont {G.~M.}\ \bibnamefont
  {Crosswhite}}, \bibinfo {author} {\bibfnamefont {A.~C.}\ \bibnamefont
  {Doherty}}, \ and\ \bibinfo {author} {\bibfnamefont {G.}~\bibnamefont
  {Vidal}},\ }\href {\doibase 10.1103/PhysRevB.78.035116} {\bibfield  {journal}
  {\bibinfo  {journal} {Phys. Rev. B}\ }\textbf {\bibinfo {volume} {78}},\
  \bibinfo {pages} {035116} (\bibinfo {year} {2008})}\BibitemShut {NoStop}%
\bibitem [{\citenamefont {Haegeman}\ \emph {et~al.}(2011)\citenamefont
  {Haegeman}, \citenamefont {Cirac}, \citenamefont {Osborne}, \citenamefont
  {Pi\ifmmode~\check{z}\else \v{z}\fi{}orn}, \citenamefont {Verschelde},\ and\
  \citenamefont {Verstraete}}]{Haegeman11}%
  \BibitemOpen
  \bibfield  {author} {\bibinfo {author} {\bibfnamefont {J.}~\bibnamefont
  {Haegeman}}, \bibinfo {author} {\bibfnamefont {J.~I.}\ \bibnamefont {Cirac}},
  \bibinfo {author} {\bibfnamefont {T.~J.}\ \bibnamefont {Osborne}}, \bibinfo
  {author} {\bibfnamefont {I.}~\bibnamefont {Pi\ifmmode~\check{z}\else
  \v{z}\fi{}orn}}, \bibinfo {author} {\bibfnamefont {H.}~\bibnamefont
  {Verschelde}}, \ and\ \bibinfo {author} {\bibfnamefont {F.}~\bibnamefont
  {Verstraete}},\ }\href {\doibase 10.1103/PhysRevLett.107.070601} {\bibfield
  {journal} {\bibinfo  {journal} {Phys. Rev. Lett.}\ }\textbf {\bibinfo
  {volume} {107}},\ \bibinfo {pages} {070601} (\bibinfo {year}
  {2011})}\BibitemShut {NoStop}%
\bibitem [{\citenamefont {Zaletel}\ \emph {et~al.}(2015)\citenamefont
  {Zaletel}, \citenamefont {Mong}, \citenamefont {Karrasch}, \citenamefont
  {Moore},\ and\ \citenamefont {Pollmann}}]{Zaletel15}%
  \BibitemOpen
  \bibfield  {author} {\bibinfo {author} {\bibfnamefont {M.~P.}\ \bibnamefont
  {Zaletel}}, \bibinfo {author} {\bibfnamefont {R.~S.~K.}\ \bibnamefont
  {Mong}}, \bibinfo {author} {\bibfnamefont {C.}~\bibnamefont {Karrasch}},
  \bibinfo {author} {\bibfnamefont {J.~E.}\ \bibnamefont {Moore}}, \ and\
  \bibinfo {author} {\bibfnamefont {F.}~\bibnamefont {Pollmann}},\ }\href
  {\doibase 10.1103/PhysRevB.91.165112} {\bibfield  {journal} {\bibinfo
  {journal} {Phys. Rev. B}\ }\textbf {\bibinfo {volume} {91}},\ \bibinfo
  {pages} {165112} (\bibinfo {year} {2015})}\BibitemShut {NoStop}%
\bibitem [{\citenamefont {{Hashizume}}\ \emph {et~al.}(2018)\citenamefont
  {{Hashizume}}, \citenamefont {{McCulloch}},\ and\ \citenamefont
  {{Halimeh}}}]{Hashizume18}%
  \BibitemOpen
  \bibfield  {author} {\bibinfo {author} {\bibfnamefont {T.}~\bibnamefont
  {{Hashizume}}}, \bibinfo {author} {\bibfnamefont {I.~P.}\ \bibnamefont
  {{McCulloch}}}, \ and\ \bibinfo {author} {\bibfnamefont {J.~C.}\ \bibnamefont
  {{Halimeh}}},\ }\href@noop {} {\bibfield  {journal} {\bibinfo  {journal}
  {arXiv e-prints}\ } (\bibinfo {year} {2018})},\ \Eprint
  {http://arxiv.org/abs/1811.09275} {arXiv:1811.09275} \BibitemShut {NoStop}%
\bibitem [{\citenamefont {Buyskikh}\ \emph {et~al.}(2016)\citenamefont
  {Buyskikh}, \citenamefont {Fagotti}, \citenamefont {Schachenmayer},
  \citenamefont {Essler},\ and\ \citenamefont {Daley}}]{Buyskikh16}%
  \BibitemOpen
  \bibfield  {author} {\bibinfo {author} {\bibfnamefont {A.~S.}\ \bibnamefont
  {Buyskikh}}, \bibinfo {author} {\bibfnamefont {M.}~\bibnamefont {Fagotti}},
  \bibinfo {author} {\bibfnamefont {J.}~\bibnamefont {Schachenmayer}}, \bibinfo
  {author} {\bibfnamefont {F.}~\bibnamefont {Essler}}, \ and\ \bibinfo {author}
  {\bibfnamefont {A.~J.}\ \bibnamefont {Daley}},\ }\href {\doibase
  10.1103/PhysRevA.93.053620} {\bibfield  {journal} {\bibinfo  {journal} {Phys.
  Rev. A}\ }\textbf {\bibinfo {volume} {93}},\ \bibinfo {pages} {053620}
  (\bibinfo {year} {2016})}\BibitemShut {NoStop}%
\bibitem [{\citenamefont {Hastings}\ and\ \citenamefont
  {Koma}(2006)}]{Hastings2006}%
  \BibitemOpen
  \bibfield  {author} {\bibinfo {author} {\bibfnamefont {M.~B.}\ \bibnamefont
  {Hastings}}\ and\ \bibinfo {author} {\bibfnamefont {T.}~\bibnamefont
  {Koma}},\ }\href {\doibase 10.1007/s00220-006-0030-4} {\bibfield  {journal}
  {\bibinfo  {journal} {Commun. Math. Phys.}\ }\textbf {\bibinfo {volume}
  {265}},\ \bibinfo {pages} {781} (\bibinfo {year} {2006})}\BibitemShut
  {NoStop}%
\bibitem [{\citenamefont {{Verdel}}\ \emph {et~al.}(2019)\citenamefont
  {{Verdel}}, \citenamefont {{Liu}}, \citenamefont {{Whitsitt}}, \citenamefont
  {{Gorshkov}},\ and\ \citenamefont {{Heyl}}}]{verdel19}%
  \BibitemOpen
  \bibfield  {author} {\bibinfo {author} {\bibfnamefont {R.}~\bibnamefont
  {{Verdel}}}, \bibinfo {author} {\bibfnamefont {F.}~\bibnamefont {{Liu}}},
  \bibinfo {author} {\bibfnamefont {S.}~\bibnamefont {{Whitsitt}}}, \bibinfo
  {author} {\bibfnamefont {A.~V.}\ \bibnamefont {{Gorshkov}}}, \ and\ \bibinfo
  {author} {\bibfnamefont {M.}~\bibnamefont {{Heyl}}},\ }\href@noop {}
  {\bibfield  {journal} {\bibinfo  {journal} {arXiv e-prints}\ } (\bibinfo
  {year} {2019})},\ \Eprint {http://arxiv.org/abs/1911.11382}
  {arXiv:1911.11382} \BibitemShut {NoStop}%
\bibitem [{\citenamefont {Eisert}\ \emph {et~al.}()\citenamefont {Eisert},
  \citenamefont {Hangleiter}, \citenamefont {Walk}, \citenamefont {Roth},
  \citenamefont {Markham}, \citenamefont {Parekh}, \citenamefont {Chabaud},\
  and\ \citenamefont {Kashefi}}]{Certification}%
  \BibitemOpen
  \bibfield  {author} {\bibinfo {author} {\bibfnamefont {J.}~\bibnamefont
  {Eisert}}, \bibinfo {author} {\bibfnamefont {D.}~\bibnamefont {Hangleiter}},
  \bibinfo {author} {\bibfnamefont {N.}~\bibnamefont {Walk}}, \bibinfo {author}
  {\bibfnamefont {I.}~\bibnamefont {Roth}}, \bibinfo {author} {\bibfnamefont
  {D.}~\bibnamefont {Markham}}, \bibinfo {author} {\bibfnamefont
  {R.}~\bibnamefont {Parekh}}, \bibinfo {author} {\bibfnamefont
  {U.}~\bibnamefont {Chabaud}}, \ and\ \bibinfo {author} {\bibfnamefont
  {E.}~\bibnamefont {Kashefi}},\ }\href@noop {} {\ }\bibinfo {note}
  {ArXiv:1910.06343},\ \Eprint {http://arxiv.org/abs/arXiv:1910.06343}
  {arXiv:arXiv:1910.06343} \BibitemShut {NoStop}%
\bibitem [{\citenamefont {Dziarmaga}(2005)}]{dziarmaga05}%
  \BibitemOpen
  \bibfield  {author} {\bibinfo {author} {\bibfnamefont {J.}~\bibnamefont
  {Dziarmaga}},\ }\href {\doibase 10.1103/PhysRevLett.95.245701} {\bibfield
  {journal} {\bibinfo  {journal} {Phys. Rev. Lett.}\ }\textbf {\bibinfo
  {volume} {95}},\ \bibinfo {pages} {245701} (\bibinfo {year}
  {2005})}\BibitemShut {NoStop}%
\bibitem [{\citenamefont {Zurek}\ \emph {et~al.}(2005)\citenamefont {Zurek},
  \citenamefont {Dorner},\ and\ \citenamefont {Zoller}}]{zurek05}%
  \BibitemOpen
  \bibfield  {author} {\bibinfo {author} {\bibfnamefont {W.~H.}\ \bibnamefont
  {Zurek}}, \bibinfo {author} {\bibfnamefont {U.}~\bibnamefont {Dorner}}, \
  and\ \bibinfo {author} {\bibfnamefont {P.}~\bibnamefont {Zoller}},\ }\href
  {\doibase 10.1103/PhysRevLett.95.105701} {\bibfield  {journal} {\bibinfo
  {journal} {Phys. Rev. Lett.}\ }\textbf {\bibinfo {volume} {95}},\ \bibinfo
  {pages} {105701} (\bibinfo {year} {2005})}\BibitemShut {NoStop}%
\bibitem [{\citenamefont {Cincio}\ \emph {et~al.}(2007)\citenamefont {Cincio},
  \citenamefont {Dziarmaga}, \citenamefont {Rams},\ and\ \citenamefont
  {Zurek}}]{cincio07}%
  \BibitemOpen
  \bibfield  {author} {\bibinfo {author} {\bibfnamefont {L.}~\bibnamefont
  {Cincio}}, \bibinfo {author} {\bibfnamefont {J.}~\bibnamefont {Dziarmaga}},
  \bibinfo {author} {\bibfnamefont {M.~M.}\ \bibnamefont {Rams}}, \ and\
  \bibinfo {author} {\bibfnamefont {W.~H.}\ \bibnamefont {Zurek}},\ }\href
  {\doibase 10.1103/PhysRevA.75.052321} {\bibfield  {journal} {\bibinfo
  {journal} {Phys. Rev. A}\ }\textbf {\bibinfo {volume} {75}},\ \bibinfo
  {pages} {052321} (\bibinfo {year} {2007})}\BibitemShut {NoStop}%
\bibitem [{\citenamefont {Cincio}\ \emph {et~al.}(2009)\citenamefont {Cincio},
  \citenamefont {Dziarmaga}, \citenamefont {Meisner},\ and\ \citenamefont
  {Rams}}]{cincio09}%
  \BibitemOpen
  \bibfield  {author} {\bibinfo {author} {\bibfnamefont {L.}~\bibnamefont
  {Cincio}}, \bibinfo {author} {\bibfnamefont {J.}~\bibnamefont {Dziarmaga}},
  \bibinfo {author} {\bibfnamefont {J.}~\bibnamefont {Meisner}}, \ and\
  \bibinfo {author} {\bibfnamefont {M.~M.}\ \bibnamefont {Rams}},\ }\href
  {\doibase 10.1103/PhysRevB.79.094421} {\bibfield  {journal} {\bibinfo
  {journal} {Phys. Rev. B}\ }\textbf {\bibinfo {volume} {79}},\ \bibinfo
  {pages} {094421} (\bibinfo {year} {2009})}\BibitemShut {NoStop}%
\bibitem [{\citenamefont {{Kennes}}\ \emph {et~al.}(2018)\citenamefont
  {{Kennes}}, \citenamefont {{Karrasch}},\ and\ \citenamefont
  {{Millis}}}]{kennes18}%
  \BibitemOpen
  \bibfield  {author} {\bibinfo {author} {\bibfnamefont {D.~M.}\ \bibnamefont
  {{Kennes}}}, \bibinfo {author} {\bibfnamefont {C.}~\bibnamefont
  {{Karrasch}}}, \ and\ \bibinfo {author} {\bibfnamefont {A.~J.}\ \bibnamefont
  {{Millis}}},\ }\href@noop {} {\bibfield  {journal} {\bibinfo  {journal}
  {arXiv e-prints}\ } (\bibinfo {year} {2018})},\ \Eprint
  {http://arxiv.org/abs/1809.00733} {arXiv:1809.00733} \BibitemShut {NoStop}%
\bibitem [{\citenamefont {Kim}\ \emph {et~al.}(2010)\citenamefont {Kim},
  \citenamefont {Chang}, \citenamefont {Korenblit}, \citenamefont {Islam},
  \citenamefont {Edwards}, \citenamefont {Freericks}, \citenamefont {Lin},
  \citenamefont {Duan},\ and\ \citenamefont {Monroe}}]{kim10}%
  \BibitemOpen
  \bibfield  {author} {\bibinfo {author} {\bibfnamefont {K.}~\bibnamefont
  {Kim}}, \bibinfo {author} {\bibfnamefont {M.~S.}\ \bibnamefont {Chang}},
  \bibinfo {author} {\bibfnamefont {S.}~\bibnamefont {Korenblit}}, \bibinfo
  {author} {\bibfnamefont {R.}~\bibnamefont {Islam}}, \bibinfo {author}
  {\bibfnamefont {E.~E.}\ \bibnamefont {Edwards}}, \bibinfo {author}
  {\bibfnamefont {J.~K.}\ \bibnamefont {Freericks}}, \bibinfo {author}
  {\bibfnamefont {G.~D.}\ \bibnamefont {Lin}}, \bibinfo {author} {\bibfnamefont
  {L.~M.}\ \bibnamefont {Duan}}, \ and\ \bibinfo {author} {\bibfnamefont
  {C.}~\bibnamefont {Monroe}},\ }\href {\doibase 10.1038/ncomms1374} {\bibfield
   {journal} {\bibinfo  {journal} {Nature}\ }\textbf {\bibinfo {volume}
  {465}},\ \bibinfo {pages} {590} (\bibinfo {year} {2010})}\BibitemShut
  {NoStop}%
\bibitem [{\citenamefont {Edwards}\ \emph {et~al.}(2010)\citenamefont
  {Edwards}, \citenamefont {Korenblit}, \citenamefont {Kim}, \citenamefont
  {Islam}, \citenamefont {Chang}, \citenamefont {Freericks}, \citenamefont
  {Lin}, \citenamefont {Duan},\ and\ \citenamefont {Monroe}}]{edwards10}%
  \BibitemOpen
  \bibfield  {author} {\bibinfo {author} {\bibfnamefont {E.~E.}\ \bibnamefont
  {Edwards}}, \bibinfo {author} {\bibfnamefont {S.}~\bibnamefont {Korenblit}},
  \bibinfo {author} {\bibfnamefont {K.}~\bibnamefont {Kim}}, \bibinfo {author}
  {\bibfnamefont {R.}~\bibnamefont {Islam}}, \bibinfo {author} {\bibfnamefont
  {M.-S.}\ \bibnamefont {Chang}}, \bibinfo {author} {\bibfnamefont {J.~K.}\
  \bibnamefont {Freericks}}, \bibinfo {author} {\bibfnamefont {G.-D.}\
  \bibnamefont {Lin}}, \bibinfo {author} {\bibfnamefont {L.-M.}\ \bibnamefont
  {Duan}}, \ and\ \bibinfo {author} {\bibfnamefont {C.}~\bibnamefont
  {Monroe}},\ }\href {\doibase 10.1103/PhysRevB.82.060412} {\bibfield
  {journal} {\bibinfo  {journal} {Phys. Rev. B}\ }\textbf {\bibinfo {volume}
  {82}},\ \bibinfo {pages} {060412(R)} (\bibinfo {year} {2010})}\BibitemShut
  {NoStop}%
\bibitem [{\citenamefont {Schneider}\ and\ \citenamefont
  {Milburn}(1998)}]{schneider98}%
  \BibitemOpen
  \bibfield  {author} {\bibinfo {author} {\bibfnamefont {S.}~\bibnamefont
  {Schneider}}\ and\ \bibinfo {author} {\bibfnamefont {G.~J.}\ \bibnamefont
  {Milburn}},\ }\href {\doibase 10.1103/PhysRevA.57.3748} {\bibfield  {journal}
  {\bibinfo  {journal} {Phys. Rev. A}\ }\textbf {\bibinfo {volume} {57}},\
  \bibinfo {pages} {3748} (\bibinfo {year} {1998})}\BibitemShut {NoStop}%
\bibitem [{\citenamefont {Friedenauer}\ \emph {et~al.}(2008)\citenamefont
  {Friedenauer}, \citenamefont {Schmitz}, \citenamefont {Glueckert},
  \citenamefont {Porras},\ and\ \citenamefont {Schaetz}}]{friedenauer08}%
  \BibitemOpen
  \bibfield  {author} {\bibinfo {author} {\bibfnamefont {A.}~\bibnamefont
  {Friedenauer}}, \bibinfo {author} {\bibfnamefont {H.}~\bibnamefont
  {Schmitz}}, \bibinfo {author} {\bibfnamefont {J.~T.}\ \bibnamefont
  {Glueckert}}, \bibinfo {author} {\bibfnamefont {D.}~\bibnamefont {Porras}}, \
  and\ \bibinfo {author} {\bibfnamefont {T.}~\bibnamefont {Schaetz}},\ }\href
  {\doibase 10.1038/nphys1032} {\bibfield  {journal} {\bibinfo  {journal}
  {Nature Phys.}\ }\textbf {\bibinfo {volume} {4}},\ \bibinfo {pages} {757}
  (\bibinfo {year} {2008})}\BibitemShut {NoStop}%
\bibitem [{\citenamefont {Acin}\ \emph {et~al.}(2018)\citenamefont {Acin},
  \citenamefont {Bloch}, \citenamefont {Buhrman}, \citenamefont {Calarco},
  \citenamefont {Eichler}, \citenamefont {Eisert}, \citenamefont {Esteve},
  \citenamefont {Gisin}, \citenamefont {Glaser}, \citenamefont {Jelezko},
  \citenamefont {Kuhr}, \citenamefont {Lewenstein}, \citenamefont {Riedel},
  \citenamefont {Schmidt}, \citenamefont {Thew}, \citenamefont {Wallraff},
  \citenamefont {Walmsley},\ and\ \citenamefont {Wilhelm}}]{Roadmap}%
  \BibitemOpen
  \bibfield  {author} {\bibinfo {author} {\bibfnamefont {A.}~\bibnamefont
  {Acin}}, \bibinfo {author} {\bibfnamefont {I.}~\bibnamefont {Bloch}},
  \bibinfo {author} {\bibfnamefont {H.}~\bibnamefont {Buhrman}}, \bibinfo
  {author} {\bibfnamefont {T.}~\bibnamefont {Calarco}}, \bibinfo {author}
  {\bibfnamefont {C.}~\bibnamefont {Eichler}}, \bibinfo {author} {\bibfnamefont
  {J.}~\bibnamefont {Eisert}}, \bibinfo {author} {\bibfnamefont
  {D.}~\bibnamefont {Esteve}}, \bibinfo {author} {\bibfnamefont
  {N.}~\bibnamefont {Gisin}}, \bibinfo {author} {\bibfnamefont {S.~J.}\
  \bibnamefont {Glaser}}, \bibinfo {author} {\bibfnamefont {F.}~\bibnamefont
  {Jelezko}}, \bibinfo {author} {\bibfnamefont {S.}~\bibnamefont {Kuhr}},
  \bibinfo {author} {\bibfnamefont {M.}~\bibnamefont {Lewenstein}}, \bibinfo
  {author} {\bibfnamefont {M.~F.}\ \bibnamefont {Riedel}}, \bibinfo {author}
  {\bibfnamefont {P.~O.}\ \bibnamefont {Schmidt}}, \bibinfo {author}
  {\bibfnamefont {R.}~\bibnamefont {Thew}}, \bibinfo {author} {\bibfnamefont
  {A.}~\bibnamefont {Wallraff}}, \bibinfo {author} {\bibfnamefont
  {I.}~\bibnamefont {Walmsley}}, \ and\ \bibinfo {author} {\bibfnamefont
  {F.~K.}\ \bibnamefont {Wilhelm}},\ }\href {\doibase 10.1088/1367-2630/aad1ea}
  {\bibfield  {journal} {\bibinfo  {journal} {New J. Phys.}\ }\textbf {\bibinfo
  {volume} {20}},\ \bibinfo {pages} {080201} (\bibinfo {year}
  {2018})}\BibitemShut {NoStop}%
\bibitem [{\citenamefont {Schauss}(2018)}]{schauss18}%
  \BibitemOpen
  \bibfield  {author} {\bibinfo {author} {\bibfnamefont {P.}~\bibnamefont
  {Schauss}},\ }\href {\doibase 10.1088/2058-9565/aa9c59} {\bibfield  {journal}
  {\bibinfo  {journal} {Quant. Sc. Tech.}\ }\textbf {\bibinfo {volume} {3}},\
  \bibinfo {pages} {023001} (\bibinfo {year} {2018})}\BibitemShut {NoStop}%
\bibitem [{\citenamefont {Schneider}\ \emph {et~al.}(2012)\citenamefont
  {Schneider}, \citenamefont {Porras},\ and\ \citenamefont
  {Schaetz}}]{Schneider12}%
  \BibitemOpen
  \bibfield  {author} {\bibinfo {author} {\bibfnamefont {C.}~\bibnamefont
  {Schneider}}, \bibinfo {author} {\bibfnamefont {D.}~\bibnamefont {Porras}}, \
  and\ \bibinfo {author} {\bibfnamefont {T.}~\bibnamefont {Schaetz}},\ }\href
  {\doibase 10.1088/0034-4885/75/2/024401} {\bibfield  {journal} {\bibinfo
  {journal} {Rep. Prog. Phys.}\ }\textbf {\bibinfo {volume} {75}},\ \bibinfo
  {pages} {024401} (\bibinfo {year} {2012})}\BibitemShut {NoStop}%
\end{thebibliography}%

\appendix
\begin{widetext}

\section{Platforms considered} 

\subsection{Rydberg atoms}
For a Rydberg atoms setup, we consider the case of an array of trapped cold neutral \ce{^{87}Rb} 70S atoms with strong, controllable interactions. In this case, the atoms are trapped by optical tweezers, and the state of the art architecture can contain up to 51 atoms\cite{Bernien17}. A two-photon process couples the ground state vector $|g \rangle =\ce{^5S_{1/2}}|\mathrm{F}=2, \mathrm{m}_F =2 \rangle$ to  a target  Rydberg state vector  $|r \rangle \ce{^{70}S_{1/2}}|\mathrm{m}_j =1-/2 \rangle$ of the Rb atoms via an intermediate state. This transition is driven by two lasers detuned from the Rydberg state. The dynamics of the system is given by \citep{Zhang17,Bernien17}
\be
H = \sum_i \frac{\Omega}{2}\sigma^{x}_i-\sum_i\Delta_in_i+\sum_{i<j}V_{i,j}n_in_j\,,
\ee
where $\Delta_i$ is a detuning factor away from the Rydberg state, $\Omega_i$ is the Rabi frequency of the atom at the $i$th position, $\sigma^{x}_i = | g_i\rangle\langle r_i |+ |r_i \rangle\langle g_i |$ is the coupling between the ground state and Rydberg state, and $n_i = |r_i\rangle\langle r_i|$. The interaction strength elements
$V_{i,j}$ can also be tuned by varying the distance between them or by coupling them to a different Rydberg state. To transform this Hamiltonian into an transverse field Ising model \cite{schauss18} we just need to identify at each site $i$, $| g_i \rangle = | \downarrow_i \rangle$ and $| r_i \rangle = | \uparrow_i \rangle$ such that $\sigma^x_i = (| \downarrow_i \rangle \langle \uparrow_i|+| \uparrow_i \rangle\langle \downarrow_i | )/2$ and  $\sigma^z_i = (| \uparrow_i \rangle\langle \uparrow_i| - | \downarrow_i \rangle\langle \downarrow_i|)/2$, with $n_i = \mathds{1}/2 + \sigma^z$. Replacing in the Hamiltonian, we obtain a transverse field Ising model with a coupling of the form
\be
H = \sum_i \frac{\Omega}{2}\sigma^{x}_i-\sum_i (1-\Delta_i)n_i+\sum_{i<j}V_{i,j}\sigma^z_i\sigma^z_j\,.
\label{Ry_Ising2}
\ee
The longitudinal field in the above expression can be suppressed by the detuning factor $\Delta$. The Ising interaction arises from van der Waals interactions between the atoms when they are both in the Rydberg state, and takes the form $V_{i,j}=-C/r_{i,j}^6$, with $r_{i,j}= |i-j|$. As such, since the Rydberg atoms are not in the ground state once in the local trap, and the experiment is carried at a finite temperature, fluctuations in the atomic positions for each atom in each cycle of the experiment are introduced \cite{Zhang17,Bernien17}. This will affect the Ising interaction, Eq.~(\ref{Ry_Ising2}). We finally note that the mean lifetime of a chain scales inversely with the number of ions, with an average of 5 minutes for a chain of 53 ions. 
\subsection{Trapped ions}
We will consider here a particular trapped ion setup, where \ce{^{171}Yb} are trapped by a linear Paul trap. A 1D spin-1/2 Ising system in the presence of a transverse field can be engineered  \cite{islam11} by the hyperfine "clock" states \ce{^{2}S_{1/2}}$|\mathrm{F}=0 ,\mathrm{m}_f=0\rangle$ and $|\mathrm{F}=1 ,\mathrm{m}_f=0\rangle$ which represent the $| \downarrow_z \rangle$ and $| \uparrow_z \rangle$ eigenstates of $\sigma_z$ and are sparated by $\nu_f$.  These ions are trapped by a linear Paul trap, and their modes of motion are cooled near the ground state.  The transverse field and spin-spin interactions can be obtained applying off-resonant laser beams \cite{kim10,edwards10}. A transverse field can be generated by uniformly illuminating the ion chain with two Raman laser beams whose difference in frequency is tuned to the hyperfine splitting $\nu_f$. This induces Rabi oscillations in the ions which are seen as AC Stark shifts in the spin states.
 On the other hand, the spin-spin interactions are created by coupling the spin states to the normal modes of motion of the ions by the Raman beams.
  The beams are made to carry beat note at frequencies $\nu_f \pm \mu$, which generates a spin-dependent force at frequency $\mu$ \cite{Schneider12}. As such, controlling the beat note frequency $\mu$, one can generate Ising interactions with a coupling given by 
\be
J_{i,j}=N\Omega_i\Omega_j\sum_{m=1}^N\frac{\eta_{im}\eta_{jm}\nu_m}{\mu^2-\nu_m^2} \approx \frac{J_{0}}{|i-j|^\alpha}\,,
\label{TI_ising}
\ee     
where $\Omega_i$ is the Rabi frequency of the $i$th ion, $\eta_{im}$ is the Lamb-Dicke parameter of the $m$th mode of the $i$th ion at frequency $\nu_m$, and the assumption of $|\mu - \nu_m| \gg \eta\Omega$ indicating that only virtual phonons are excited. These interactions are not necessarily uniform, since the Rabi frequency $\Omega_i$ can vary across the chain and other vibrational models can also contribute considerably.
 Furthermore, by tuning $\mu$ one can tune $\alpha$ between $0 <\alpha < 3$.

\section{Numerical results for position fluctuations and random fields of Rydberg atoms}
\label{random_results}
\begin{figure}[H]
\includegraphics[scale=0.45]{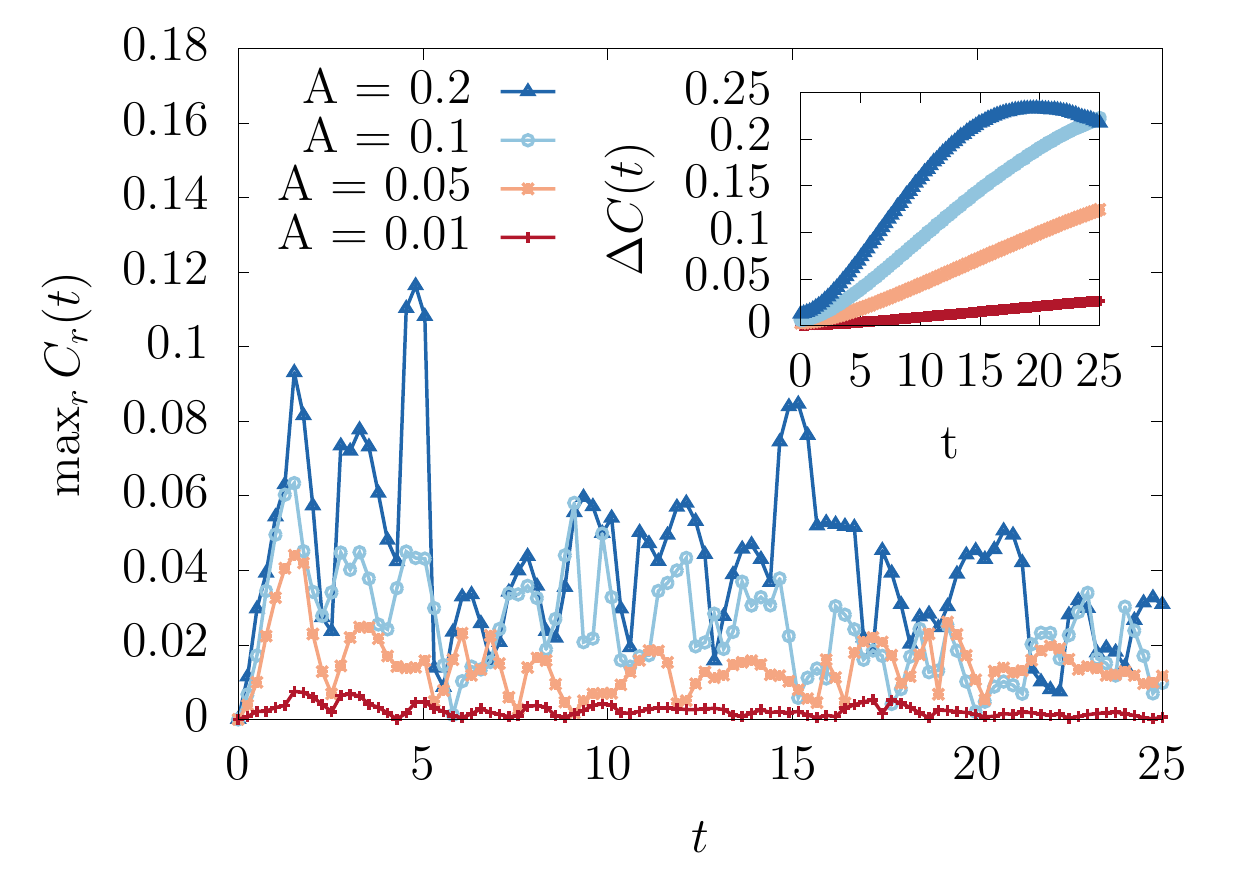}
\includegraphics[scale=0.45]{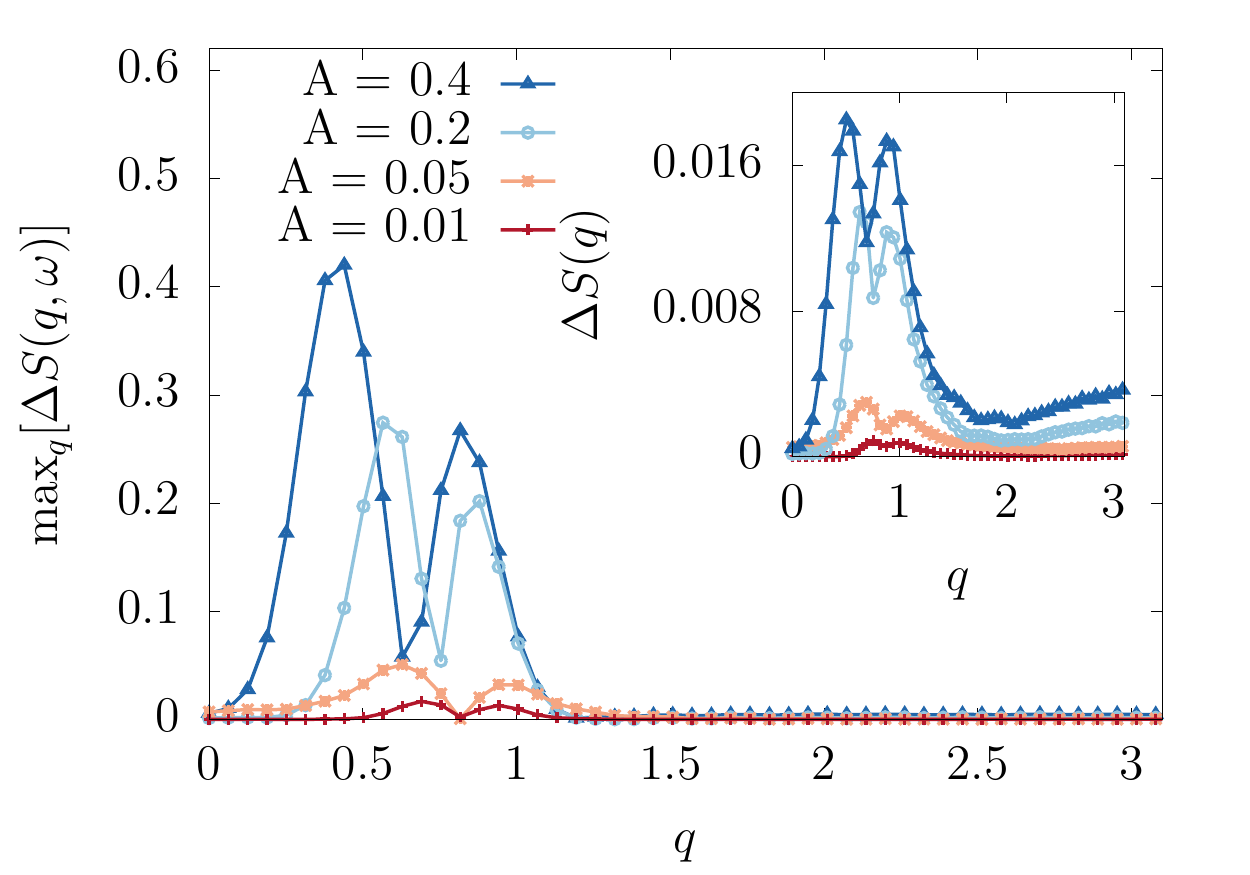}
\includegraphics[scale=0.45]{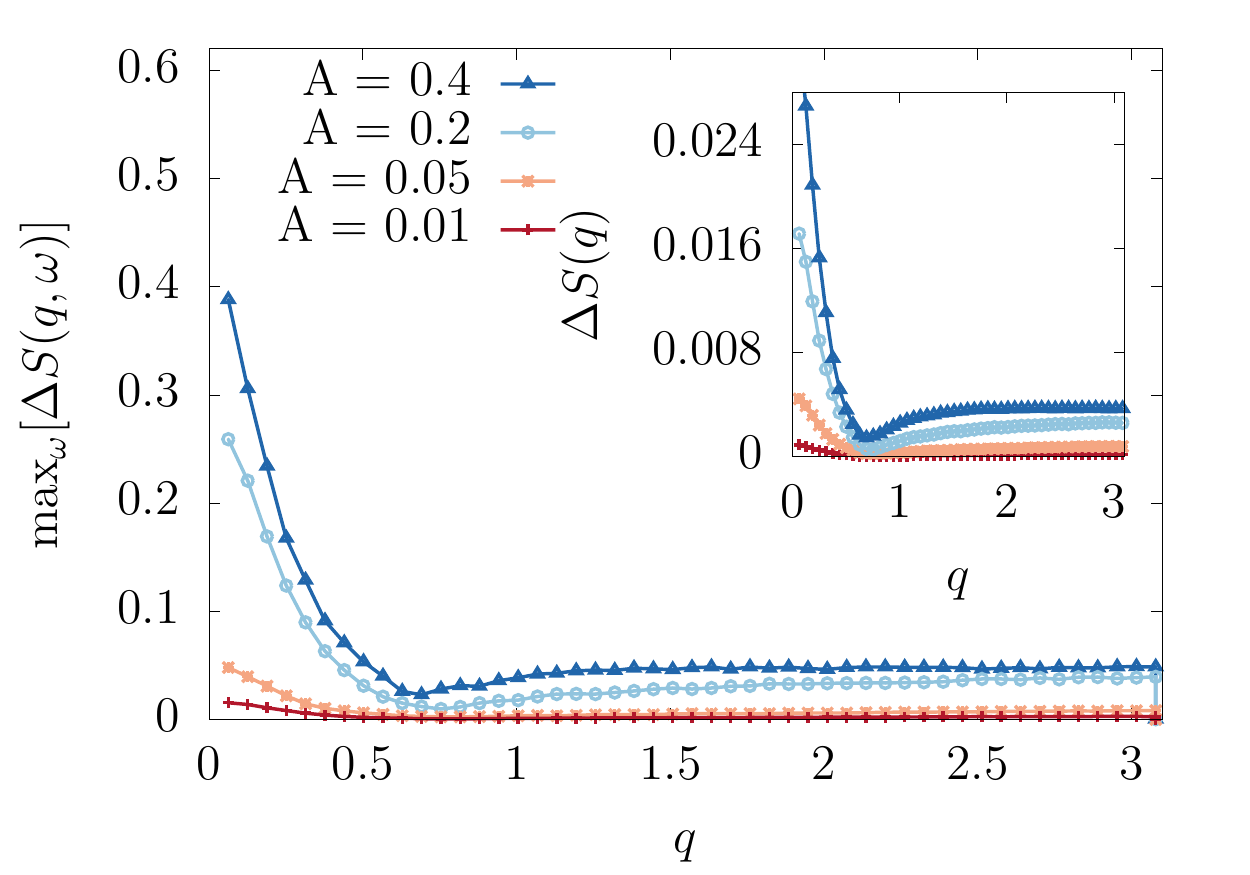}
\caption{ {\it Error analysis of the influence of lattice imperfections on DSF for the case of random Ising interactions.}
{\it (Left panel)} Maximum of the absolute error for the two point correlators integrated in real space. {\it Inset:} Average absolute error of the correlators integrated over real space.
 {\it (Middle panel)} Maximum of the absolute error for the DSF integrated over reciprocal space. {\it Inset:} Average absolute error of the correlators integrated over reciprocal space. 
 {\it (Right panel)} Maximum of the absolute error for the DSF integrated over frequency. {\it Inset:} Average absolute error of the correlators integrated over frequency. 
 The DSF results show that the error is concentrated around the gap, which closes as the value of $A$ is increased. 
 At the same time the correlators have errors which increase in average with time, indicating that measurements at long times might be unreliable for this quantity.}
\label{error_rand_int}
\end{figure}
\begin{figure}[H]
\includegraphics[scale=0.45]{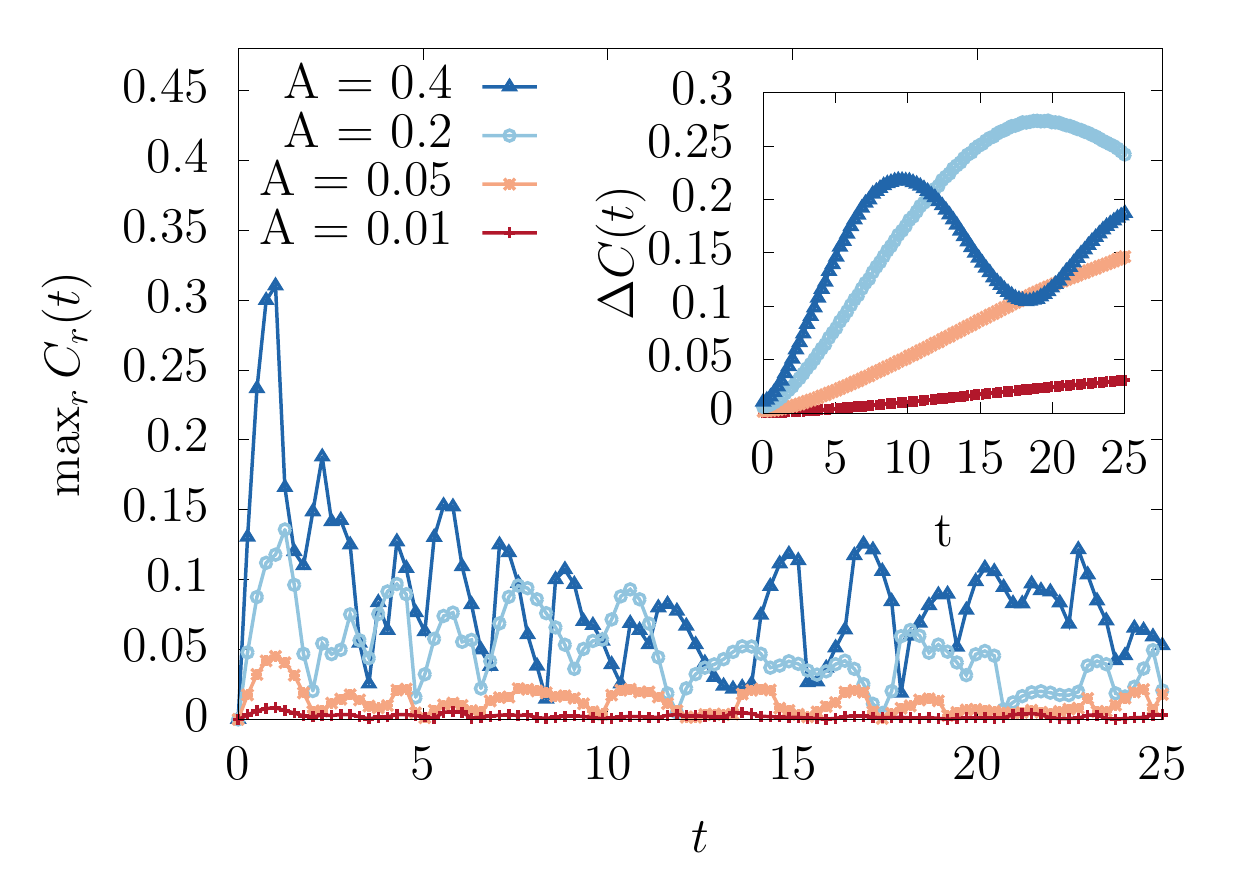}
\includegraphics[scale=0.45]{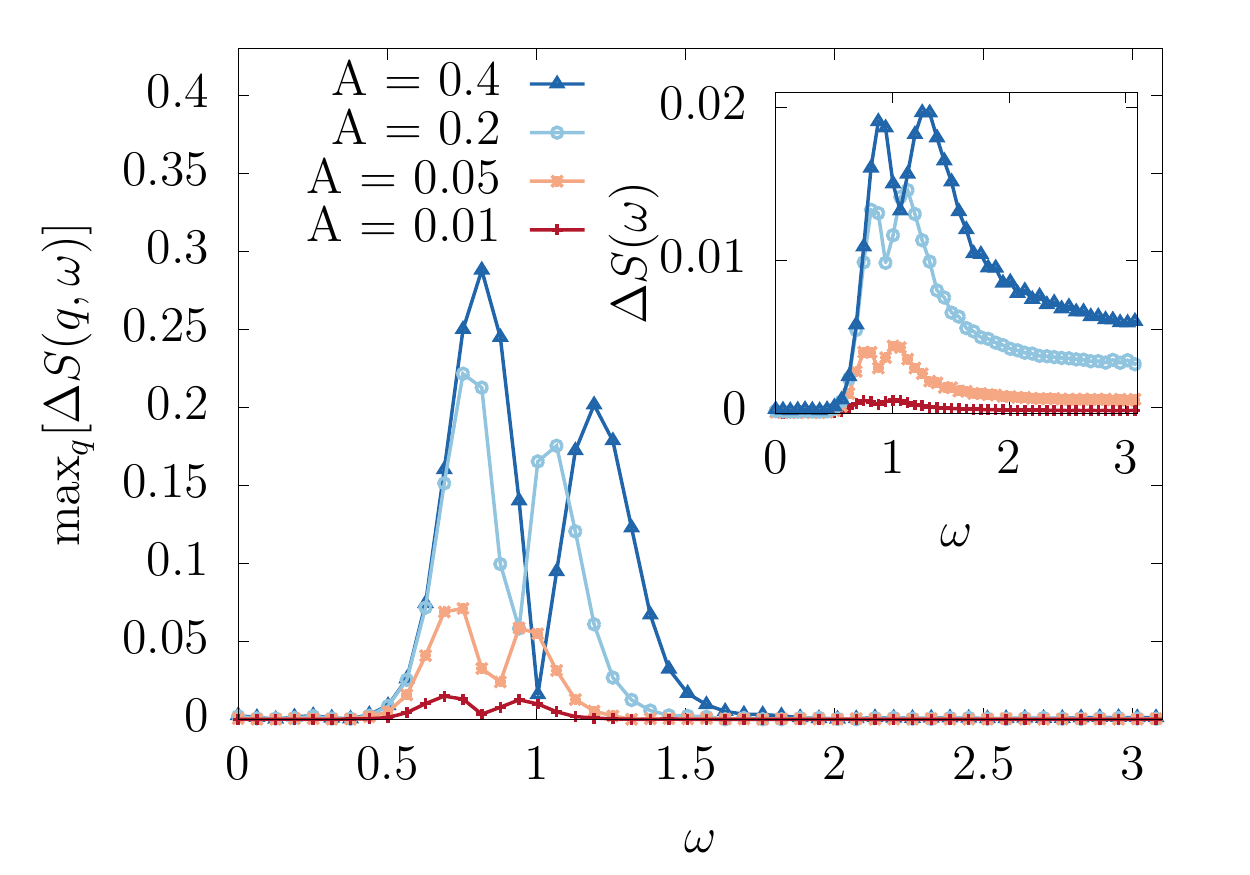}
\includegraphics[scale=0.45]{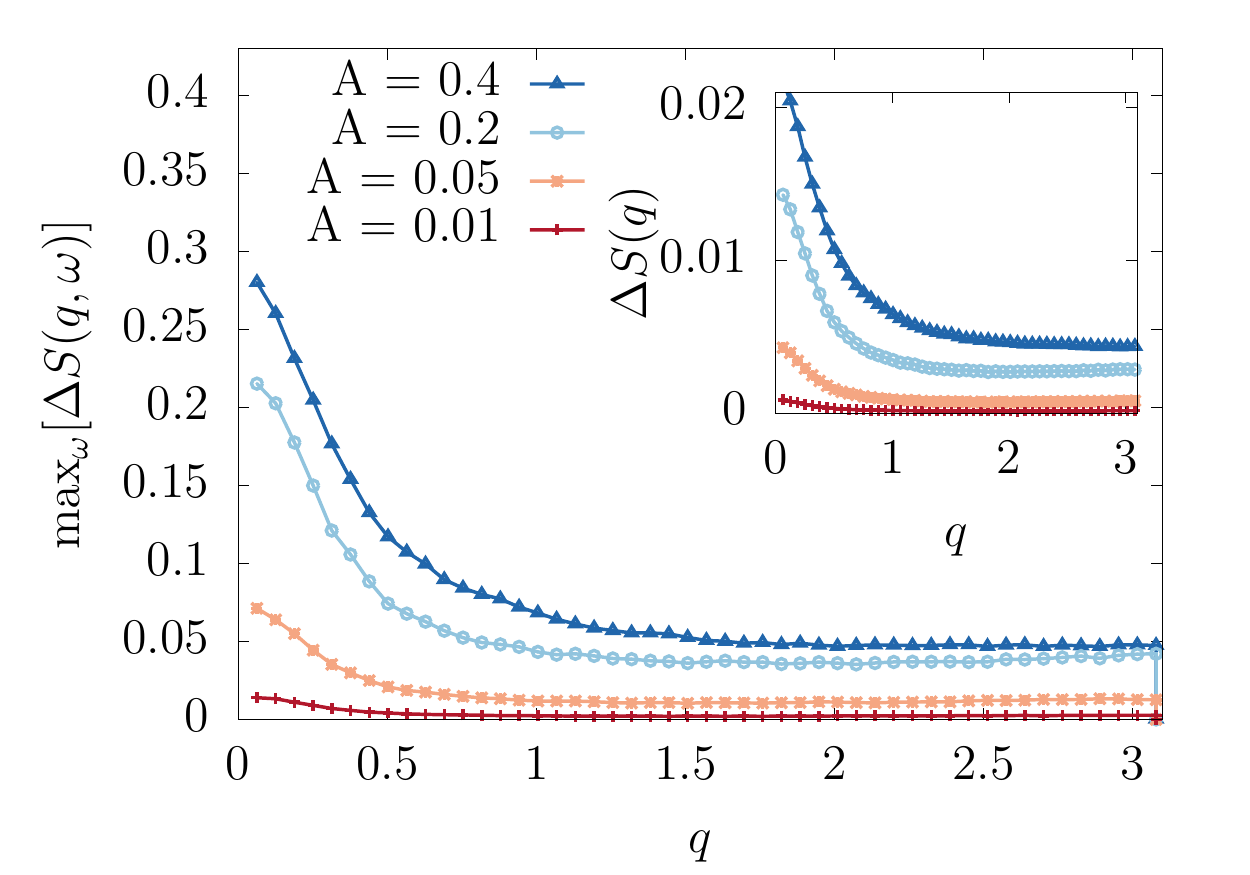}
\caption{ {\it Error analysis of the influence of lattice imperfections on DSF for the case of random transverse fields.} 
{\it (Left panel)} Maximum of the absolute error for the two point correlators integrated in real space. {\it Inset:} Average absolute error of the correlators integrated over real space.
 {\it (Middle panel)} Maximum of the absolute error for the DSF integrated over reciprocal space. {\it Inset:} Average absolute error of the correlators integrated over reciprocal space. 
 {\it (Right panel)} Maximum of the absolute error for the DSF integrated over frequency. {\it Inset:} Average absolute error of the correlators integrated over frequency.
 The DSF results show that the error is concentrated around the gap, which opens as the value of $A$ is increased. As with the previous figures, the average error increases with time, which indicates that measurements at long times might be unreliable for this quantity for this type of imperfection. }
\label{error_rand_field}
\end{figure}

\section{Dynamical structure factors for the short range transverse field Ising}\label{appendix:shortrangenoise}
\label{DSF_noise}
\begin{figure}[H]
\centering
\includegraphics[scale=0.45]{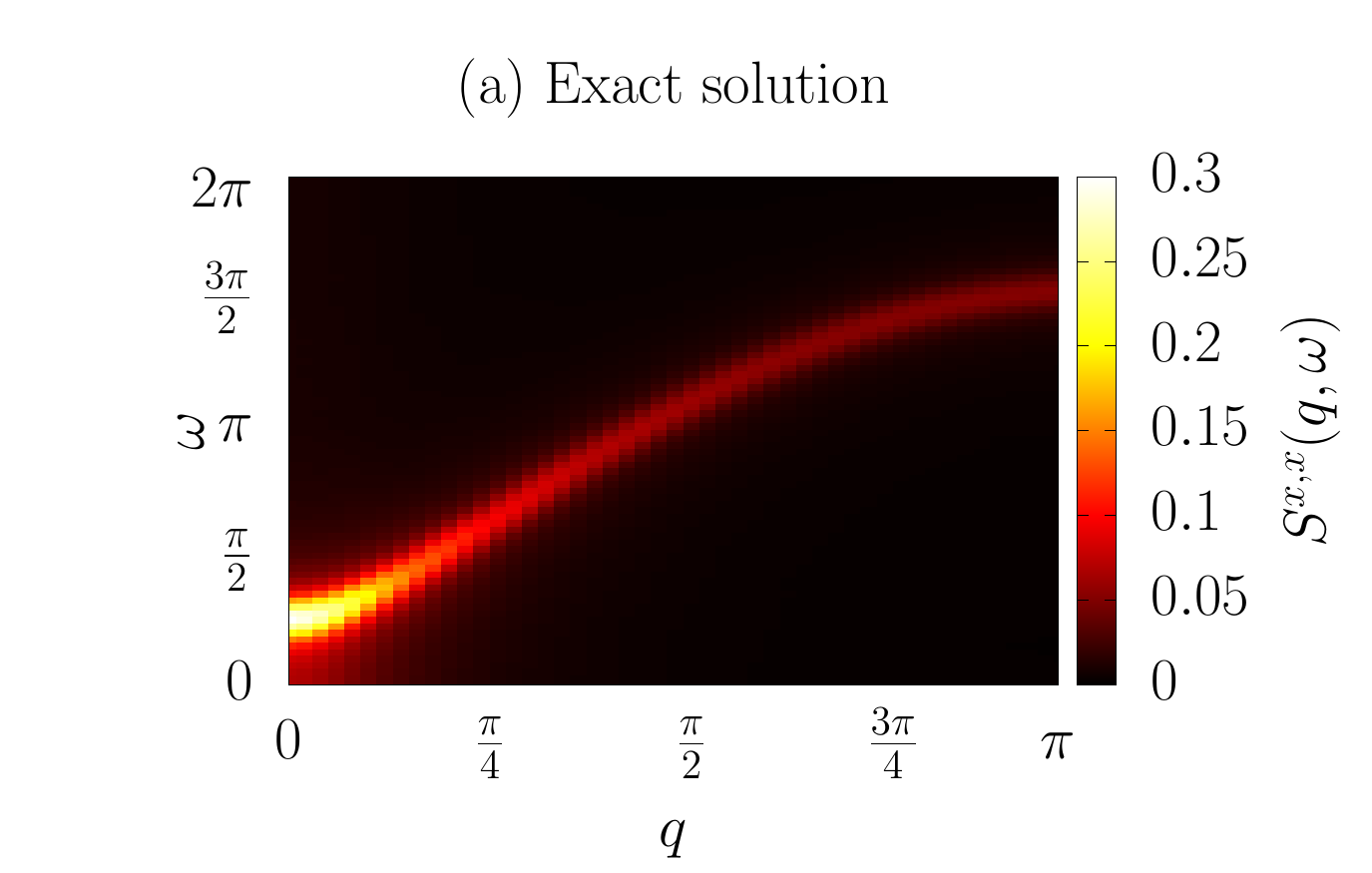}
\includegraphics[scale=0.45]{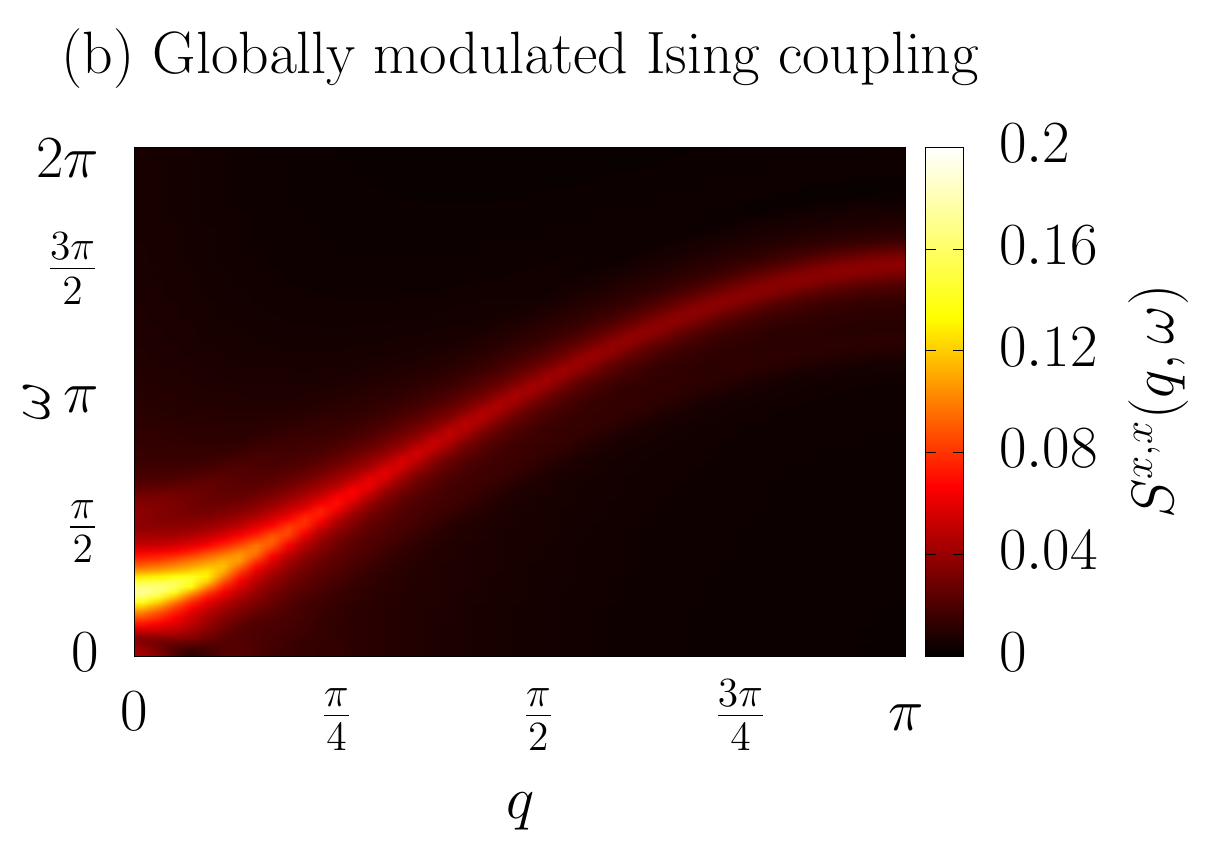}
\includegraphics[scale=0.45]{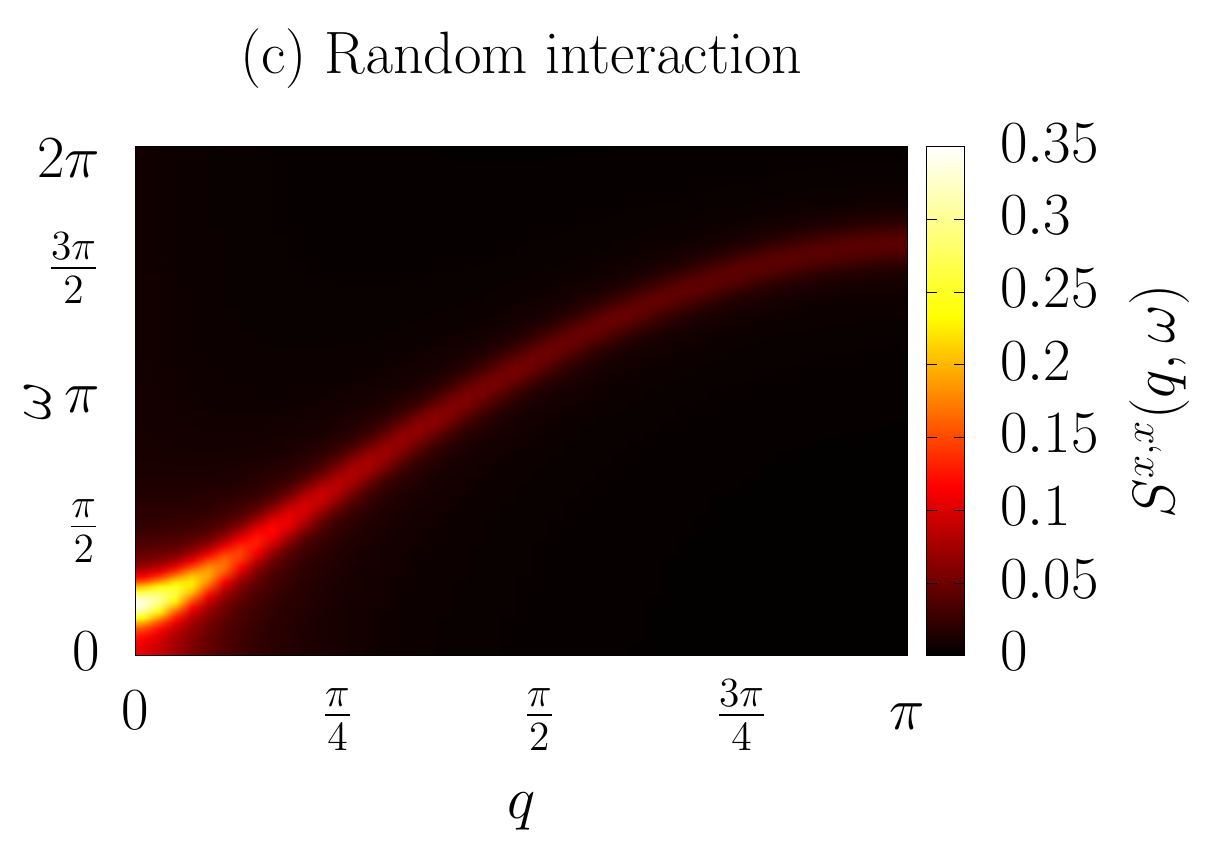}
\caption{{\it DSF for the different experimental imperfections for the short range TFIM.} 
{\it (a)} Exact solution as obtained from our quasi-free fermionic calculation for the TFIM without imperfections, the gap is located at $q = 0$ and $\omega = \pi / 4$, and we observe a clear two particle continuum for $q >0$.
 {\it (b)} Exact solution for the case of globally fluctuating Ising couplings. In this case we observe that the shape of the low $\omega$ sector is slightly modified, but the gap and the two particle continuum can still be observed, though at a lower intensity than for (a).
 {\it (c)} Exact solution for the case of random Ising interactions. We see how the gap has closed, with the maxima located at $q=0$ and $\omega <\pi/4$, }
\label{DSF_noise_models}
\end{figure}
\begin{figure}[H]
\centering
\includegraphics[scale=0.45]{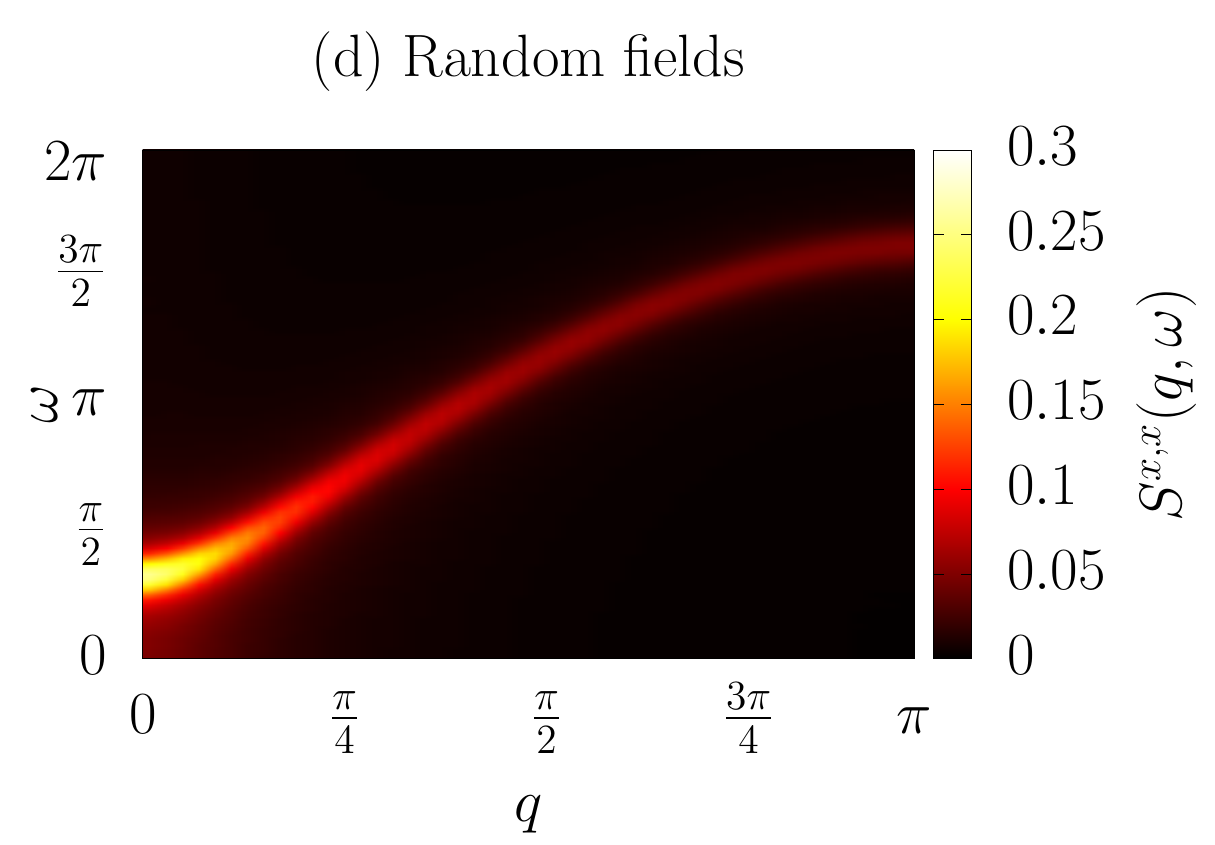}
\includegraphics[scale=0.45]{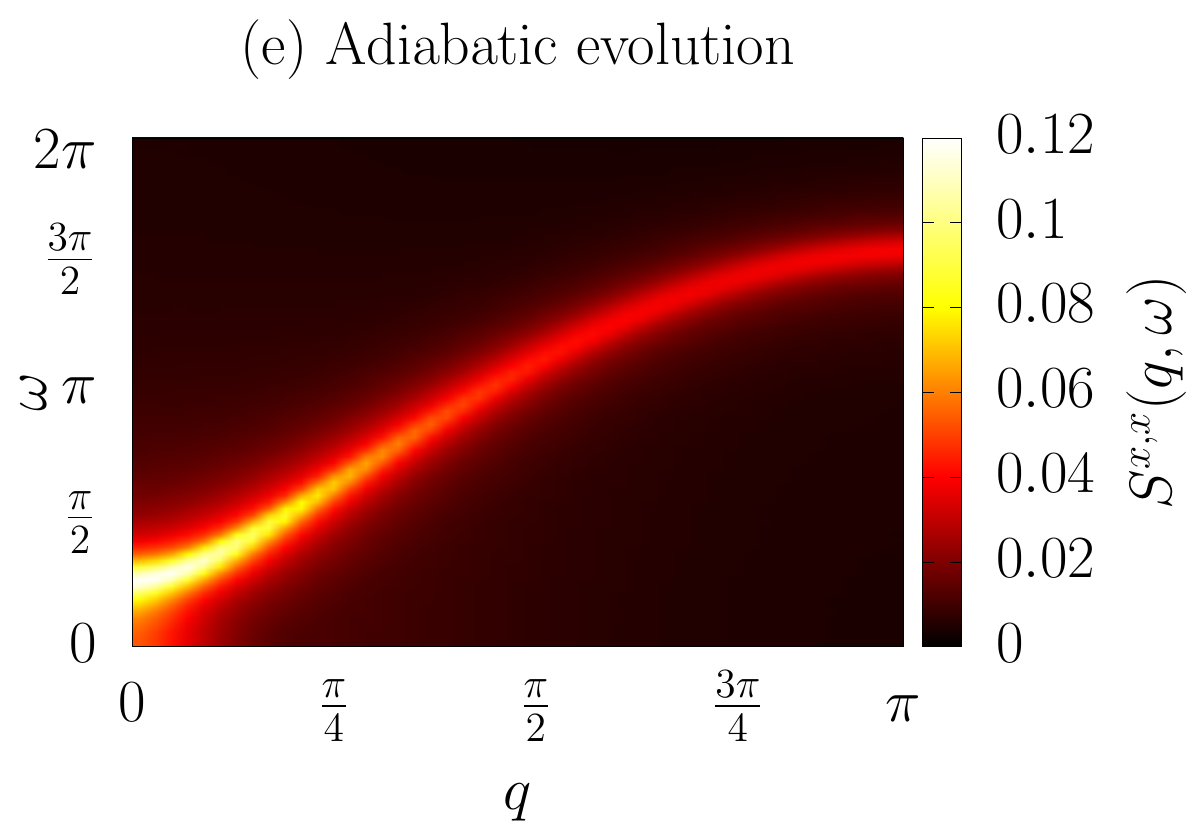}
\caption{{\it DSF for the different experimental imperfections for the short range TFIM. }
 {\it (d)} Exact solution for the case of random transverse fields. We see how the gap has opened, with the maxima located at $q=0$ and $\omega > \pi / 4$.
 {\it (e)} Solution for case in which the ground state has been prepared by an adiabatic evolution for a time $\tau_Q = 0.5$, well below the adiabatic regime. In this case the intensity of the DSF is greatly reduced.}
\label{DSF_noise_models_2}
\end{figure}
\begin{figure}[H]
\centering
\includegraphics[scale=0.5]{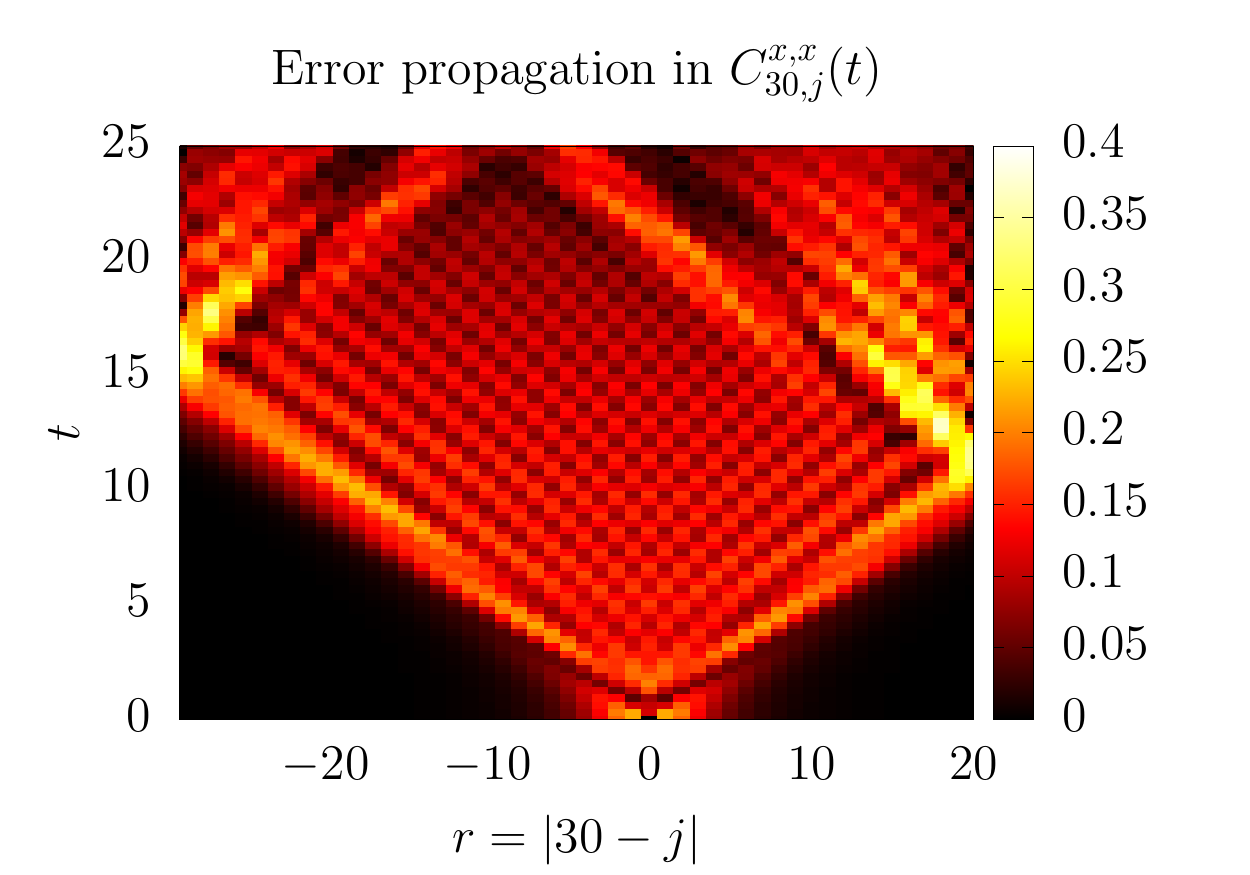}
\caption{{\it Propagation of errors in the correlators}. 
We plot the error in the correlator with respect to a site in the middle of the chain, $C^{x,x}_{30,j}$, as a function of time.
 We observe the propagation of this error with a maximal velocity, creating a Lieb-Robinson cone, with an interference pattern within the cone.
 Indicating that errors propagate and delocalize at long times, making a direct study of many-body excitations through correlation function challenging. }
\label{corr_error_prop}
\end{figure}

\section{Dynamical structure factors for the long range transverse field Ising}\label{appendix:longrange}
\begin{figure}[H]
\centering
\includegraphics[scale=0.45]{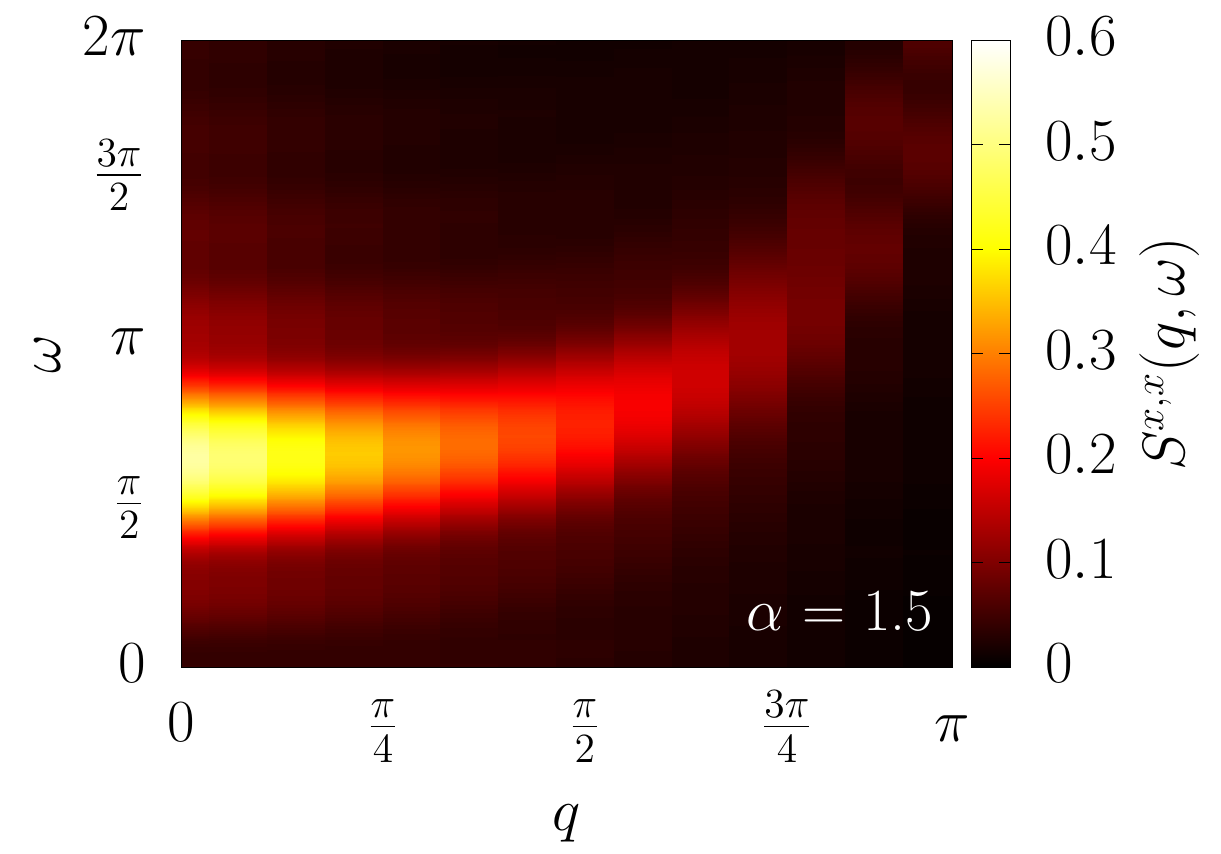}
\includegraphics[scale=0.45]{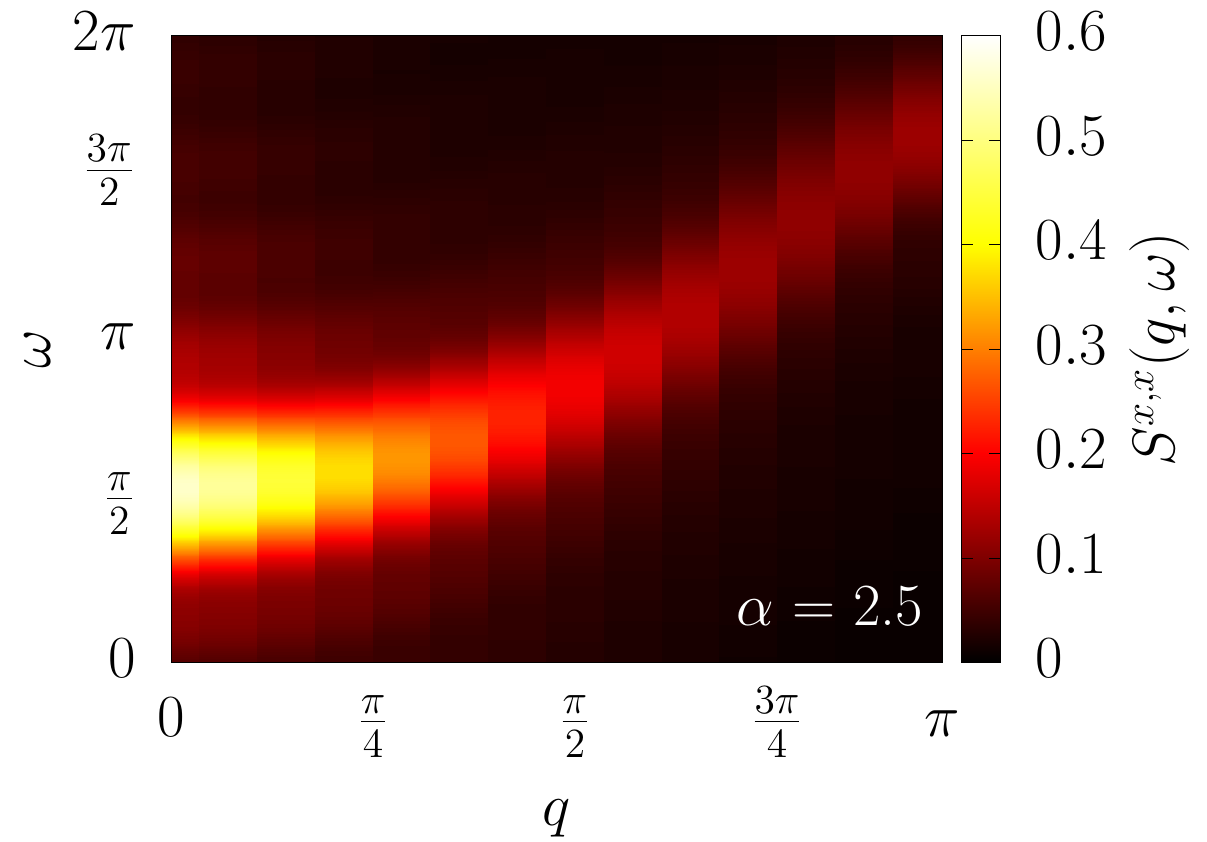}
\includegraphics[scale=0.45]{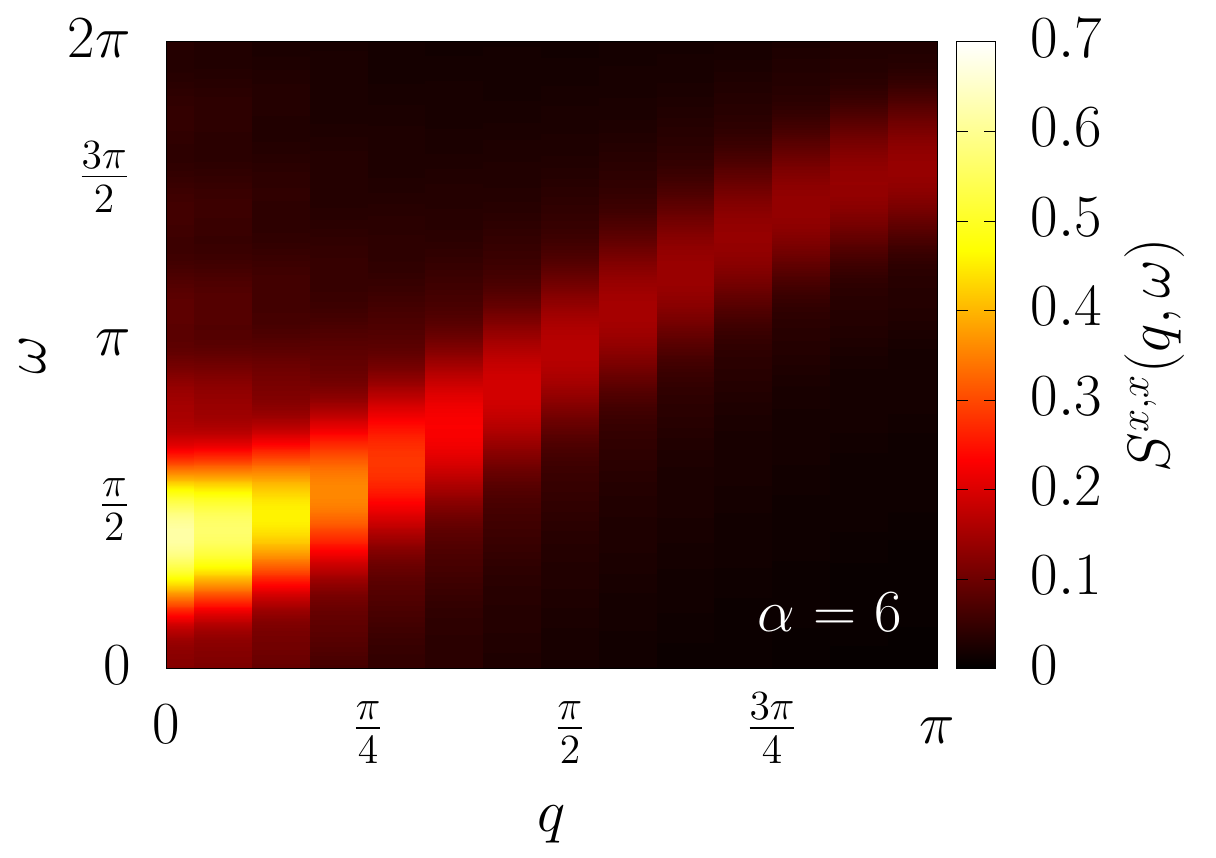}
\caption{{\it DSF for the long range TFIM for the values of $\alpha = 1.5$ (left), $2.5$ (middle) and $6$ (right).}
We see that as the value of $\alpha$ increases, the two particle continuum approaches the cosine form which is expected for the short range TFIM.
 For $\alpha < 2$ the continuum has almost no $\omega$ dependence.
 This can be connected to the presence of excitation confinement in this model, in which the fermionic excitations do not propagate at short times.
 Please note two important points. First, as recently reported, it is expected that the excitations relax at long times \cite{verdel19} but the time range studied in this work is not sufficiently long to see this effect.
 Second, unlike in previous studies \cite{liu19, lerose19,verdel19}, we observe the signatures of confinement in both the DSF and correlators in equilibrium, i.e., without the need of quenching the Hamiltonian. }
\label{DSF_long_range_2}
\end{figure}

\section{Unequal time correlation function for the long range transverse field Ising}\label{appendix:unequal}
\begin{figure}[h!]
\includegraphics[scale=0.4]{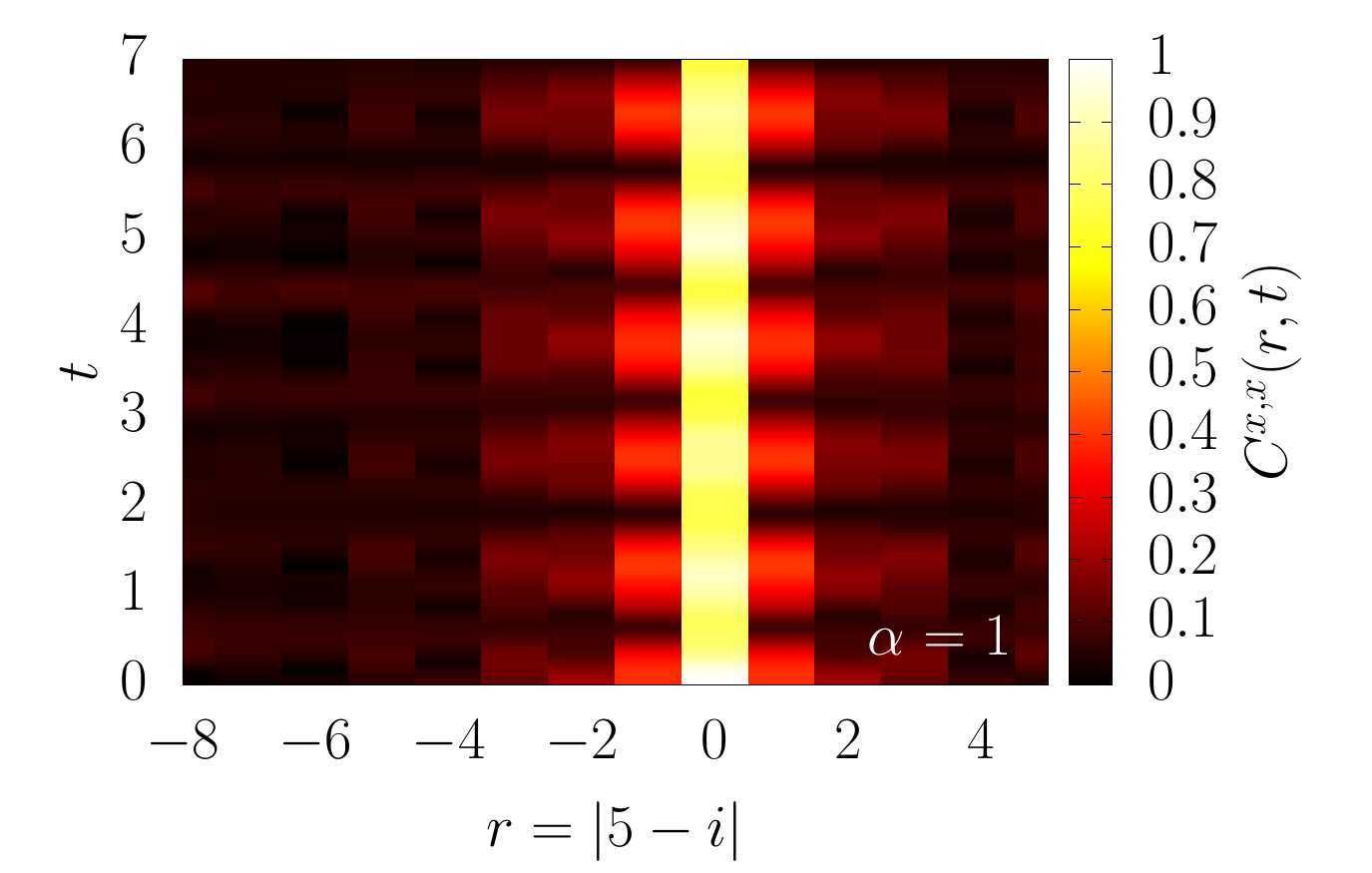}
\includegraphics[scale=0.4]{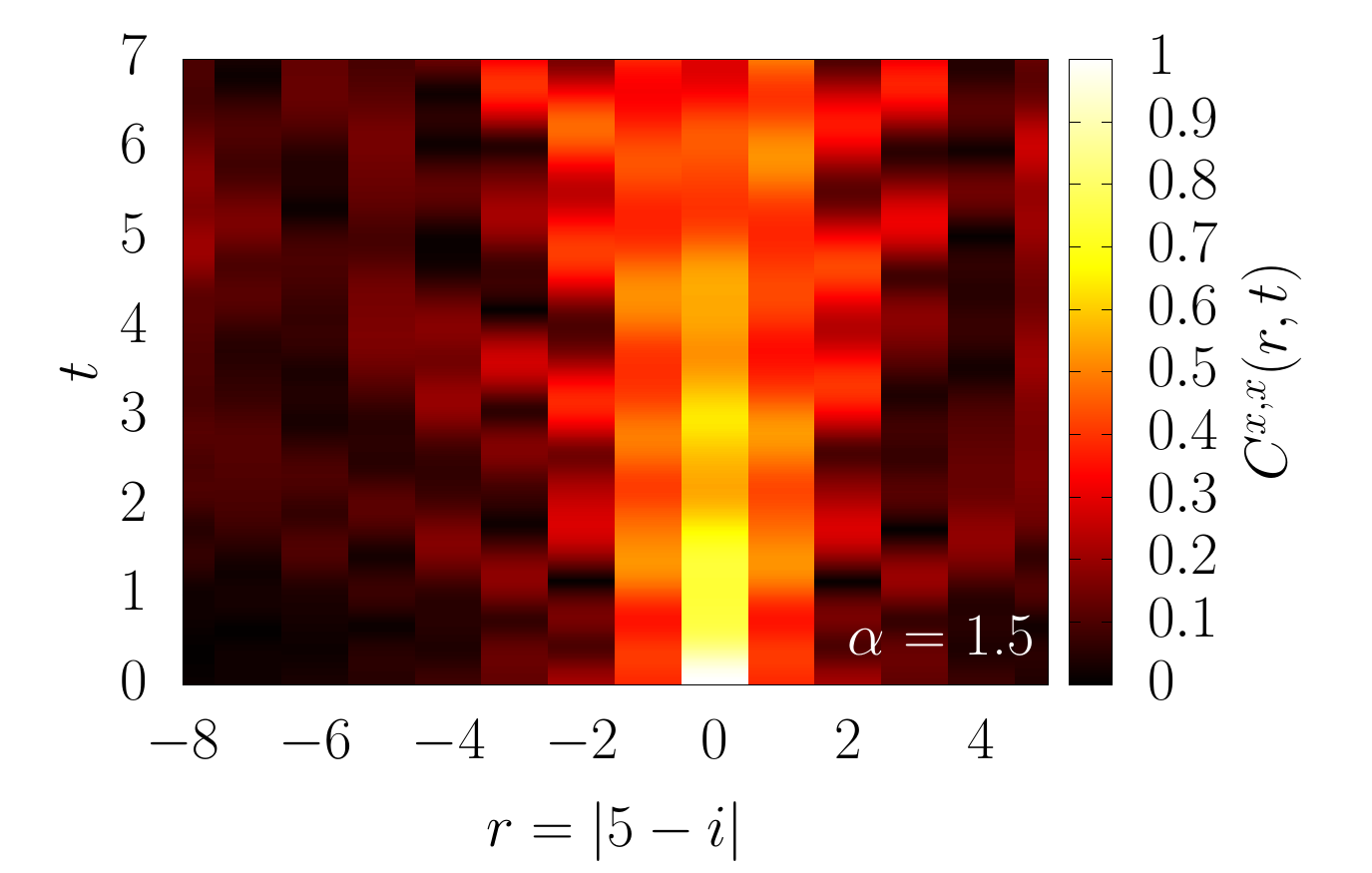}
\includegraphics[scale=0.4]{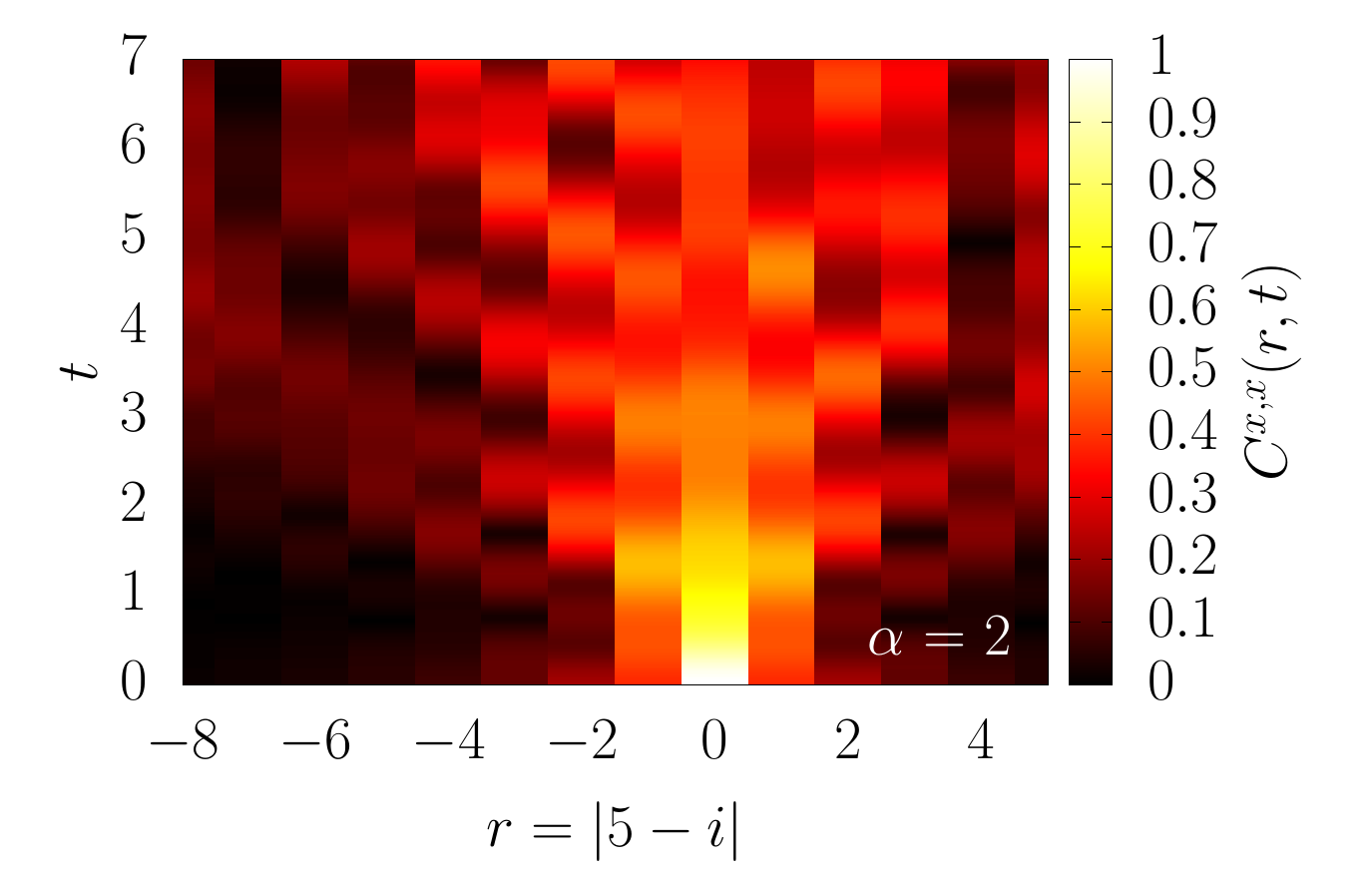}
\caption{\it Correlation function $C^{x,x}_{i,5}(t)$ for the values of $\alpha = 1$ {\it (top)}, $1.5$ {\it (middle)}, and $2$ 
{\it (bottom)}.}
\label{Correlators_long_range}
\end{figure}

\begin{figure}[h!]
\includegraphics[scale=0.4]{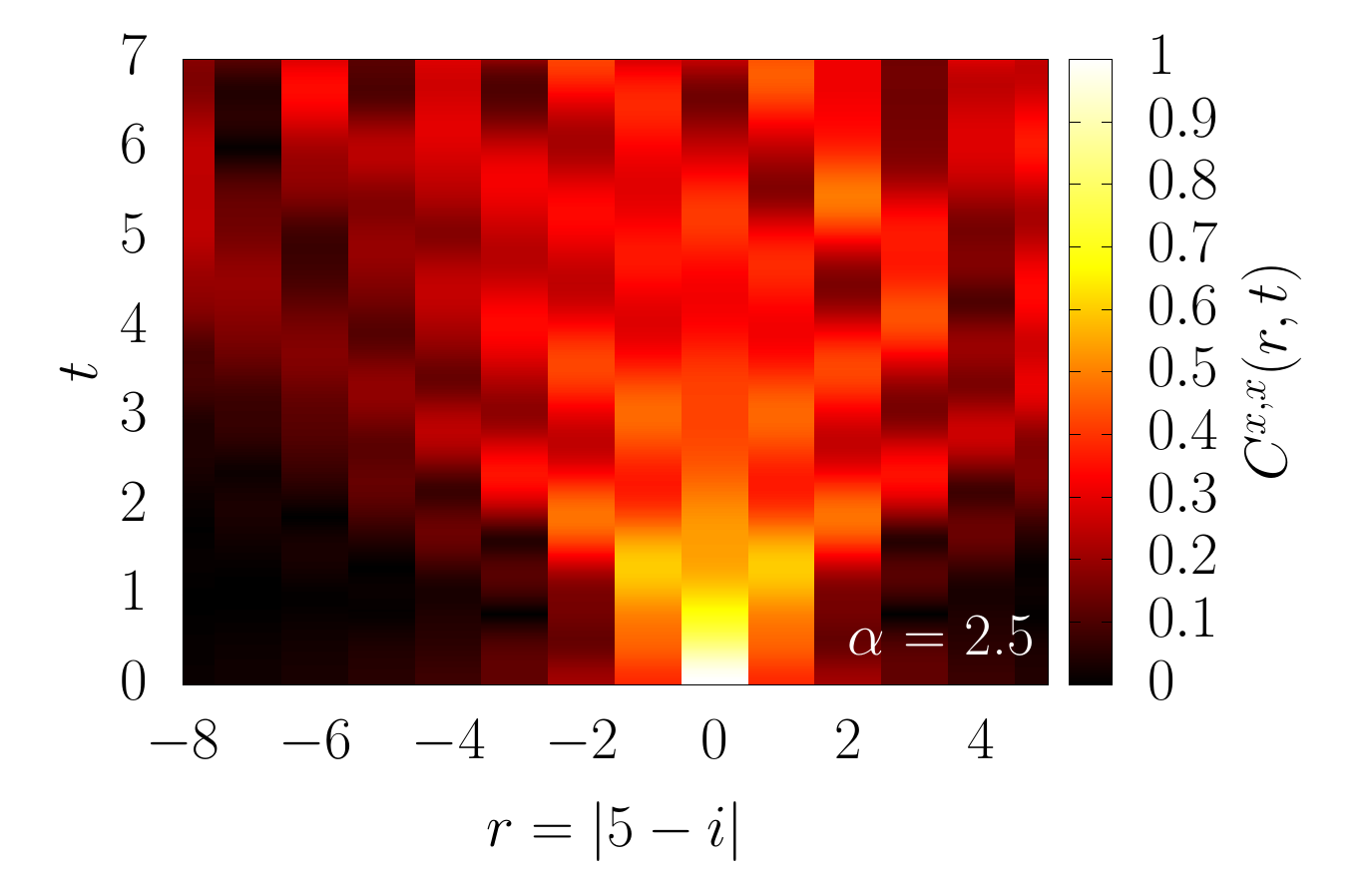}
\includegraphics[scale=0.4]{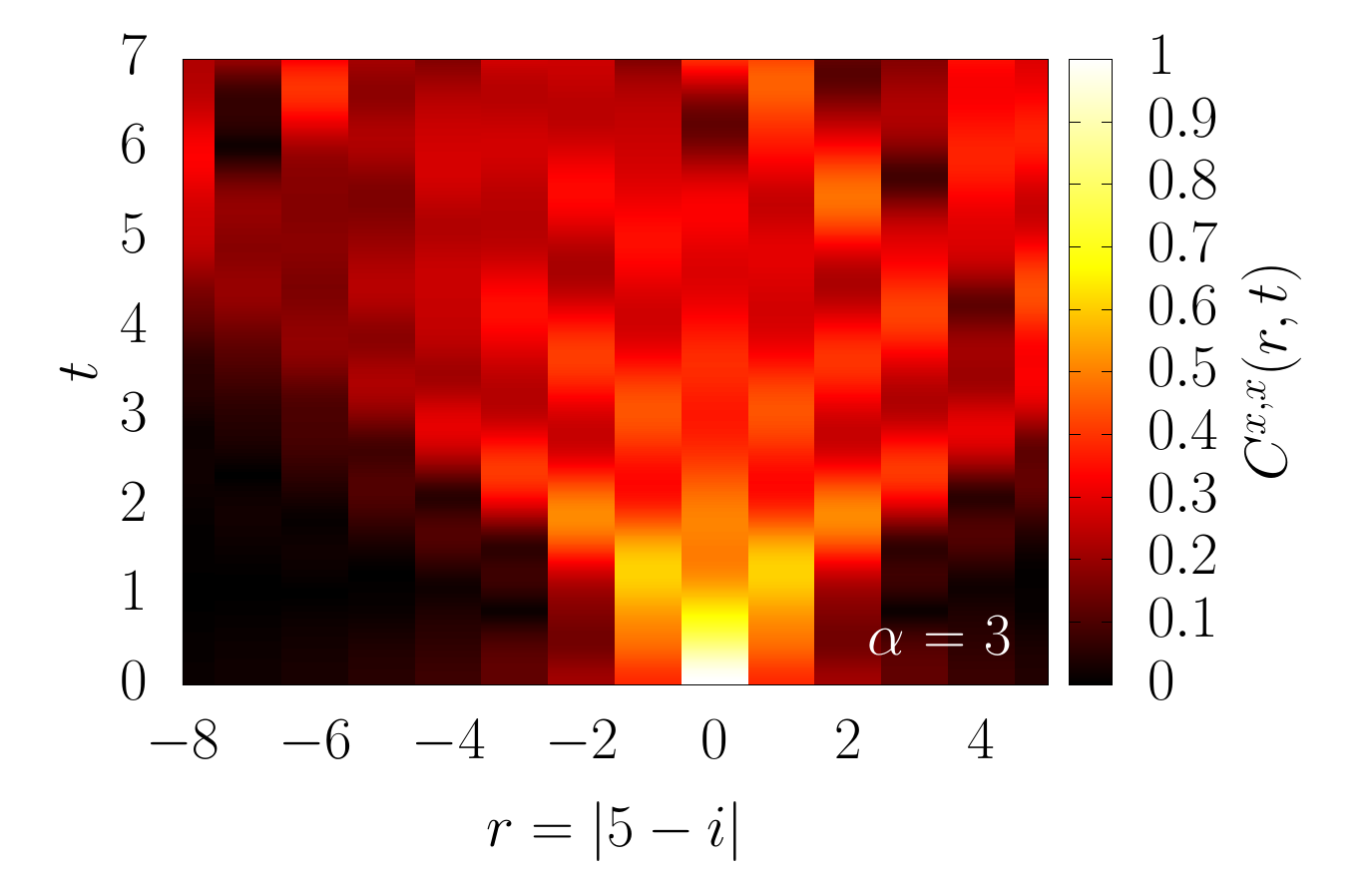}
\includegraphics[scale=0.4]{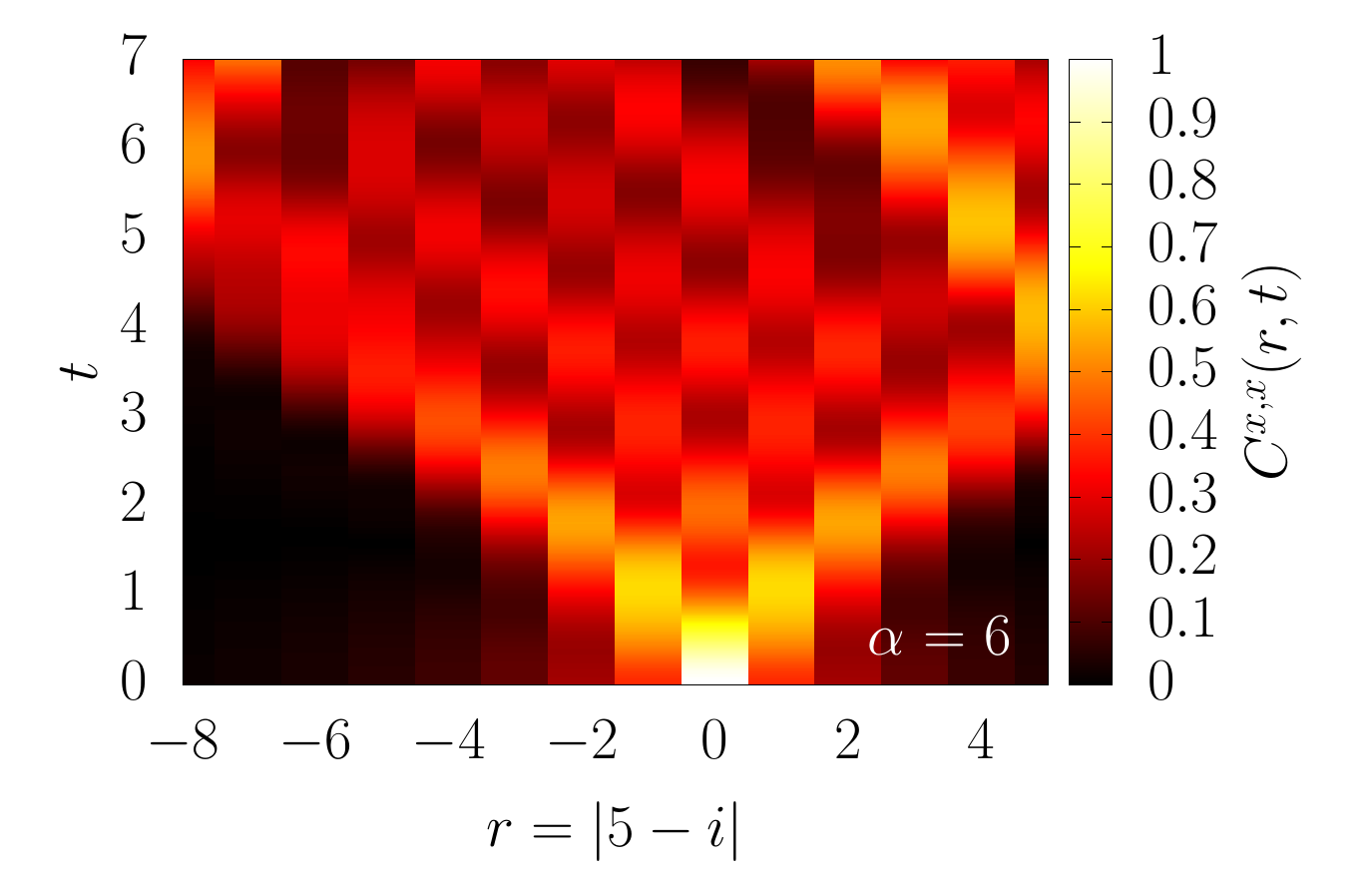}
\caption{\it Correlation function $C^{x,x}_{i,5}(t)$ for the values of $\alpha = 2.5$ {\it (top)}, $3$ {\it (middle)}, 
and $6$ {\it (bottom)}.}
\label{Correlators_long_range_2}
\end{figure}

\section{Numerical results for position fluctuations in the long range TFIM}
\label{ap:num_LR}

\begin{figure}[H]
\centering
\includegraphics[scale=0.6]{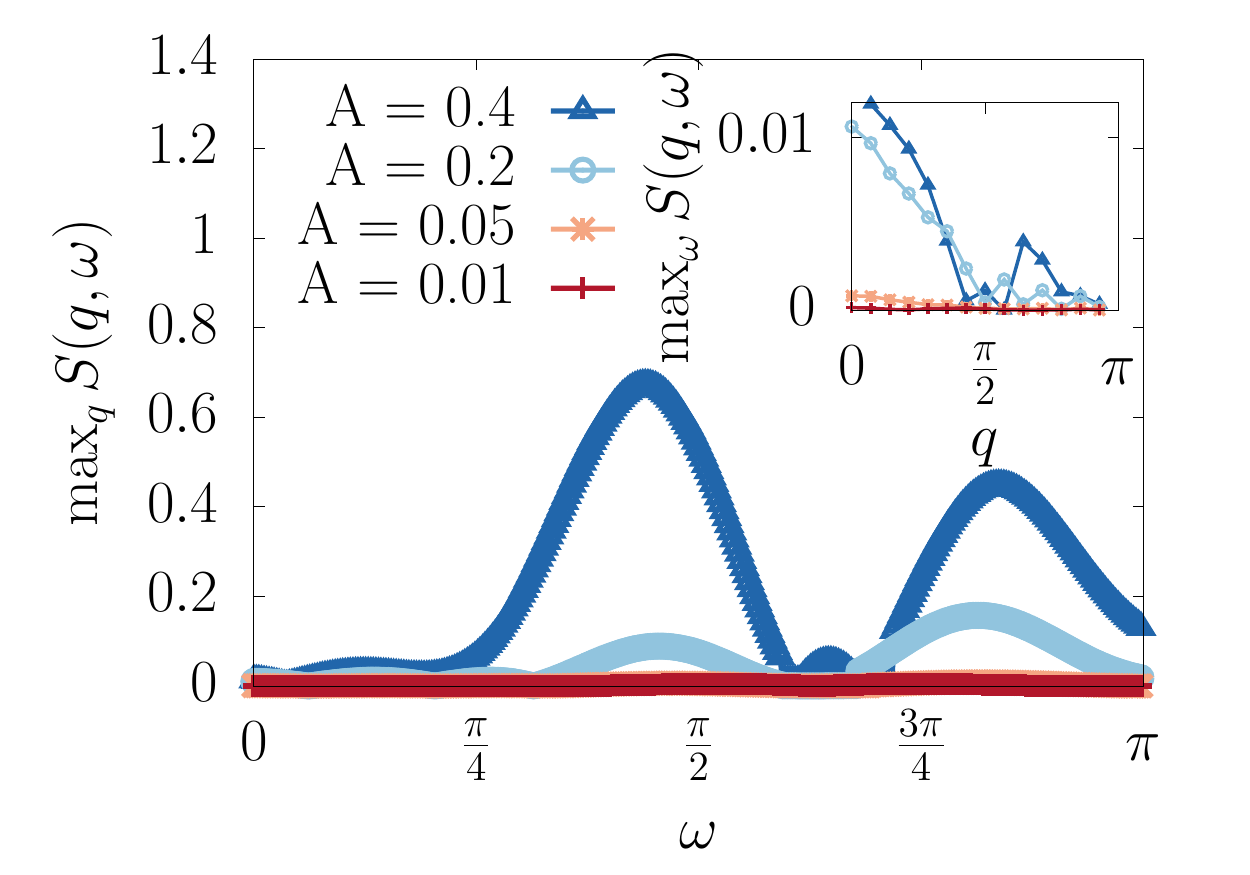}
\includegraphics[scale=0.6]{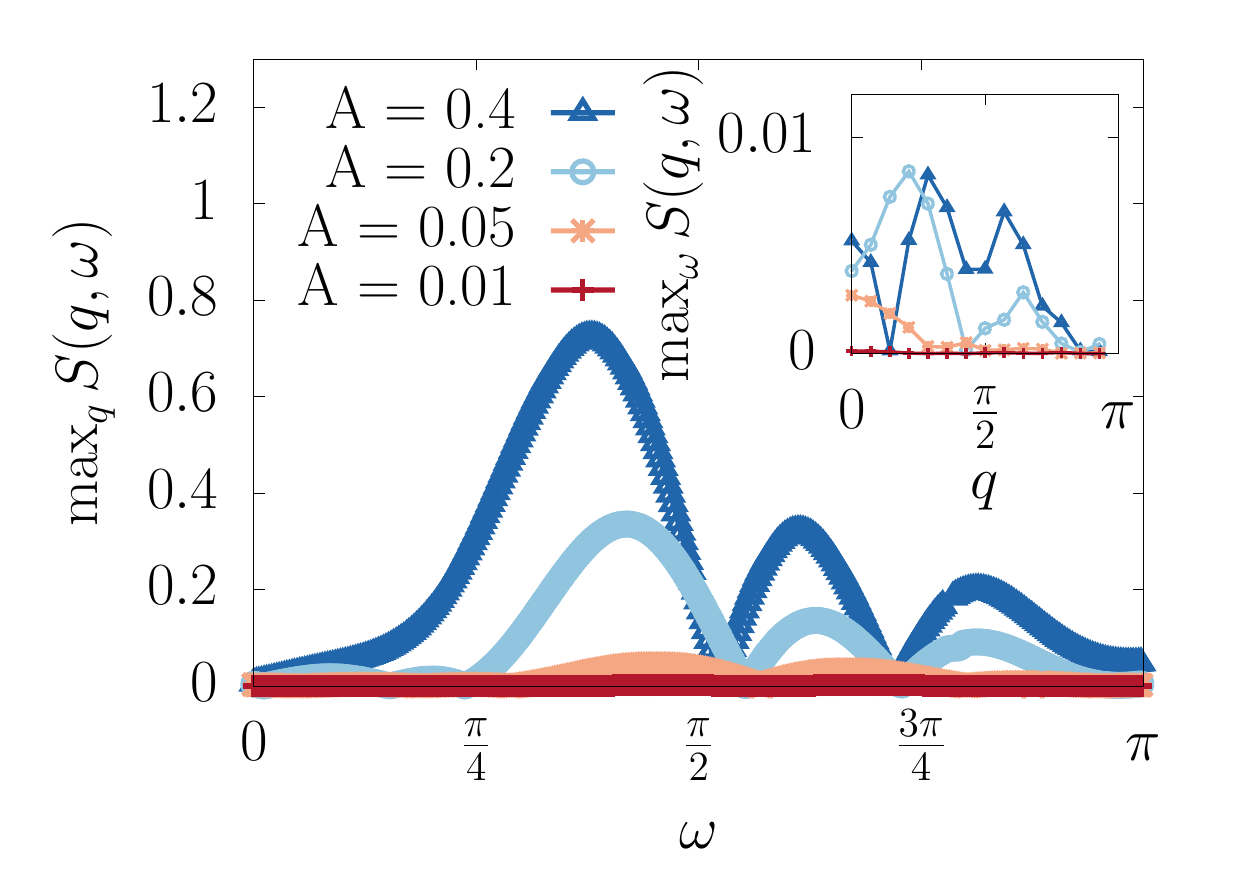}
\includegraphics[scale=0.6]{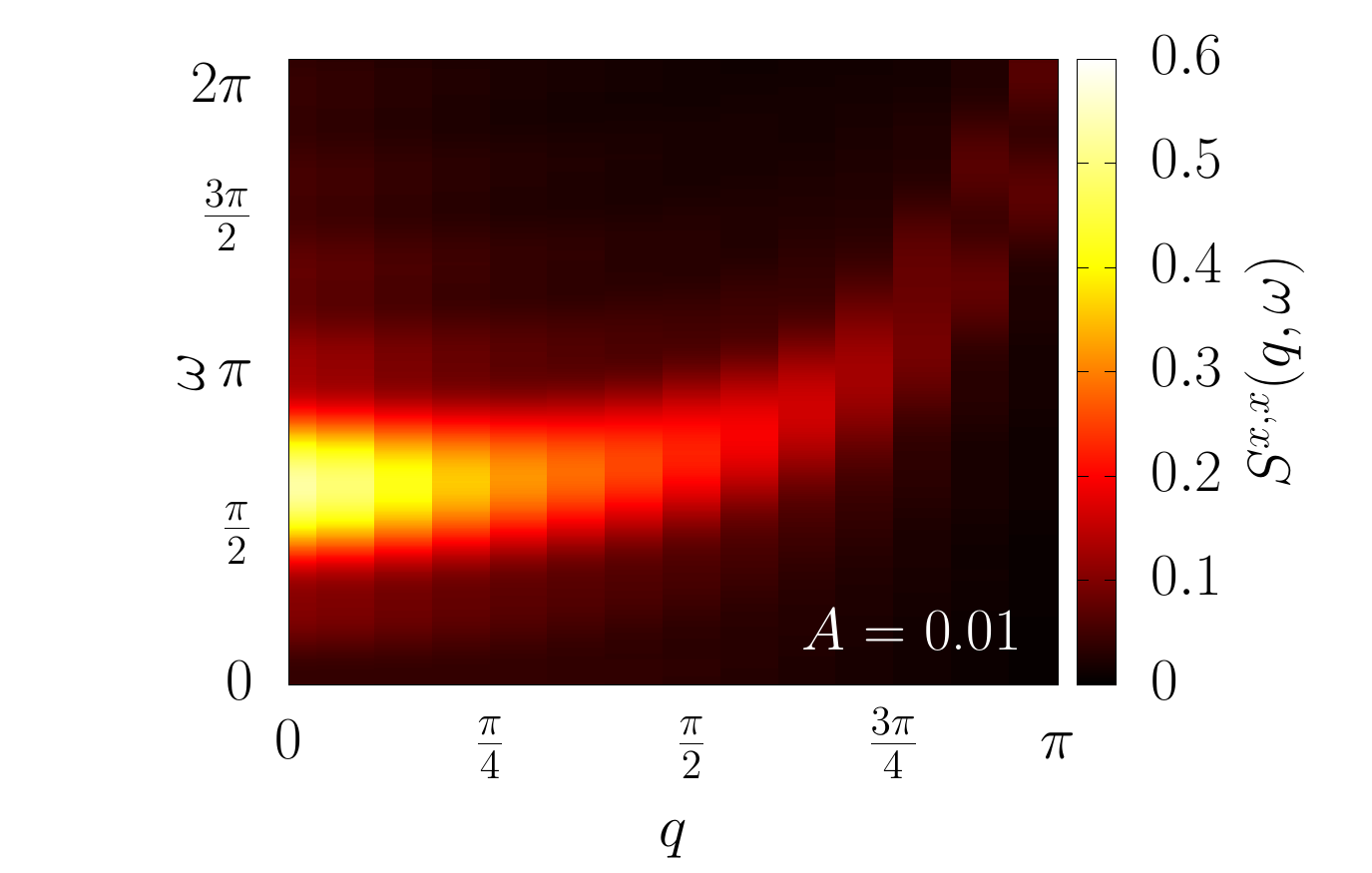}
\includegraphics[scale=0.6]{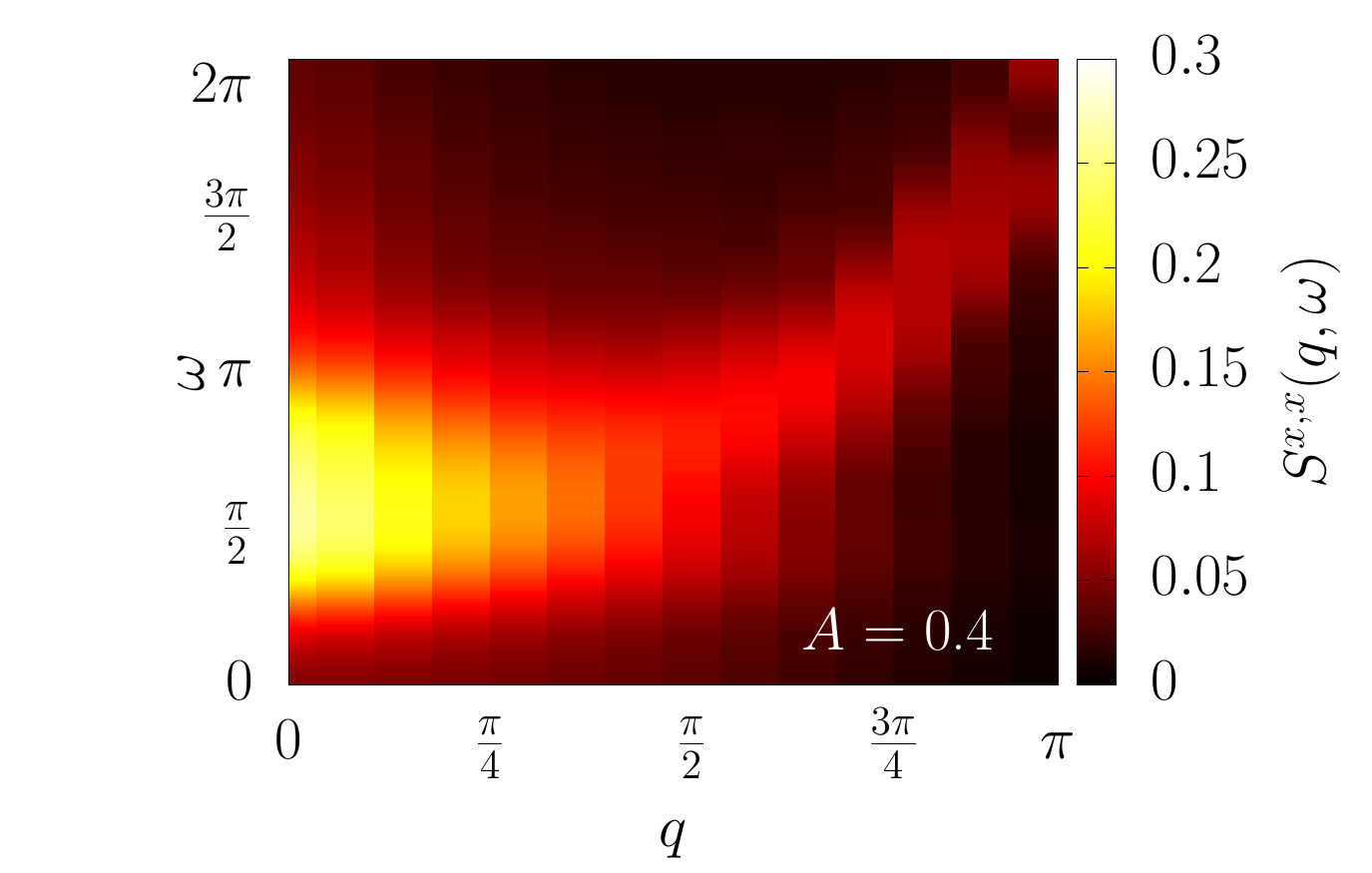}
\caption{{\it Numerical results  for random fields in the long range TFIM.} 
 {\it Top figures:} Maximal error in the DSF for the long range TFIM with random fields, as obtained from full ED, for the cases $\alpha =1.5$ (top left), and $\alpha = 2$ (top right). For imperfection levels within current experimental reach, $A = 0.01$ and $0.05$ the error is small and well behaved on the entire $q-\omega$ domain. For imperfection levels above $0.05$ the DSF is deformed, as evidenced by the maxima in the blue curves. We can compare this results to the bottom figures.
 {\it Bottom figures:} DSF for the long range TFIM in the presence of random fields, for the case $\alpha = 1.5$, $A = 0.01$ {\it (bottom left)} and $A = 0.4$ {\it (bottom right)}. If we compare these DSFs to the clean case (left DSF in Fig.~\ref{DSF_long_range_2}) we observe no discrepancies between the case $A = 0.01$ and the clean case. For the DSF corresponding to an imperfection level of $40\%$, $A = 0.4$ we see that the overall intensity has decreased, and the maxima has broadened considerably. The broadening of the maxima in the DSF is exhibited in the top left figure as strong and broad peaks in the maximal error as a function of frequency.   }
\end{figure}

\end{widetext}

\end{document}